\documentclass[usegraphicx,useAMS,usenatbib]{mn2e}
\usepackage{graphicx}
\usepackage{ulem}
\usepackage{amssymb}
\usepackage{amsmath}
\usepackage{soul}
\usepackage{color}

\def\tilde{~}
\def\fig{Fig.\,}
\def\eq{Eq.\,}
\def\sec{Section~}

\def\rfive{\,R_{500}}

\def\rtwo{\,R_{200}}

\def\Zsun{\rm{\,Z_{\odot}}}
\def\msun{\rm{\,M_{\odot}}}

\def\ks{\rm{\,ks}}

\def\kev{\rm{\,keV}}
\def\cm{\rm{\,cm}}
\def\kms{\rm{\,km/s}}
\def\kpc{\rm{\,kpc}}

\def\Log10{{\rm~Log_{10}}}

\newcommand{\ltsima}{$\; \buildrel < \over \sim \;$}
\newcommand{\simlt}{\lower.5ex\hbox{\ltsima}}
\newcommand{\gtsima}{$\; \buildrel > \over \sim \;$}
\newcommand{\simgt}{\lower.5ex\hbox{\gtsima}}
\title[AC-SPH: impact on the ICM properties in galaxy clusters]{The role of the artificial conductivity in SPH simulations of galaxy clusters: effects on the ICM properties}

\author[V.~Biffi \& R.~Valdarnini]{
  V. Biffi$^{1}${\thanks{E-mail: biffi@sissa.it (VB)}} and
  R. Valdarnini$^{1,2}$
  \\
  $^{1}$SISSA --- Scuola Internazionale Superiore di Studi Avanzati, Via Bonomea 265, I-34136 Trieste, Italy\\
  $^{2}$INFN --- Iniziativa Specifica QGSKY, Via Valerio 2, I-34127 Trieste, Italy
}
\begin{document}

\pagerange{\pageref{firstpage}--\pageref{lastpage}} \pubyear{...}
\maketitle
\label{firstpage}

\begin{abstract}
We study the thermal structure of the intra-cluster medium (ICM) in a
set of cosmological hydrodynamical cluster simulations performed with an SPH
numerical scheme employing an artificial conductivity (AC) term.  We
explore the effects of this term
on the ICM temperature and entropy profiles, thermal distribution,
velocity field and expected X-ray emission.  We find that in adiabatic
runs the artificial conductivity favours (i) the formation of an
entropy core, raising and flattening the central entropy profiles,
in better agreement with findings from Eulerian codes; and (ii)
a systematic reduction of the cold gas component.
In fact, the cluster large-scale structure and
dynamical state are preserved across different runs, but the improved
gas mixing enabled by the AC term
strongly increases the stripping rate of gas from the cold clumps
moving through the ICM.
This in turn reduces the production of turbulence generated
by the instabilities which develop
because of the interaction between clumps and ambient ICM.
We then find that turbulent motions,
enhanced by
the time-dependent artificial viscosity scheme we use,
are rather damped by the AC term.
The ICM synthetic X-ray emission substantially mirrors the changes in
its thermo-dynamical structure, stressing the robustness of the
AC impact. All these effects are softened by the
introduction of radiative cooling but still present, especially a
partial suppression of cold gas.
Therefore, not only the physics accounted for, but also the numerical
approach itself can have an impact
in shaping the ICM thermo-dynamical structure and ultimately in
the use of SPH cluster simulations for cosmological studies.
\end{abstract}

%
%
\section{Introduction}\label{sec:intro}
Within the commonly accepted hierarchical scenario of cosmic structure
formation, clusters of galaxies are the latest and largest
gravitationally-bound systems formed. For this reason, they most
likely still carry the imprints of the underlying cosmological model
driving the expansion of the Universe and represent therefore
interesting targets for cosmological investigations.  During the
formation process, baryonic matter, mainly in the form of gas,
collapses in the halo potential well shaped by the dominating dark
matter (DM) component, cools, fragments and originates the visible
structures we observe, such as galaxies \cite[e.g.][for a recent review]{KB2012}.

Within galaxy clusters, however, the
largest fraction of baryons
is in the form of diffuse hot
plasma --- the intracluster medium (ICM) --- that emits mainly in the
X-rays, with a characteristic temperature which reflects the depth of
the cluster potential well.  Despite this simplistic description of
clusters, observations in several bands (X-ray, radio, optical) have
shown evidences for a more complex thermo-dynamical structure,
characterized for instance by substantial bulk motions, merger-driven
shocks, turbulence, cold gas clumps.  Processes such as gas cooling,
metal diffusion, accretion of pristine gas or smaller haloes, galactic
winds, feedback from active galactic nuclei (AGN), interaction between
AGN jets and the ambient medium can be sources of non-thermal velocity
patterns, local inhomogeneities and hydrodynamical instabilities, that
are highly non-linear phenomena. \\
Hence, an accurate description of
clusters through the modeling of such complex physical processes
necessarily requires a numerical approach and simulations have indeed
proved to be extremely helpful.

Pure N-body codes have been extensively studied and tested and are
capable of properly follow the formation and evolution of DM-only
structures, where gravity is the only force at place.  Instead, the
numerical treatment of the hydrodynamics still presents some
uncertainties, partly reflecting the complexity of the physics involved
and partly due to the computational methods themselves.
\\
In the last decades a number of codes have been designed and employed
to perform hydrodynamical simulations of both cosmological volumes and
isolated cosmic structures, among which clusters of galaxies. Two
different approaches are usually followed: the Lagrangian description
is the base for the mesh-free, particle-based smoothed particle
hydrodynamics \cite[SPH:][]{Rosswog2009,SP10},
while the Eulerian grid-based approach is instead
employed by adaptive mesh refinement methods \cite[AMR:][]{be89,K98,nor99,fry20,TEY02}.

Both methods have been widely used to study in a self-consistent manner the
hydrodynamics that takes place during the ICM evolution.
 However, there is now a general agreement in the literature that the two
numerical methods produce results which
differ significantly in a number of hydrodynamical test cases
\cite[][]{Ag07,Task08,Wad08,Mi09,Read2010,Junk10,Val10,Mc12}.
In particular, in the SPH approach,
the formation  of  Kelvin-Helmholtz  instabilities (KHI) in shear flows
is strongly suppressed at the fluid interfaces \cite[][]{Ag07,Read2010}.
Moreover, there is a significant discrepancy in the level of core entropy found
in simulations of non-radiative clusters performed using the two methods
\cite[][]{Wad08,Mi09},
namely their central entropies are
systematically higher by a factor $\sim2$
in the case of the AMR runs than in the SPH ones.
The origin of these differences are due to two distinct causes which significantly
affect the standard SPH method.
The first problem is the inconsistency of standard SPH in the treatment of
contact discontinuities which in turn suppress the growth of KHI at
fluid interfaces \cite[][]{Ag07,Price2008,Read2010,Val10}.
The other difficulty of classic SPH relies in the sampling errors
associated with the momentum equation \cite[][]{I02,Read2010}.
This error can not be reduced by arbitrary increasing the neighbor number, as
for the standard SPH cubic spline kernel the onset of clumping instability
significantly degrades the convergence rate.

However, the SPH method possesses several properties which render its use
particularly appealing in many astrophysical problems.
For instance, being the method Lagrangian, it has very good conservation
properties and, moreover, is  naturally adaptive.
For this reason, many authors have recently proposed several improvements
in the numerical scheme to solve the problems encountered by standard SPH in handling
fluid instabilities \cite[][]{Price2008,Cha10,HS10,Ab11,MU11,Read12,Garcia12,
valdarnini2012,Pow13,Sai13,Hu14,hopkins2014}.
In particular, \cite{Price2008} suggested to include in the SPH thermal energy
equation  an artificial  conduction (AC) term
with the purpose of smoothing the thermal energy at fluid interfaces and
in turn to  ensure  a smooth entropy transition at
contact discontinuities.
A similar term was introduced by \cite{Wad08} with the aim of
mimicking the effects of  turbulence diffusion, which in SPH is
inhibited by the Lagrangian nature of the method.
The hydrodynamical performances of this approach were later investigated
in detail\cite[][]{valdarnini2012},
indicating a general improvement over
standard SPH in a number of hydrodynamical test problems.
In particular, the final levels of core
entropies in cosmological simulations of galaxy clusters were found consistent
with those  found  using AMR codes.
This result is in accord with previous studies \cite[][]{Wad08,Mi09}, showing
that in standard SPH it is the lack of diffusion and the subsequent damping
of entropy mixing which are responsible for the differences in central entropies
between AMR and SPH galaxy cluster simulations.

Motivated by these findings we investigate here the impact of the new AC-SPH
scheme on the ICM thermodynamical properties of a set of simulated galaxy clusters.
The baseline sample consists of eight different initial conditions for the
simulated clusters, extracted from a larger cosmological ensemble.
 We construct an ensemble of different test cases by performing, for the same set
of initial conditions, simulations with different settings in the hydrodynamical
part of the code.
 The code implements a time-dependent artificial viscosity (AV) scheme and
 we consider runs with different AV parameters as well as simulations  which
 incorporate or not the AC term.
In addition, we consider both non-radiative simulations and runs in which
the gas can cool radiatively.

The principal aim of this investigation is to study in detail
how the proposed hydrodynamical scheme modifies the thermodynamic of
the ICM and in turn its impact on cluster X-ray observational
properties.

The paper is organized as follows.
In Section~\ref{sec:sims} we introduce the main hydrodynamical
modifications of the SPH scheme used to run the various simulations,
and
the construction of the sample of simulated clusters is then described
in Section~\ref{sec:clu}.
In Section~\ref{sec:stat} we provide the statistical tools used to
explore the dynamical state of the clusters, in terms of morphology
and gas velocity field.
Results on how the numerical improvements of SPH impact the final ICM
thermo-dynamical properties are presented and extensively discussed in
Section~\ref{sec:results}.
Finally we summarize our findings and draw our conclusions in
Section~\ref{sec:discuss}.

\section{Simulations}\label{sec:sims}
%
We now outline here the fundamentals of the hydrodynamical method
--- we refer the reader to
\cite{Rosswog2009} and \cite{Price2012},
for comprehensive reviews.
%
\subsection{Basic equations}\label{subsec:method}
According to the SPH method, the fluid is described by a set of
particles with mass $m_i$, velocity $\vec v_i$, density $\rho_i$,
thermal energy per unit mass $u_i$ and entropy $A_i$.  The latter is
related to the particle pressure $P_i$ by
$P_i=A_i\rho_i^{\gamma}=(\gamma-1) \rho_i u_i$, where $\gamma=5/3$ for
a mono-atomic gas.  The density at the particle position $\vec r_i$ is
given by
 \begin{equation}
 \rho_i=\sum_j m_j W(|\vec r_{ij}|,h_i),
    \label{rho.eq}
 \end{equation}
where $W(|\vec r_i-\vec r_j|,h_i)$ is the cubic spline kernel
that has compact support and is zero for $|\vec r_i-\vec r_j|\geq2h_i$
\citep{Price2012}. The sum in Eq. (\ref{rho.eq}) is over a finite number of
particles and the smoothing length $h_i$  is a variable
that is implicitly determined by
 \begin{equation}
h_i=\eta (m_i/\rho_i)^{1/D}~,
  \label{hzeta.eq}
 \end{equation}
so that  for the adopted spline
$N_{sph}= 4 \pi \rho_i (2h_i)^3 /3={4 \pi (2 \eta)^3 }/{3}$
is the number of neighboring particles  of particle $i$ within a radius
$2h_i$.
The Euler equation can be derived by a Lagrangian formulation
\citep{Rosswog2009,Price2012}:
   \begin{equation}
  \frac {d \vec v_i}{dt}=-\sum_j m_j \left[
  \frac{P_i}{\Omega_i \rho_i^2}
  \vec \nabla_i W_{ij}(h_i) +\frac{P_j}{\Omega_j \rho_j^2}
   \vec \nabla_i W_{ij}(h_j)
\right]~,
  \label{fsph.eq}
   \end{equation}
with the coefficients $\Omega_i$ being given by
   \begin{equation}
   \Omega_i=\left[1-\frac{\partial h_i}{\partial \rho_i}
   \sum_k m_k \frac{\partial W_{ik}(h_i)}{\partial h_i}\right]~.
    \label{fh.eq}
   \end{equation}

   To properly treat the effects of shocks, the term in the rhs of
   Eq. (\ref{fsph.eq}) must be complemented by the viscous force
   \begin{equation}
   \left (\frac {d \vec v_i}{dt}\right )_{AV}=-\sum_i m_j \Pi_{ij} \vec \nabla_i \bar W_{ij}~,
    \label{fvis.eq}
   \end{equation}
   where the term $\bar W_{ij}=
   \frac{1}{2}(W(r_{ij},h_i)+W(r_{ij},h_j))$ is the symmetrized kernel
   and $\Pi_{ij}$ is the AV tensor.

Following \cite{Mon97}, the latter is written as
   \begin{equation}
\Pi_{ij} =
 -\frac{\alpha_{ij}}{2} \frac{v^{AV}_{ij} \mu_{ij}} {\rho_{ij}} f_{ij}~,
  \label{pvis.eq}
 \end{equation}
where a pair of subscripts denotes arithmetic averages, and the signal
 velocity $v^{AV}_{ij}$ is given by
   \begin{equation}
v^{AV}_{ij}= c_i +c_j - 3 \mu_{ij}~,
  \label{vsig.eq}
 \end{equation}
 with $c_i$ being the sound velocity, $\mu_{ij}= \vec v_{ij} \cdot
 \vec r_{ij}/|r_{ij}|$ if $ \vec v_{ij} \cdot \vec r_{ij}<0$ but zero
 otherwise and $\vec v_{ij}= \vec v_i - \vec v_j$.  The parameter
 $\alpha_i$ regulates the amount of AV, and $f_i$ is a damping factor
 which is introduced to reduce the strength of AV when shear flows are
 present.  For this term \cite{Bals95} proposed the expression
   \begin{equation}
  f_i=\frac {|\vec \nabla \cdot \vec v|_i}
  {|\vec \nabla \cdot \vec v|_i+|\vec \nabla \times \vec v|_i}~,
   \label{fdamp.eq}
 \end{equation}
 where $(\vec \nabla \cdot \vec v)_i$ and $(\vec \nabla \times \vec
 v)_i$ are the standard SPH estimates for divergence and curl.

For the viscosity parameter $\alpha_i$,
it is assumed $\alpha_i=\mathrm{constant}\equiv \alpha_0=1$, in standard SPH,
whilst \cite{MM97}
suggested to let the parameter to vary with time so that the strength
of the AV is reduced away from shocks. In such a case the
time-evolution of $\alpha_i$ is given by
\begin{equation}
 \frac {d \alpha_i}{dt} =-\frac{\alpha_i-\alpha_{min}}{\tau_i} +{\tilde S}_i~,
  \label{alfa.eq}
\end{equation}
where
\begin{equation}
  \tau_i=\frac{h_i}{c_i ~l_d}
    \label{tau.eq}
  \end{equation}
  is a decay time scale which is controlled by the dimensionless decay
  parameter $l_d$, $\alpha_{min}$ is a floor value and $S_i$ is a
  source term.  This is constructed so that it increases in the
  presence of shocks:
   \begin{equation}
 {\tilde S}_i=f_i S_0 {max}\big(-\big(\vec \nabla \cdot \vec v\big)_i,0\big)
(\alpha_{max}-\alpha_i)\equiv S_i (\alpha_{max}-\alpha_i).
    \label{salfa.eq}
   \end{equation}
\pagebreak

   Here $S_0$ is unity for $\gamma=5/3$.  Recommended values for the
   parameters $\alpha_{max},\alpha_{min}$, and $l_d$ are $1.5,0.05$,
   and $0.2$, respectively \citep{Rosswog2009}.  In principle, one can
   reduce the presence of AV by decreasing $\tau_i$ as much as
   possible.  Nonetheless, a lower limit to the timescale $\tau_i$ is
   set by the minimum time taken to propagate through the resolution
   length $h_i$, so that the value $l_d=1$ sets an upper limit to the
   parameter $l_d$.

   For mild or weak shocks and very short decaying timescales, the
   peak value of $\alpha_i$ at the shock front might be however below
   the value necessary to avoid inconsistencies in the integration.
   In order to maintain the same shock resolution capabilities when
   $l_d\geq0.2$, a prefactor $\xi$ is introduced in
   Eq.~(\ref{salfa.eq}) by substituting $\alpha_{max}$ with
   $\alpha_{max}\rightarrow \xi \alpha_{max}$
   (\citealp{valdarnini2011}; hereafter V11).  The correction factor
   $\xi$ was calibrated in a number of hydrodynamical tests, obtaining
   $\xi=(l_d/0.2)^{0.8}$ for $ l_d\geq0.2$~(V11).

   In the following, simulation runs performed according to the
   standard AV scheme will be denoted by AV$_0$, whereas the labels
   AV$_2$ and AV$_5$ will indicate simulations employing a
   time-dependent AV scheme with parameters
   $\{\alpha_{min},\alpha_{max},l_d\} = \{0.1,1.5,0.2\}$ and
   $\{0.01,1.5,1\}$, respectively.

   For the number of neighbors $N_{sph}$ typical values lie in the range
   $N_{sph}\sim 33-50$, here we use $N_{sph}=33 ~(\eta \sim 1.06) $.
   Setting $N_{sph}=64$ would have improved the accuracy in density estimates
   below a  few per cent \cite[see Table 3 of][]{valdarnini2012} at the price of doubling
   the computational cost of SPH summations.

%
\subsection{The artificial conductivity scheme}\label{subsec:ac}
In the SPH code we use, we also adopt
an entropy-conserving
approach \citep{SH02}, in which
entropy is generated at a rate
   \begin{equation}
  \frac {d A_i}{dt} =\frac{\gamma-1}{\rho_i^{\gamma-1}}\{
   Q_{AV} +Q_R +Q_{AC}\}~,
    \label{aen.eq}
   \end{equation}
where the terms in brackets refer to different sources.
 In particular,  $Q_{AV}$ is the source term due to viscosity:
   \begin{equation}
  Q_{AV} =  \left ( \frac {d u_i}{dt} \right)_{AV} =
  \frac{1}{2}
  \sum_j m_j \Pi_{ij} \vec v_{ij}\cdot \nabla_i \bar W_{ij}~.
    \label{avis.eq}
   \end{equation}

The term $Q_{AC}$ in Eq. (\ref{aen.eq}) represents an artificial conduction
term and has been introduced by \cite{Price2008} with the purpose of avoiding
inconsistencies in the treatment of thermal energy at contact discontinuities.
 This term can be written as
   \begin{equation}
  \left ( \frac {d u_i}{dt} \right)_{AC} =
\sum_j \frac{m_j v^{AC}_{ij}}{\rho_{ij}}
\left[ \alpha^C_{ij}(u_i-u_j) \right ] \vec {e_{ij}}\cdot \vec {\nabla_i}
 \bar W_{ij}~,
  \label{duc.eq}
   \end{equation}
where $ v^{AC}_{ij}$ is the signal velocity,
$\vec e_{ij} \equiv \vec r_{ij}/r_{ij}$, and $\alpha^C_{i}$ is the AC parameter.
 In analogy with the AV scheme, the AC parameter evolves in time according to
 \begin{equation}
\frac {d \alpha^C_i}{dt} =-\frac{\alpha^C_i-\alpha^C_{min}}{\tau^C_i} +{S^C}_i~,
    \label{alfac.eq}
 \end{equation}
where $\tau^C_i={h_i}/{0.2 c_i }$, and  the source term is given by
 \begin{equation}
{S^C}_i= f_C h_i\frac{|\nabla^2 u_i|}{\sqrt{u_i+\varepsilon}}
\left(\alpha^C_{max}-\alpha^C_i\right).
    \label{salfac.eq}
 \end{equation}

Here the Laplacian is calculated according to
 \begin{equation}
\nabla^2 u_i=2\sum_j m_j \frac{u_i-u_j}{\rho_j}\frac{\vec e_{ij}\cdot
\vec {\nabla} W_{ij}}{r_{ij}}~.
 \label{udii.eq}
   \end{equation}

In what follows we set $f_C=1$, $\alpha^C_{min}=0$  and $\alpha^C_{max}=1.5$,
$\varepsilon=10^{-4} u_i$. For the signal velocity  we use
 \begin{equation}
v^{AC}_{ij} = |(\vec v_i-\vec v_j)\cdot \vec r_{ij}|/r_{ij}~.
 \label{vsgv.eq}
 \end{equation}
This choice guarantees the absence of thermal diffusion for
self-gravitating systems in hydrostatic equilibrium
and its reliability has been tested in a number of hydrodynamical
tests \citep{valdarnini2012} in which gravity is present.

Finally, for the cooling runs, the term $Q_{R}= -\Lambda(\rho_i,T_i)$
--- $T_i$ being the particle temperature --- accounts for the radiative
losses.  For these simulations, the physical modeling of the gas
includes radiative cooling, star formation, energy feedback, and metal
enrichment that follow from supernova explosions,
\cite[see][for a detailed description of the implemented procedures]{V06}.

The code including the AC modification will be also referred to as AC-SPH, in the
following.

\begin{table*}
\centering
\caption{\label{clu.tab}%
Main cluster properties and simulation parameters of the considered
sample. From left to the right:
parent cosmological sample,
cluster id,
cluster mass $M_{200}$ (within $R_{200}$) in units of $h^{-1} \msun$,
cluster radius $R_{200}$ is units of Mpc,
number of gas ($N_{gas}$) and dark matter ($N_{DM}$) particles
inside the initial high-resolution sphere,
mass of the gas ($m_{gas}$) and dark matter ($m_{DM}$) particles in $\msun$,
gravitational softening parameter for the gas in kpc.}
\begin{tabular}{ccccccccc}
\hline
 sample & index &$M_{200}[h^{-1}\msun]$ &
$R_{200}$\,[Mpc]  & $N_{gas}$ & $N_{DM}$ & $m_{gas}[\msun]$ &
 $m_{dm}[\msun]$ & $\varepsilon_{gas}$\,[kpc]   \\
\hline
  ${\mathrm S_8}$ & $1$ & $6\cdot10^{14}$ &  $1.96$ &
$220175 $ &$220144 $ & $3\cdot10^9$& $1.5\cdot10^{10}$ & $25.4$  \\
  ${\mathrm S_4}$ & $5$ &  $2.4\cdot10^{14}$ &  $1.45$ &
$ 221007$ &$ 220976$ & $2\cdot10^9$& $1.1\cdot10^{10}$ & $22.2$ \\
  ${\mathrm S_4}$ &$11$ & $5.7\cdot10^{14}$ &  $1.94$ &
$ 220759$ &$220728 $ & $1.5\cdot10^9$& $7.6\cdot10^{9}$ & $20$  \\
  ${\mathrm S_4}$  &$16$ &  $2\cdot10^{14}$ &  $1.37$ &
$ 221167$ &$ 221136$ & $1.7\cdot10^9$& $8.7\cdot10^{9}$ & $21$ \\
  ${\mathrm S_4}$ &$19$ & $7\cdot10^{14}$ &  $2.07$ &
$ 219631$ &$219600 $ & $1.7\cdot10^9$& $8.7\cdot10^{9}$ & $21$ \\
  ${\mathrm S_2}$ &$13$ & $2.4\cdot10^{14}$ &  $1.44$ &
$ 221823$ &$221792 $ & $6.4\cdot10^8$& $3.3\cdot10^{9}$ & $15.1$  \\
  ${\mathrm S_2}$ &$105$ & $6\cdot10^{13}$ &  $0.91$ &
$ 221391$ &$221360 $ & $2.5\cdot10^8$& $1.3\cdot10^{9}$ & $11.1$  \\
  ${\mathrm S_2}$ &$110$ & $9.4\cdot10^{13}$ &  $1.05$ &
$ 222111$ &$222080 $ & $2.5\cdot10^8$& $1.3\cdot10^{9}$ & $11.1$ \\
\hline
\end{tabular}
\end{table*}
\section{Sample construction }\label{sec:clu}
To construct our ensemble of hydrodynamical cluster simulations we use
a baseline sample consisting of eight different initial conditions for
the simulated clusters.
These objects have been extracted from a set of cosmological simulations,
for which we
assume a standard $\Lambda$CDM cosmological model with present
matter density $\Omega_\mathrm{m}=0.3$, cosmological constant density
parameter $\Omega_\mathrm{\Lambda}=0.7$, baryonic density
$\Omega_\mathrm{b}=0.0486$, power spectrum normalization
$\sigma_\mathrm{8}=0.9$, primeval power spectrum index $n=1$ and
Hubble constant $H_0=70\equiv 100h$\,km\,s$^{-1}$\,Mpc$^{-1}$.

In order to construct the cluster subsample we first run a dark matter
only simulation with comoving box size $L_2=200h^{-1}$\,Mpc.  The dark
matter halos are then identified at $z=0$ using a friends-of-friends
(FoF) algorithm, so as to detect overdensities in excess of
$\sim200 \Omega_\mathrm{m}^{-0.6}$ within a radius $R_{200}$.
The corresponding mass is defined as $M_{200}$, where
\begin{equation}
  M_{\Delta}= (4 \pi/3) \, \Delta\, \rho_\mathrm{c} \, R_{\Delta}^3
\end{equation}
denotes the mass contained in a sphere of radius $R_{\Delta}$
with mean density $\Delta$ times the critical density
$\rho_\mathrm{c}$.

The halos are then sorted according to the value of their mass
$M_{200}$ to generate a catalog of $N_2=120$ simulated clusters,
which have been then re-simulated individually using
both the standard and
the entropy-conserving AC-SPH code described in Section~\ref{sec:sims}.
The re-simulations were
performed using a zoom-in method (see V11 for more details) so that at
the initial redshift each cluster comprises $\sim 220,000$ gas and
dark matter particles within a sphere of comoving radius $\propto
R_{200}$.
The gravitational softening parameter of the particles is set
according to the scaling $\varepsilon_i \propto m_i^{1/3}$, where
$m_i$ is the mass of particle $i$. The relation is normalized by
$\varepsilon_i =15~( m_i/6.2\cdot10^8 \msun)^{1/3}$\,kpc.
The set of simulated clusters
constructed in this way is referred to as sample $S_2$.

With a similar procedure, we
generate the cluster samples $S_4$ and $S_8$,
from cosmological simulations with box
sizes $L_4=400h^{-1}$\,Mpc and $L_8=800h^{-1}$\,Mpc. The
$S_8$ and $S_4$ sets consist of $N_8=10$ and $N_4=33$ clusters, respectively,
and have been constructed such that the threshold in mass for the $S_4$ haloes
is greater than the maximum $M_{200}$ mass of the $S_2$ sample,
and a similar criterion holds for $S_8$, with respect to $S_4$.
The final catalog ($S_{all}$) is built combining the three aforementioned ones.

The eight clusters, whose initial conditions are used for the present analysis,
have been finally selected as a representative subsample
of the mass range and dynamical states of the whole set $S_{all}$
(similarly to V11).

The main properties of these clusters are reported in Table~\ref{clu.tab}.
%
\section{Statistical tools}\label{sec:stat}
In this section we describe the statistical tools used to measure the
cluster morphology and the spectral properties of the ICM velocity
field. The former is quantified using the power ratios and the
centroid shifts of the simulated cluster, whereas the longitudinal and
solenoidal components of the velocity power spectrum are analyzed
separately
to obtain statistical measurements of the ICM velocity field.
The implementation of the spectral analysis is the same as in
V11, to which we refer for more details.
\subsection{Indicators of the cluster dynamical state}
\label{subsec:pwr}
We quantify the dynamical state of the simulated clusters using two
methods: the power ratio \citep{Bu95} and the centroid shift
\citep{Mo93,OH06}.  In the power ratio method, the projected X-ray
surface brightness $\Sigma_X(\vec x)$ is the source term of the pseudo
potential $\Psi(\vec x)$ which satisfies the 2-D Poisson equation. A
multipole expansion of the solution gives the moments
 \begin{equation}
\alpha_m = \int_{R^{'}\leq R}  d ^2 x^{'}  \Sigma_X(\vec x^{'}) {R^{'}}^m
\cos(m\varphi^{'})
\label{am}
 \end{equation}
and
 \begin{equation}
\beta_m = \int_{R^{'}\leq R}  d ^2 x^{'}  \Sigma_X(\vec x^{'}) {R^{'}}^m
\sin(m\varphi^{'}),
\label{bm}
 \end{equation}
 where the integral extends over a circular aperture of radius
 $R\equiv R_{ap}$ (i.e. the aperture radius). Then the power ratios are defined
 according to
 \begin{equation}
\Pi^{(m)} (R_{ap}) = \log_{10} (P_m/P_0)~,
\label{pm}
 \end{equation}
where
\begin{eqnarray}
P_m (R_{ap})& =& {1 \over 2 m^2} (\alpha_m^2 + \beta_m^2) ~~ m>0,\\
P_0 &= &[ \alpha_0 \ln(R_{ap}/{\rm kpc}) ]^2.
\end{eqnarray}
In the following, we will use the quantity $\Pi_3(R_\mathrm{ap})$ as
an indicator of the cluster dynamical state since it provides an
unambiguous detection of asymmetric structure. Moreover, we minimize
projection effects by introducing the average quantity $\bar
{\Pi}_3(R_{ap})=\log_{10} (\bar {P}_3/\bar {P}_0)$, where $\bar {P}_m$
is the {\it rms}
average of the moments $P_m$ evaluated along
the three orthogonal lines of sight.

Another useful measure commonly employed to quantify cluster
morphology is the centroid shift. This method is based on the
displacement between the cluster centre and the centroid calculated
using the first moment of the X-ray surface brightness.
Following \cite{BO10}, we then calculate the centroids within 9
apertures with radii $r=0.1 \times i \times R_{ap}$, with $i=2,3,
\ldots,10$.  We excise from the computation a central region of radius
$r=0.05R_{ap}$. The centroid shift is then defined as the standard
deviation of the centroids in units of $R_{500}$:
\begin{equation}
  w = \left [ \frac{1}{N-1} \sum (\Delta_i-< {\Delta}>)\right ]^{1/2}
  \times \frac{1}{R_{500}},
  \label{wx}
\end{equation}
where $\Delta_i$ is the centroid corresponding to the $i-$th aperture.

For the present analysis, we evaluate both power ratios and centroid
shifts for the regions enclosed within three typical aperture radii:
$R_{ap}=R_{2500}$,~$R_{500}$,~$R_{200}$.

To compute the centroid shift a common choice for the coordinate
origin is the X-ray peak of the surface brightness. However, for the
numerical study undertaken here, this choice is not free of possible
ambiguities.  Because of the presence of the new AC term in the
thermal energy equation, there might be significant variations in the
final gas thermal properties between simulations performed for the
same test cluster, depending on whether the AC term is included
or not.  This in turn might imply variations in the
location of the
gas emission peak between different runs and a difficulty to
consistently compare quantities for which their definition depends on
the choice of the origin.

In order to avoid this difficulty and to unambiguously identify the same
emission peak among different runs, we have used the following
procedure to determine the cluster center. The latter is defined as
the maximum of the gas density,
calculated iteratively as the
center of mass of the gas particles contained within spheres of
shrinking radius \citep{V06}. However, the origin of the first sphere
is located at the maximum of the dark matter density, which is
determined using the same approach.

We found this procedure to produce quite stable cluster center
identifications, thereby allowing to unambiguously compare results
from different runs.  The cluster center defined in this way is
consistently used also in the computation of the power ratios as well
as in that of spherically averaged radial profiles of several
hydrodynamic variables.  Finally, to remove cold clumps from the
computation of the power ratios and centroid shifts, we consider
only gas particles with temperatures above $0.5keV$.
\subsection{ Velocity-field diagnostics}\label{subsec:ek}
The spectral properties of the ICM
velocity field are quantified using the velocity power spectrum $E(k)$.  This
is computed by first determining $\vec {\tilde u_{w}}^d(\vec k)$, the
discrete Fourier transform of the density-weighted velocity field
$\vec u_{w}(\vec x)\equiv w(\vec x) \vec u(\vec x)$, where $w(\vec
x)\propto {\rho(x)}^{1/2}$.

To obtain the transform $\vec {\tilde u_{w}}^d(\vec k)$, a cube of size
$L_{sp}=R_{200}$ with $N_g^3=128^3$ grid points is placed at the
cluster center and the velocity field is computed at the grid points
$\vec x_p$ according to the SPH prescription. The discrete transforms
of $\vec u_{w}(\vec x_p)$ are then found using fast Fourier transforms
and used to define a dimensionless velocity power spectrum
\begin{equation}
E(k)=\frac{1}{L_{sp} \sigma^2_v}\left [ 2 \pi k^2 \mathcal{P}^d(k)
\left (\frac{L_{sp}}{2\pi} \right )^3 \right ],
\label{pow.eq}
\end{equation}
where $\sigma_v=\sqrt {G M_{200}/r_{200}}$, $k=|\vec k|$ and
$\mathcal{P}^d(k)=< |\vec {\tilde u_{w}}^d(\vec k)|^2$ is the
spherically averaged discrete power spectrum.

In addition, we address separately the longitudinal and solenoidal
components of the power spectrum by introducing in the $\vec k-$space
the corresponding velocity field components
\begin{eqnarray}
\vec {\tilde u}(\vec k)_{shear}&=& \frac{\vec k \times \vec {\tilde u}(\vec k)}  {|\vec k|}~,\\
 \vec {\tilde u}(\vec k)_{comp}&=& \frac{\vec k \cdot \vec {\tilde u}(\vec k)}
  {|\vec k|} \,.
   \label{pvisc.eq}
\end{eqnarray}

The velocity power spectrum is then defined as $E(k)=E_s(k)+E_c(k)$
\citep{KI09}.
%
\section{Results}\label{sec:results}
In this section we present and discuss the main results of our analysis.
The notation that will be adopted to distinguish the various simulation runs can be
sketched as follows:
\begin{itemize}
\item depending on the physics included in the simulations, ``ar'' will
refer to adiabatic runs, ``cr'' will refer to radiative runs
(comprising cooling, star formation and metal production);
\item depending on the particular modification of the viscosity scheme
(AV$_2$, AV$_5$; see \sec\ref{sec:sims}), we will refer to the adiabatic
runs as ``ar2, ar5'' and to the radiative runs as ``cr2, cr5'';
\item ``ar0'' and ``cr0'' runs will be always used as terms of comparison,
being the reference simulations with standard SPH and fixed viscosity
(NOAC, AV$_0$);
for the other runs, instead, we consider both the standard SPH version
(NOAC) and the modification including the artificial conductivity
term (AC).
\end{itemize}
\subsection{Adiabatic runs}
Here, we discuss the results on radial and global properties
of the simulated clusters, for different adiabatic runs at redshift
$z=0$. Differences in the simulations consist in the modifications to
the classical SPH approach presented in Sections~\ref{sec:sims} and
\ref{subsec:ac}.  In this section, we consider as reference case the
adiabatic run: ``ar0'' (NOAC, AV$_0$).
%
\subsubsection{Entropy and temperature profiles}\label{sec:rad_prof}
\begin{figure*}
\centering
\includegraphics[width=0.45\textwidth]{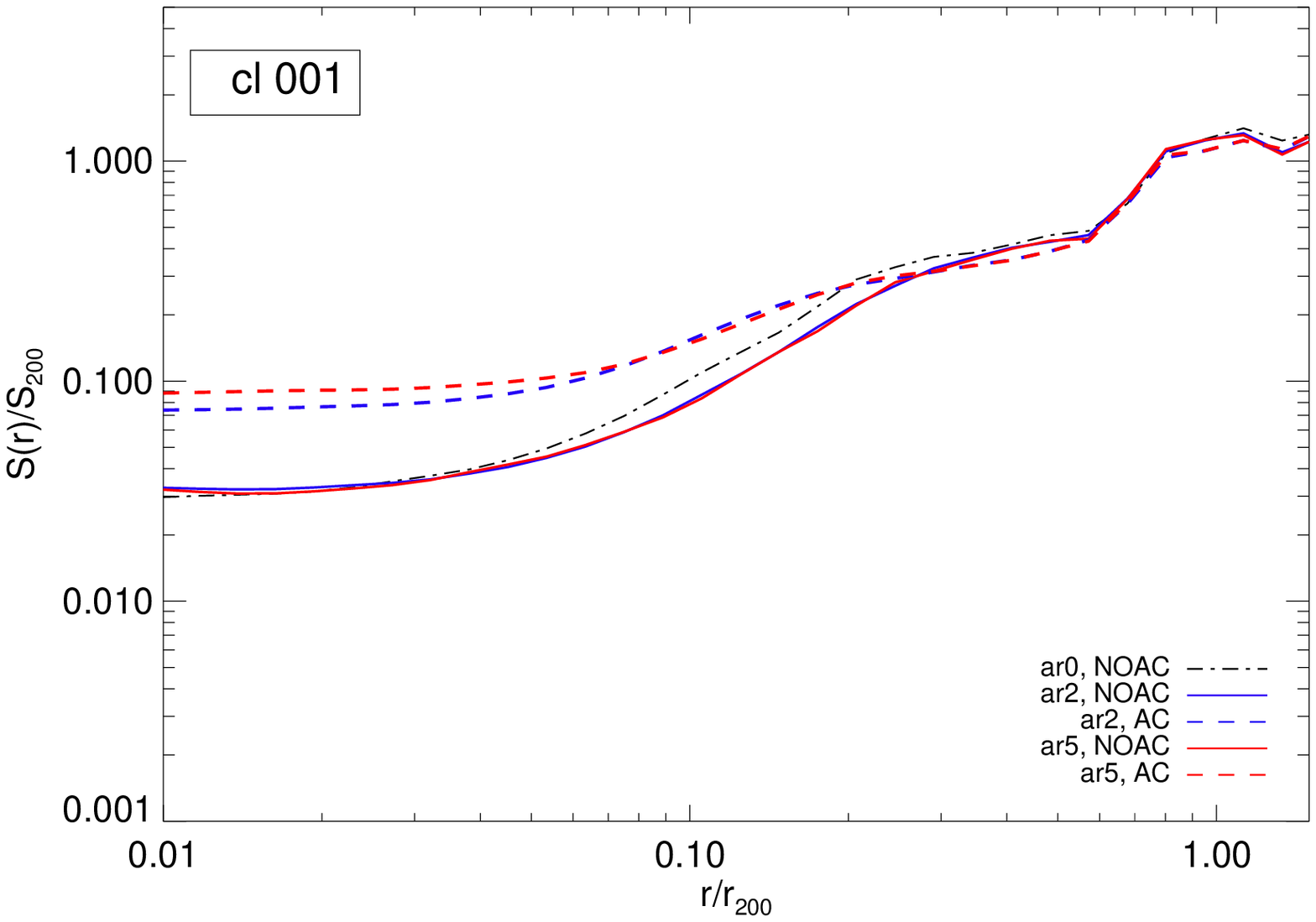}
\includegraphics[width=0.45\textwidth]{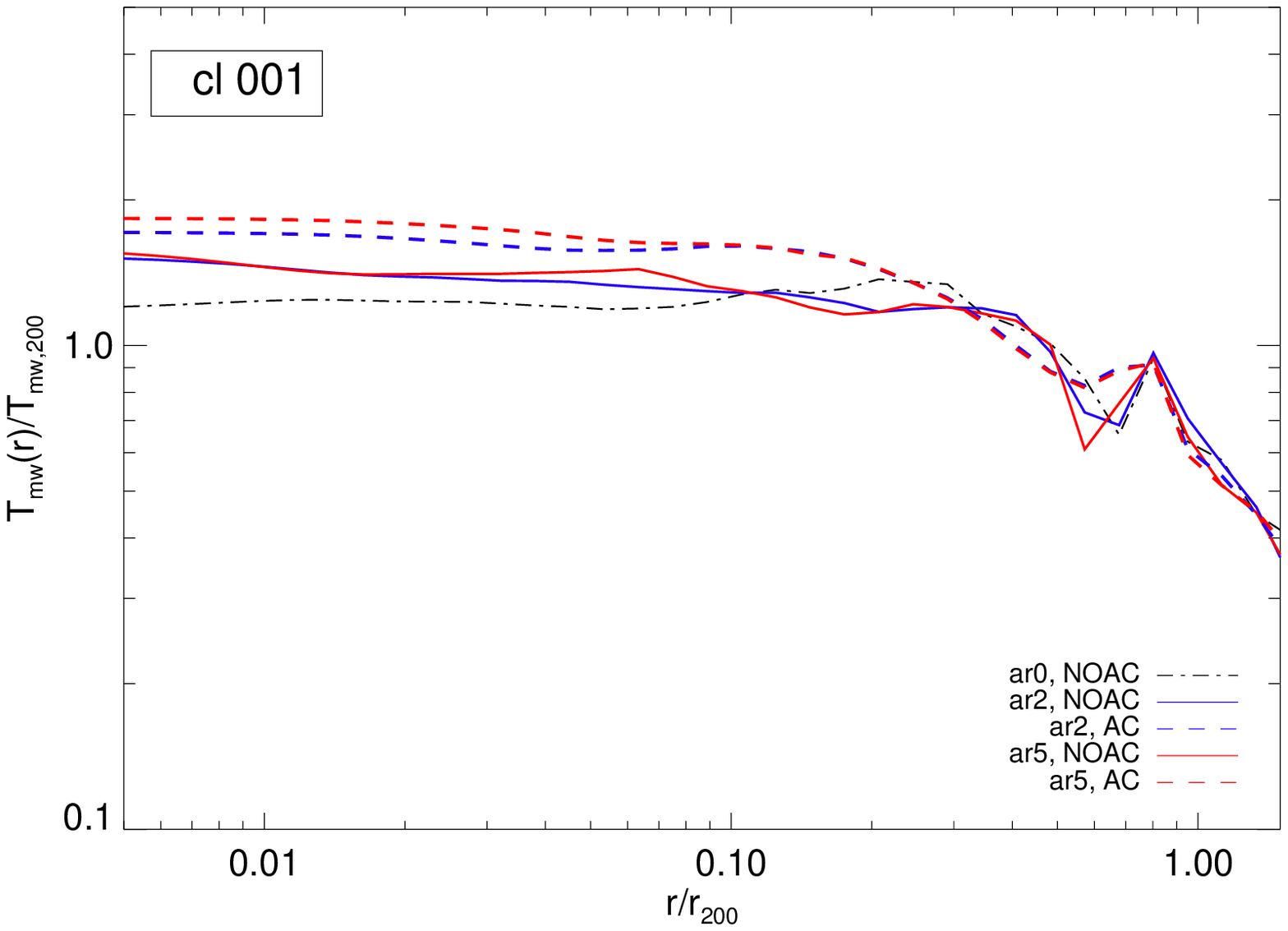}\\
\includegraphics[width=0.45\textwidth]{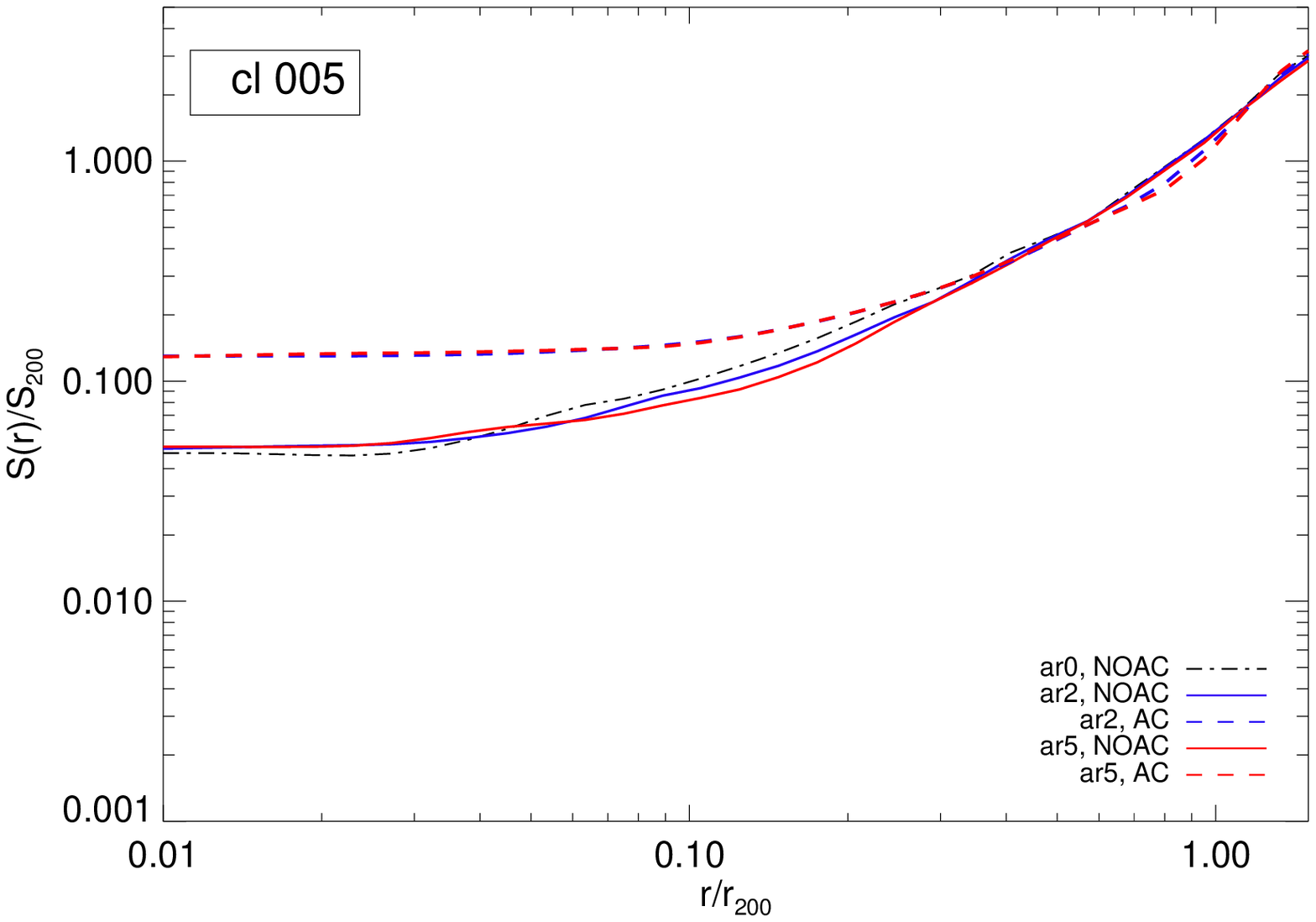}
\includegraphics[width=0.45\textwidth]{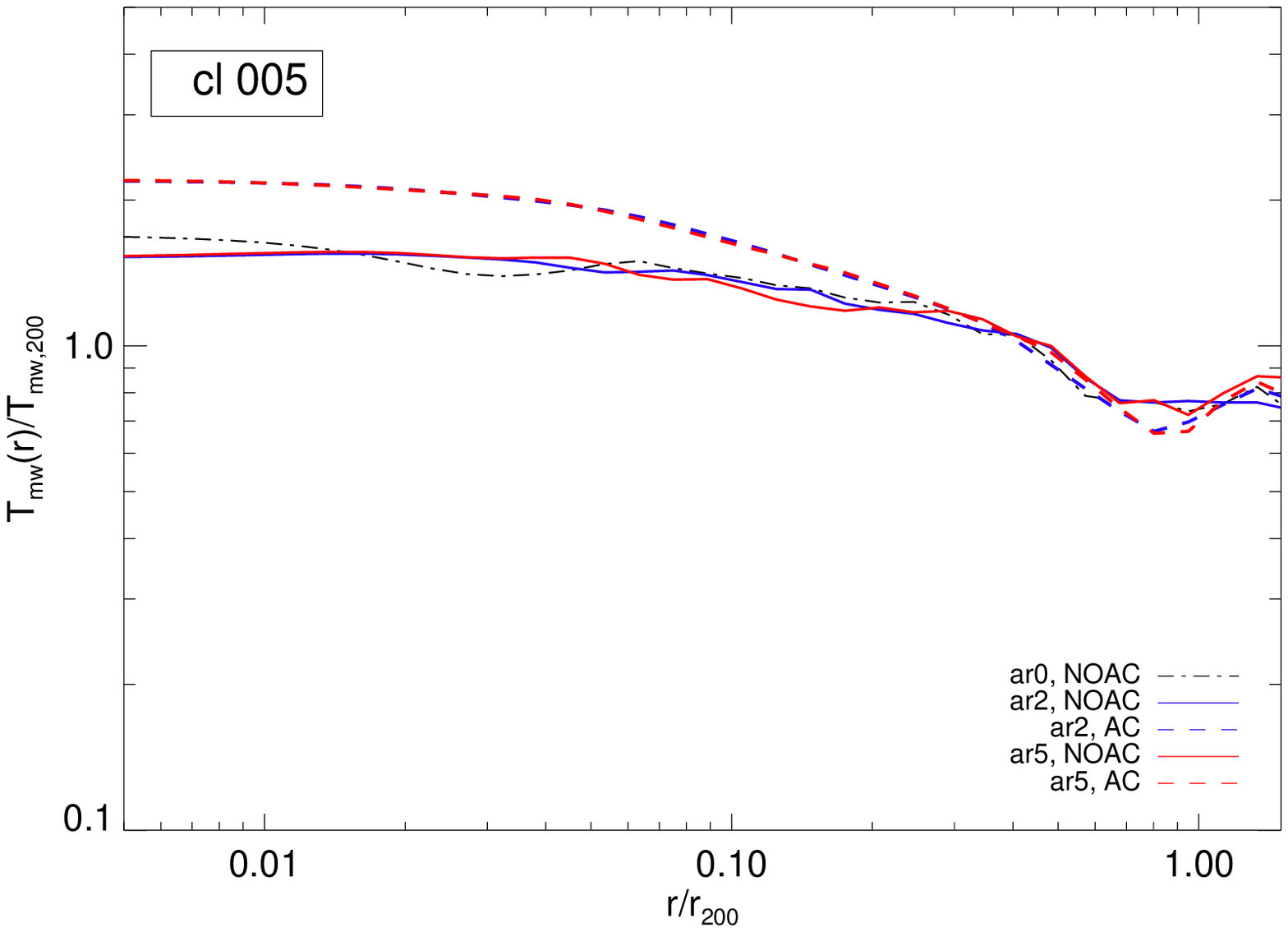}\\
\includegraphics[width=0.45\textwidth]{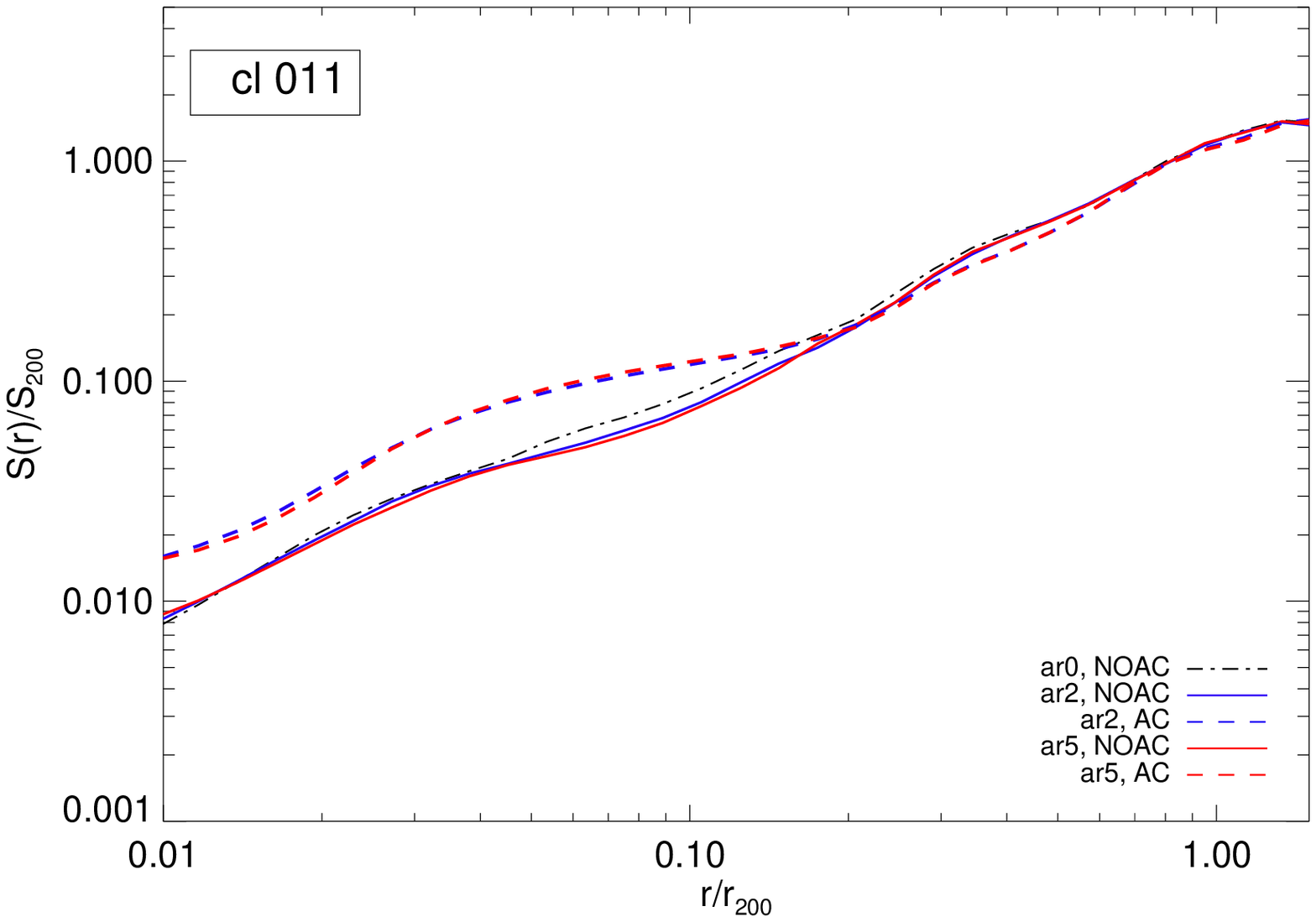}
\includegraphics[width=0.45\textwidth]{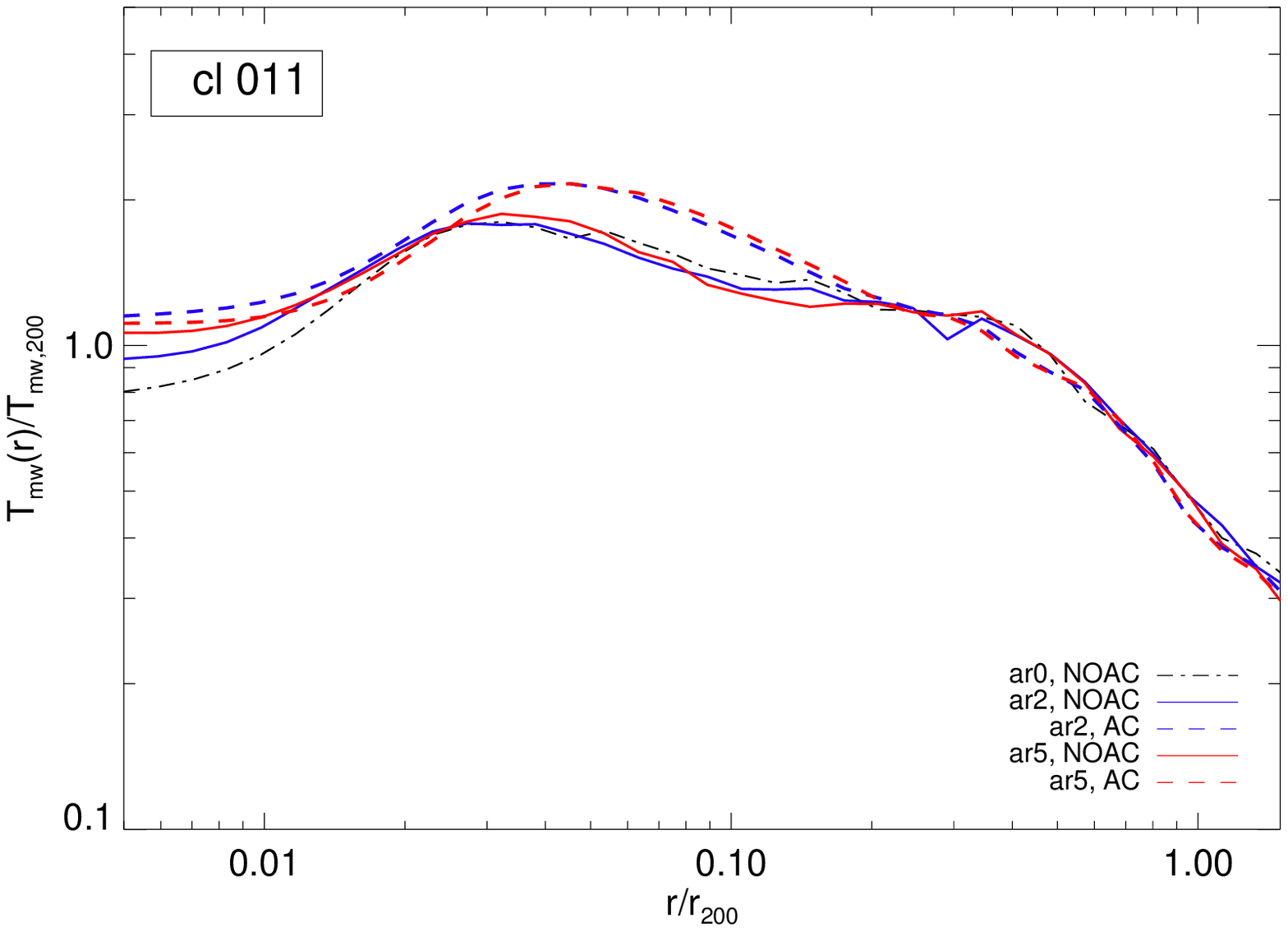}\\
\includegraphics[width=0.45\textwidth]{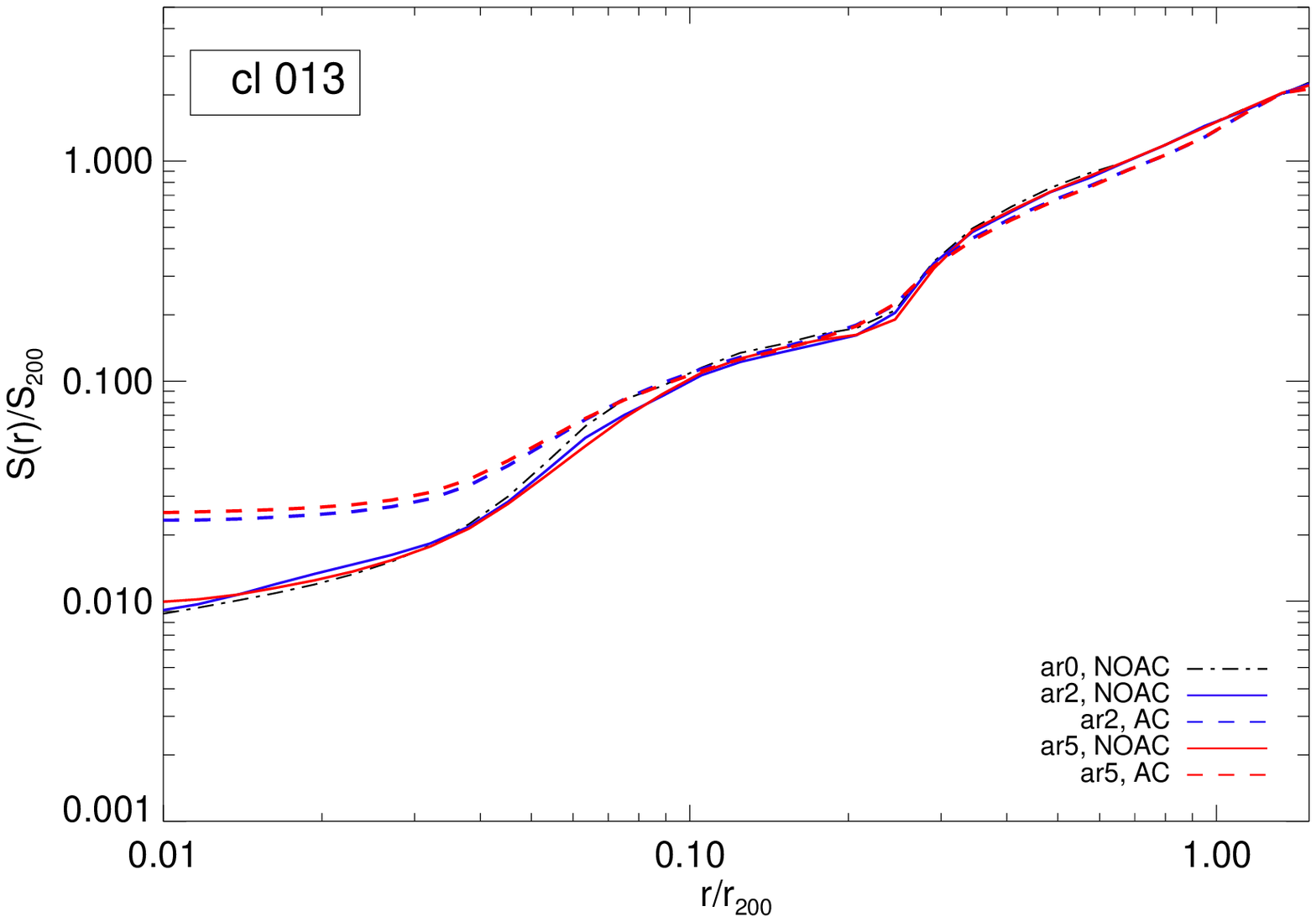}
\includegraphics[width=0.45\textwidth]{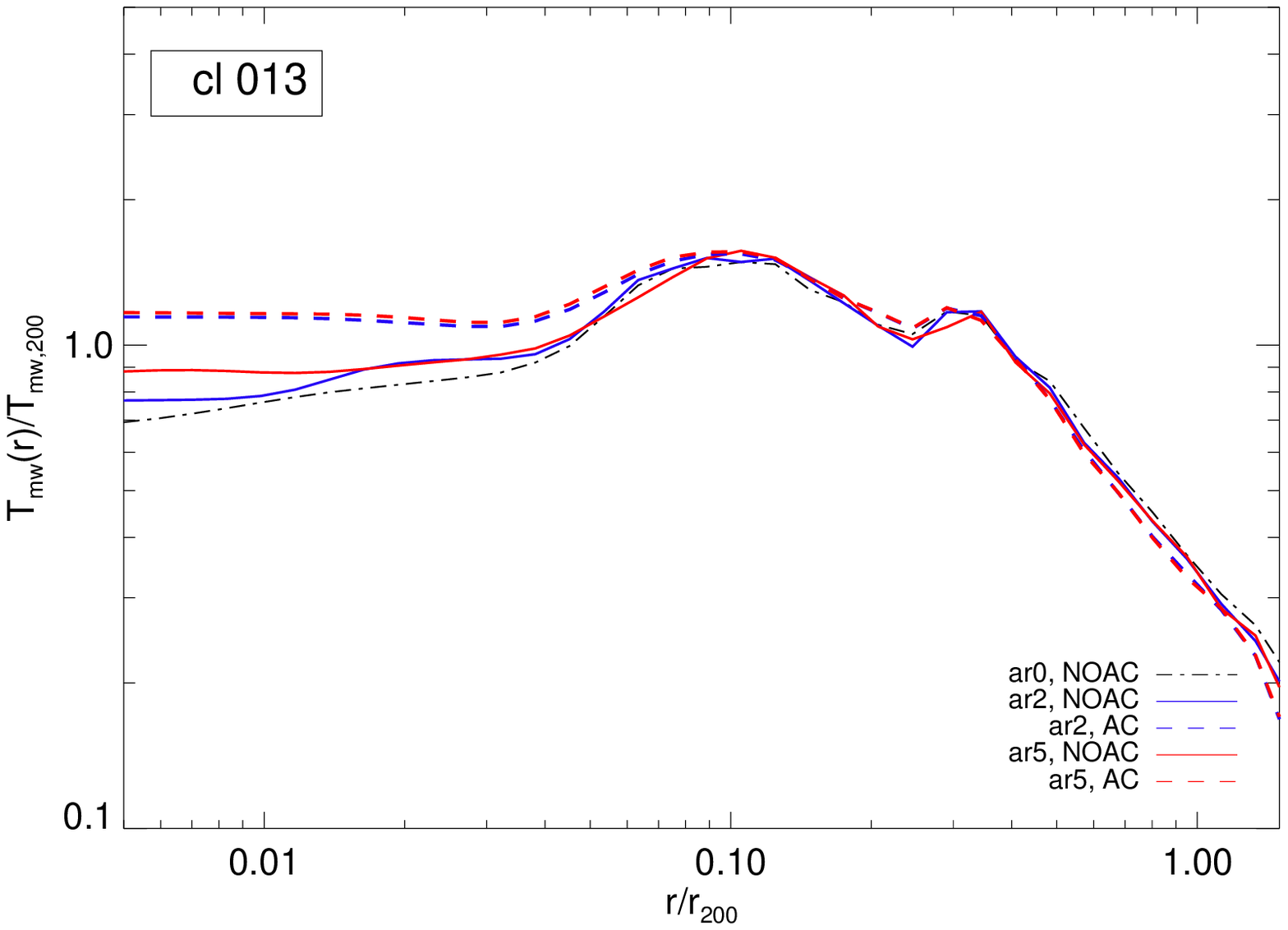}\\
\caption{Entropy and temperature profiles of the clusters in the
  sample, for the adiabatic simulations at $z=0$. Different lines
  refer to different runs: AC-SPH (dashed) and standard SPH (solid),
  for the two viscosity schemes AV$_2$ (blue) and AV$_2$ (red). The
  dot-dashed black line marks the reference run (ar0; NOAC - AV$_0$).
  \label{fig:entr}}
\end{figure*}

\begin{figure*}
\centering
\includegraphics[width=0.45\textwidth]{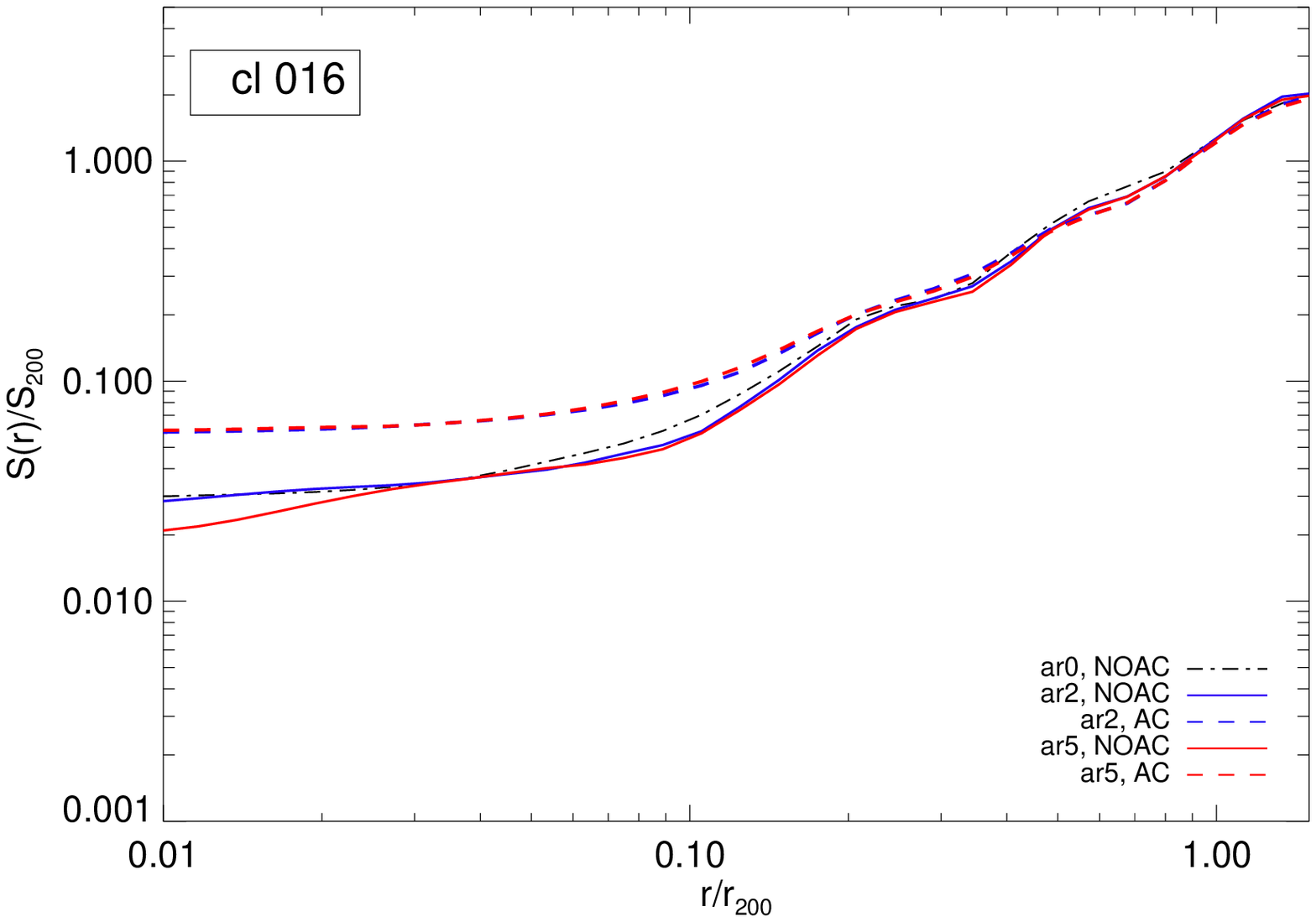}
\includegraphics[width=0.45\textwidth]{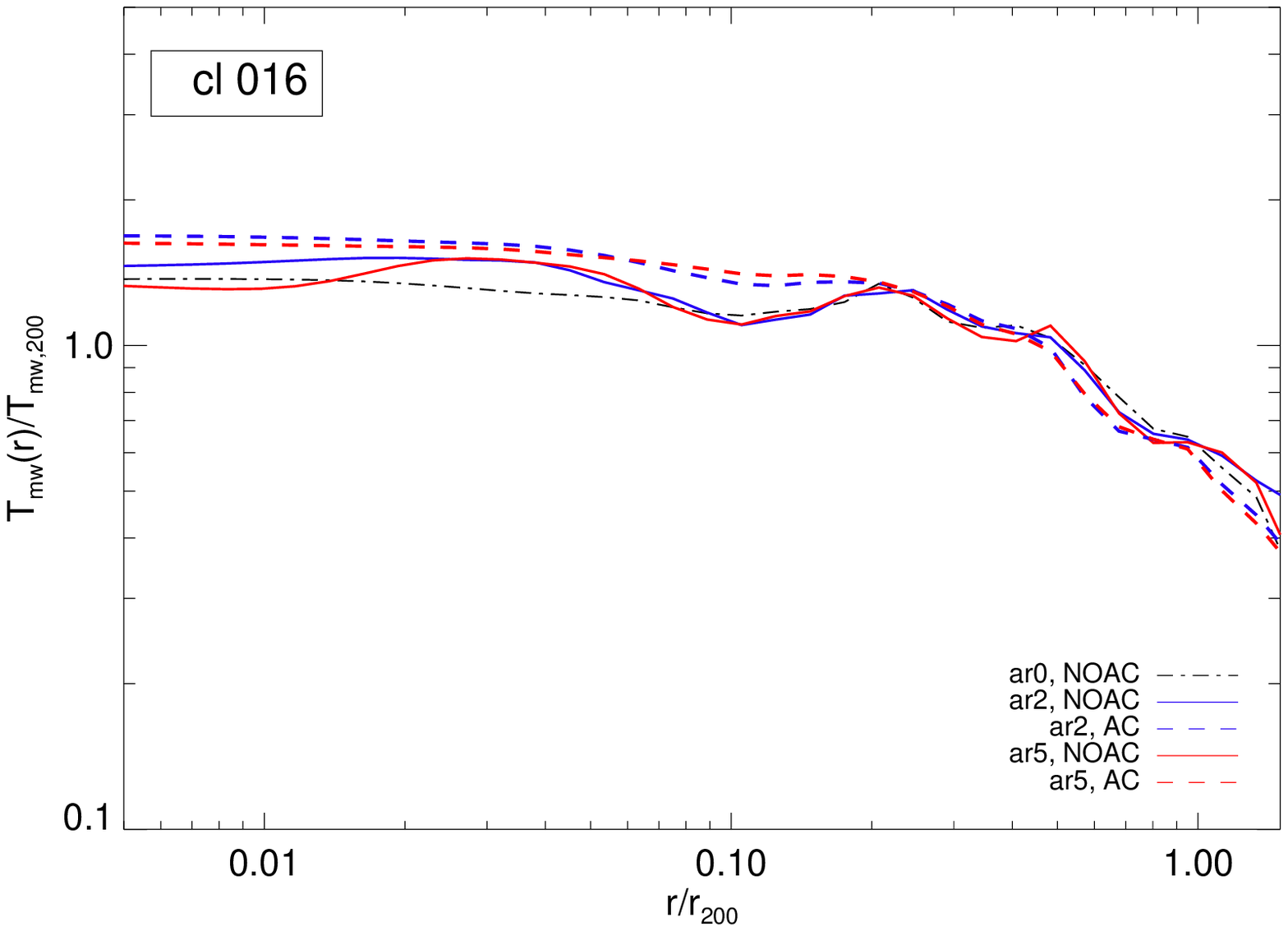}\\
\includegraphics[width=0.45\textwidth]{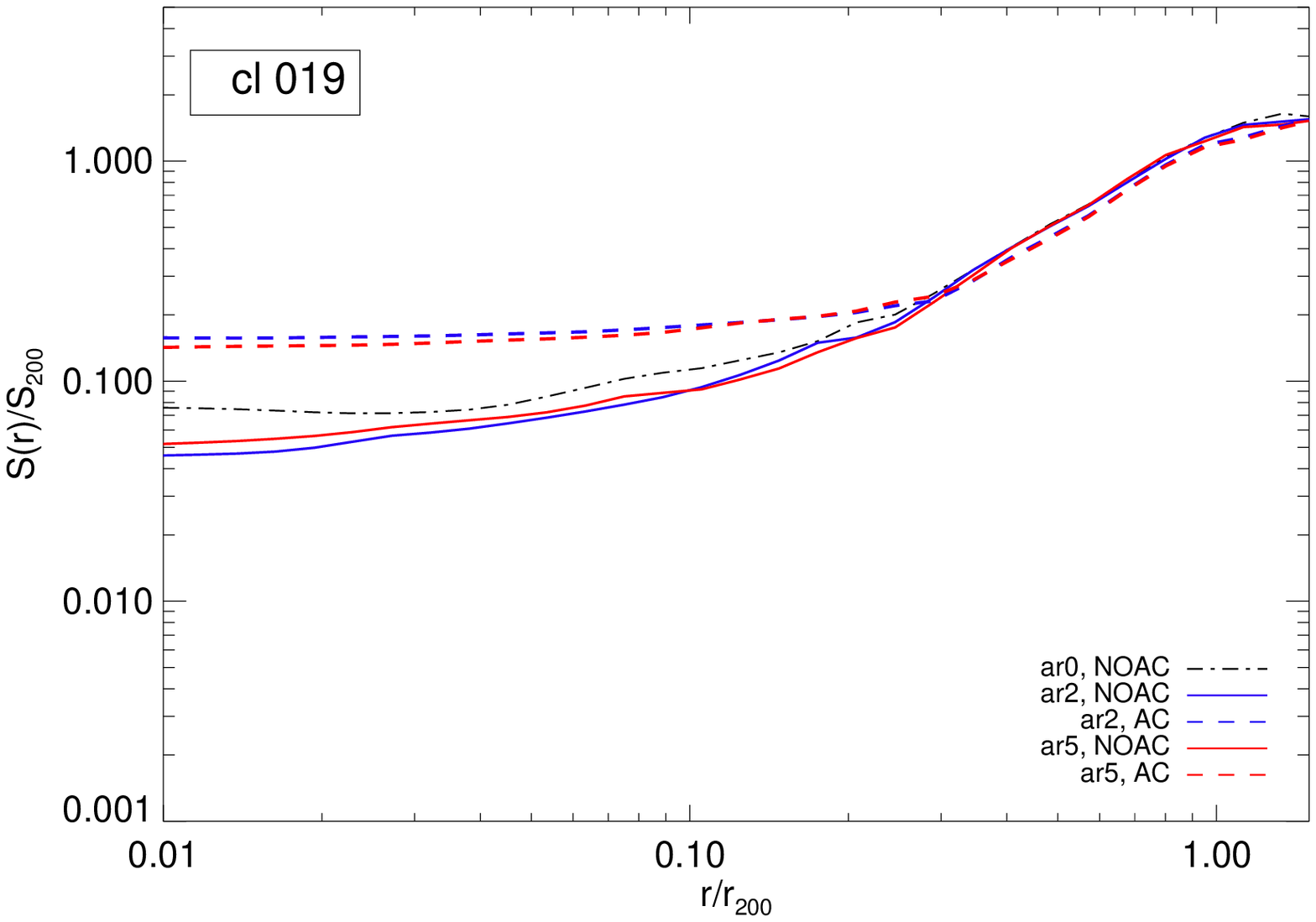}
\includegraphics[width=0.45\textwidth]{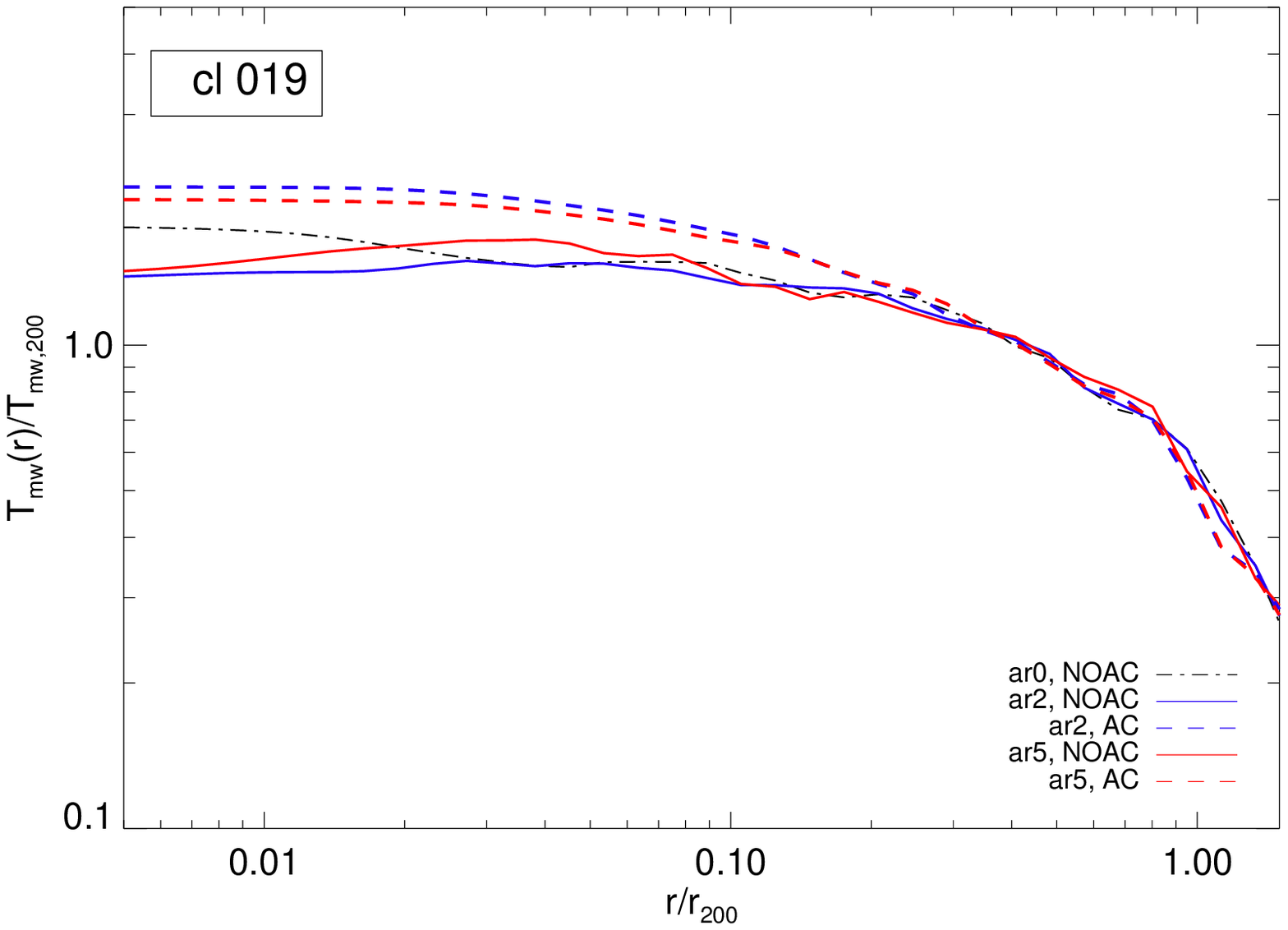}\\
\includegraphics[width=0.45\textwidth]{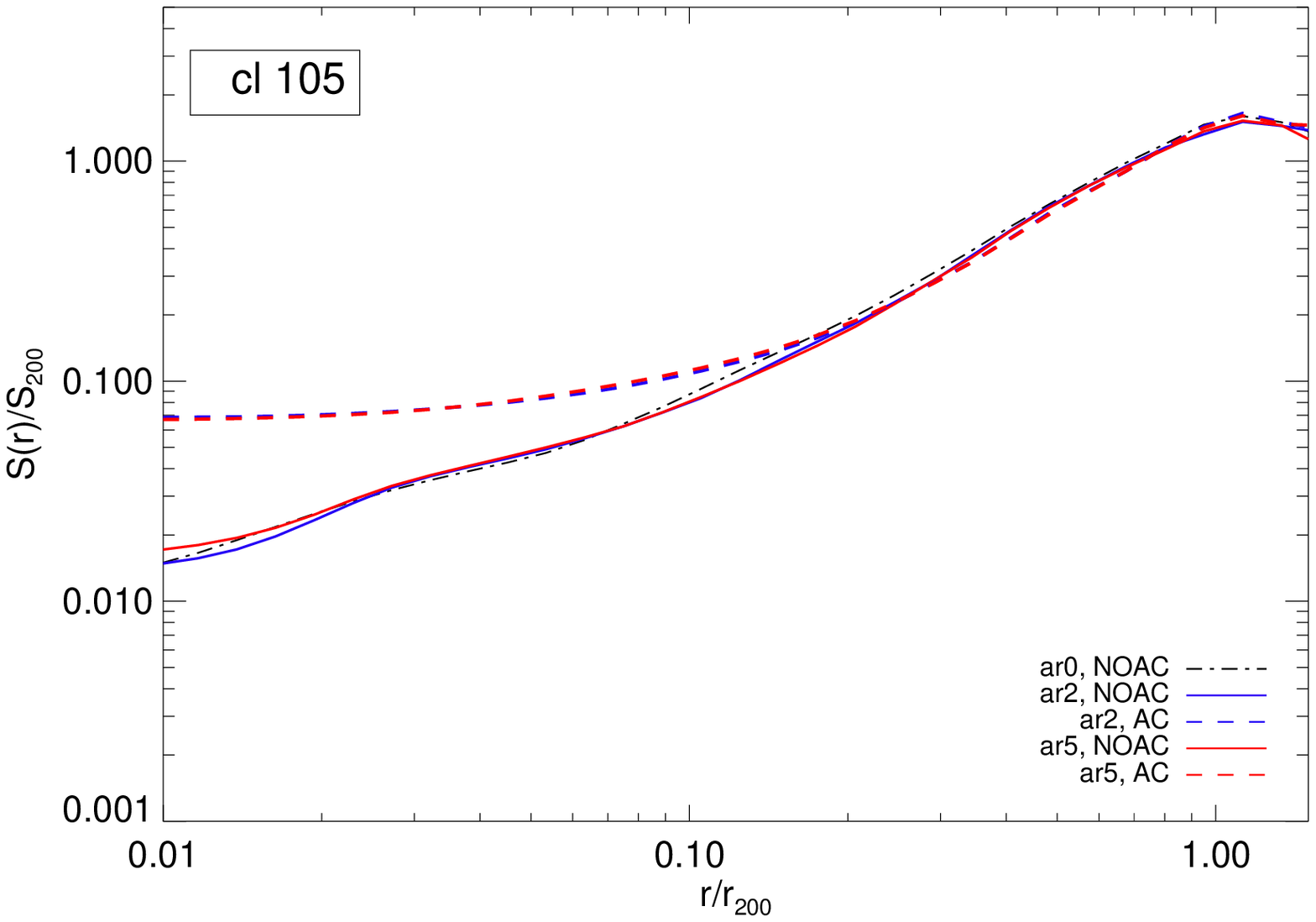}
\includegraphics[width=0.45\textwidth]{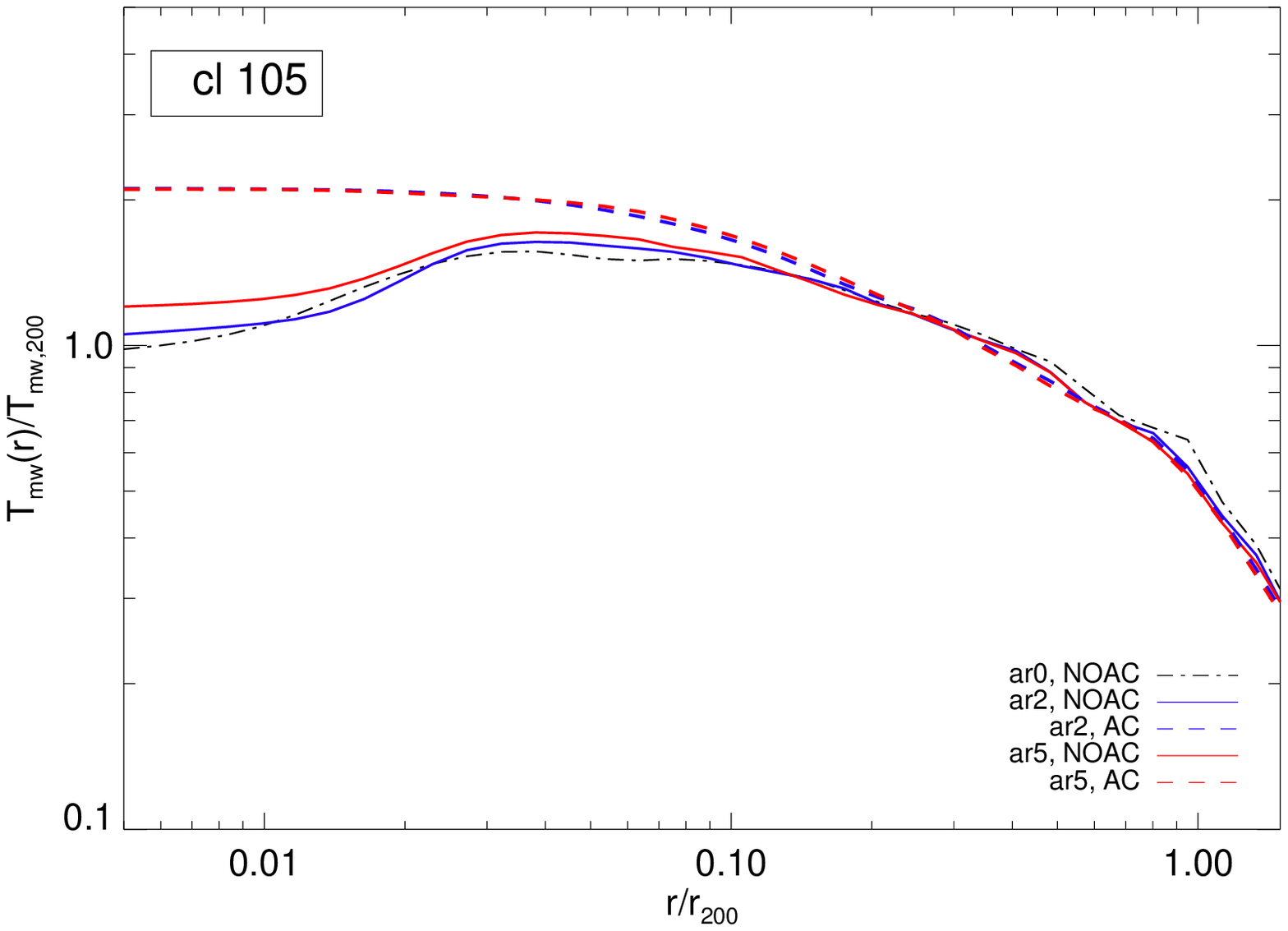}\\
\includegraphics[width=0.45\textwidth]{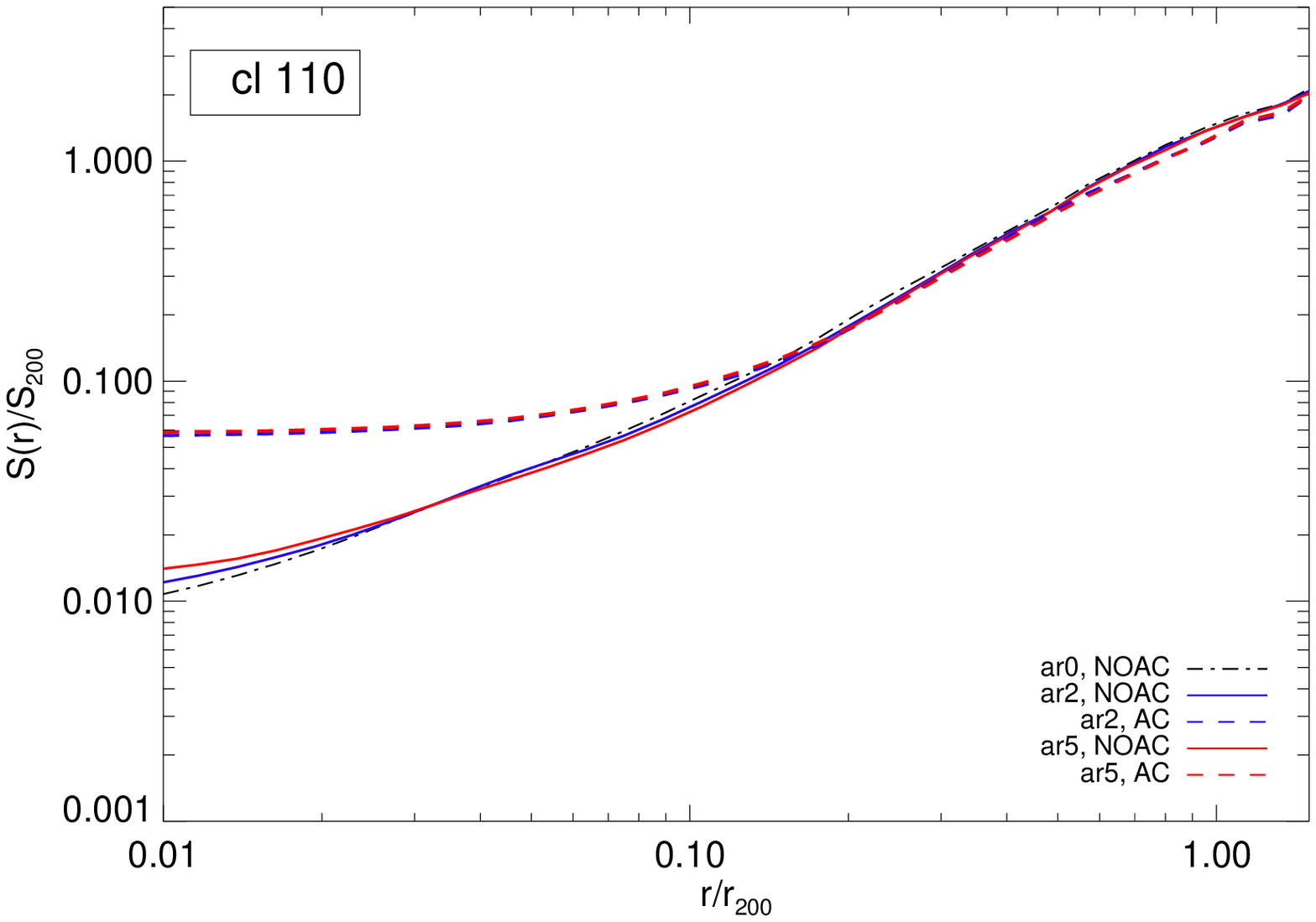}
\includegraphics[width=0.45\textwidth]{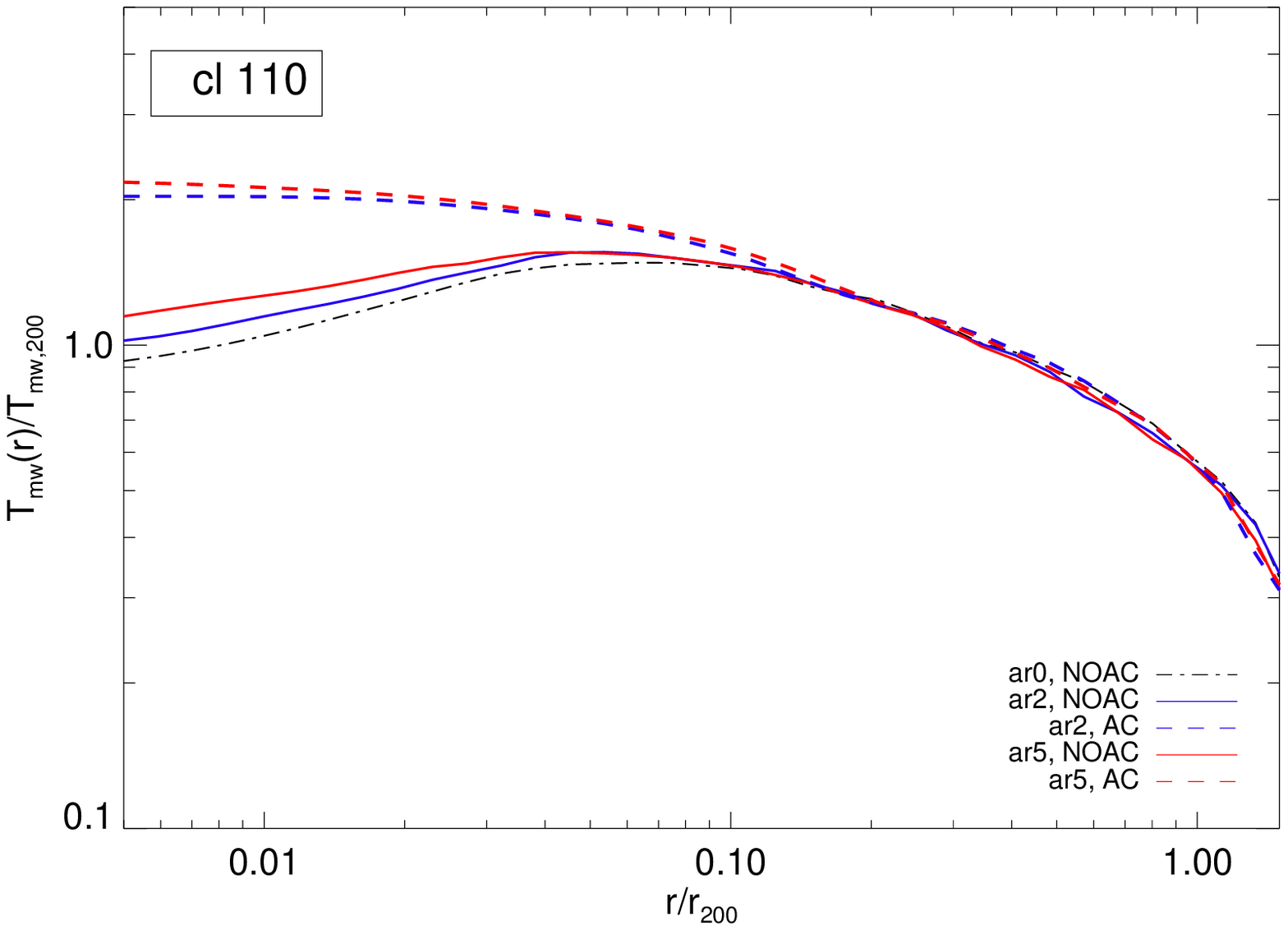}\\
\caption{$z=0$ --- continues from \fig\ref{fig:entr}.}\label{fig:entr2}
\end{figure*}

We explore the different properties of the ICM in the adiabatic runs
by investigating the spherically-averaged radial entropy and
temperature profiles of the sample clusters, as shown in
\fig\ref{fig:entr} and \fig\ref{fig:entr2}.  In order to allow for a
fair comparison among different clusters,
the profiles are shown as a function of $r/R_{200}$. The temperature
profiles are expressed in units of $T_{200}$, and employ the
mass-weighted temperature, which represents the ``dynamical''
temperature of the gas, without any bias due to its thermal phase
(e.g. emissivity).  For the entropy profiles we rescale the values to
   \begin{equation}
S_{200}=\frac{1}{2}\left [\frac{2 \pi}{15} \frac {G^2 M_{200}}{f_bH_0}\right]^{2/3}~,
  \label{entr.eq}
   \end{equation}
 where $f_b=\Omega_\mathrm{b}/\Omega_\mathrm{m}$
is the global baryon fraction.

Despite the different configurations of individual objects, the
results in \fig\ref{fig:entr} and \fig\ref{fig:entr2} indicate that
the introduction of the AC term (coloured, dashed curves in the plots) tends in
general to produce cluster cores with higher entropy and higher
temperature with respect to their standard-SPH counterparts (NOAC; coloured,
solid lines).
 The AC term basically facilitates the redistribution of
internal energy produced in shocks and enhances the entropy mixing,
thereby increasing the entropy level in the core. The major
differences between AC and NOAC entropy profiles --- up to a factor of
$\sim4$ --- are in fact visible at $r/\rtwo<0.1$, whereas there is
reasonable agreement at larger distances from the center.  In some
individual cases, the entropy profile even presents a real plateau
towards the cluster center,
approaching the features obtained by AMR codes and observed in real
clusters.  Temperature profiles (right-hand-side column in the
figures) also show similar behaviour at small radii, where the AC
profiles are noticeably higher than the NOAC ones.
Different line colors in Figures~\ref{fig:entr} and~\ref{fig:entr2}
refer to the viscosity scheme adopted in the run, while the black, dot-dashed line always
indicates to the reference simulation (ar0). From the comparison, we observe that the
particular implementation of the viscosity scheme alone can already
have a certain impact on the level of gas mixing, hence mildly raising
the entropy of the core region (within $\sim 10\%$ of $\rtwo$), in the
NOAC runs (see results by V11).  However, we remark here that the
introduction of the AC term seems to dominate over the numerical
treatment of the viscosity, causing an increase in the central cluster
entropy which is definitely more significant, in the direction
indicated by the results from AMR codes.
%
%
%
%
%
%
%
\subsubsection{Velocity power spectra}
As previously investigated in V11,
the implementation of a time-dependent viscosity scheme in SPH simulations
of galaxy clusters can affect the characteristics of ICM motions, due
to the better modeling of the artificial viscosity depending on the
local conditions of the shock strength. This in fact can have an
additional impact on the gas mixing in cluster simulations, which is
typically inhibited in standard SPH codes causing also a lower central
entropy.
In the present work,
we additionally investigate
the effects of
artificial conductivity on the gas motions.
From the study of entropy and temperature radial profiles we have
already pointed out a significant improvement in the cluster cores due
to the introduction of the AC term, which in fact favours directly the
gas mixing.

\begin{figure*}
\centering
\includegraphics[width=0.48\textwidth,height=0.23\textheight]{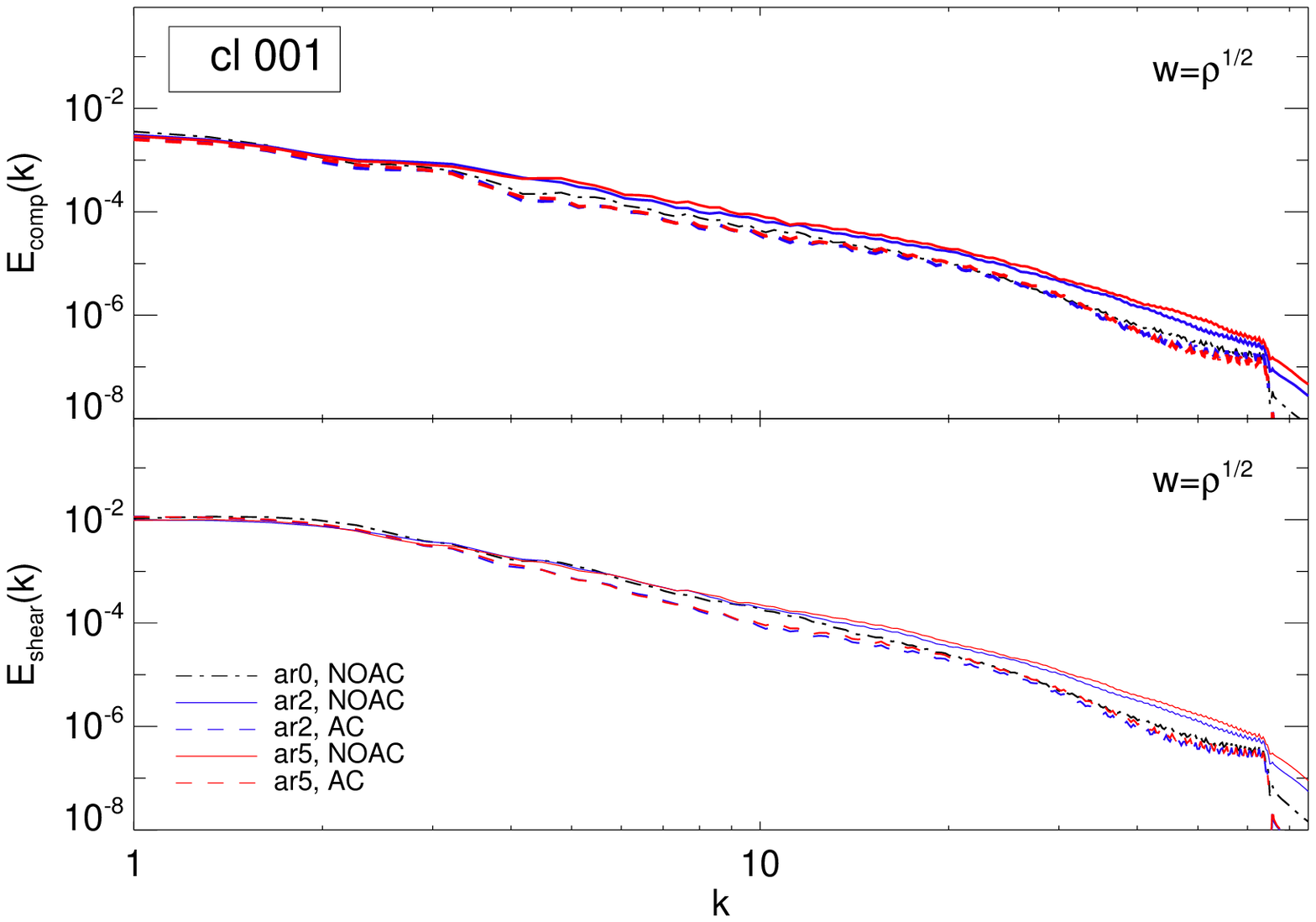}
\includegraphics[width=0.48\textwidth,height=0.23\textheight]{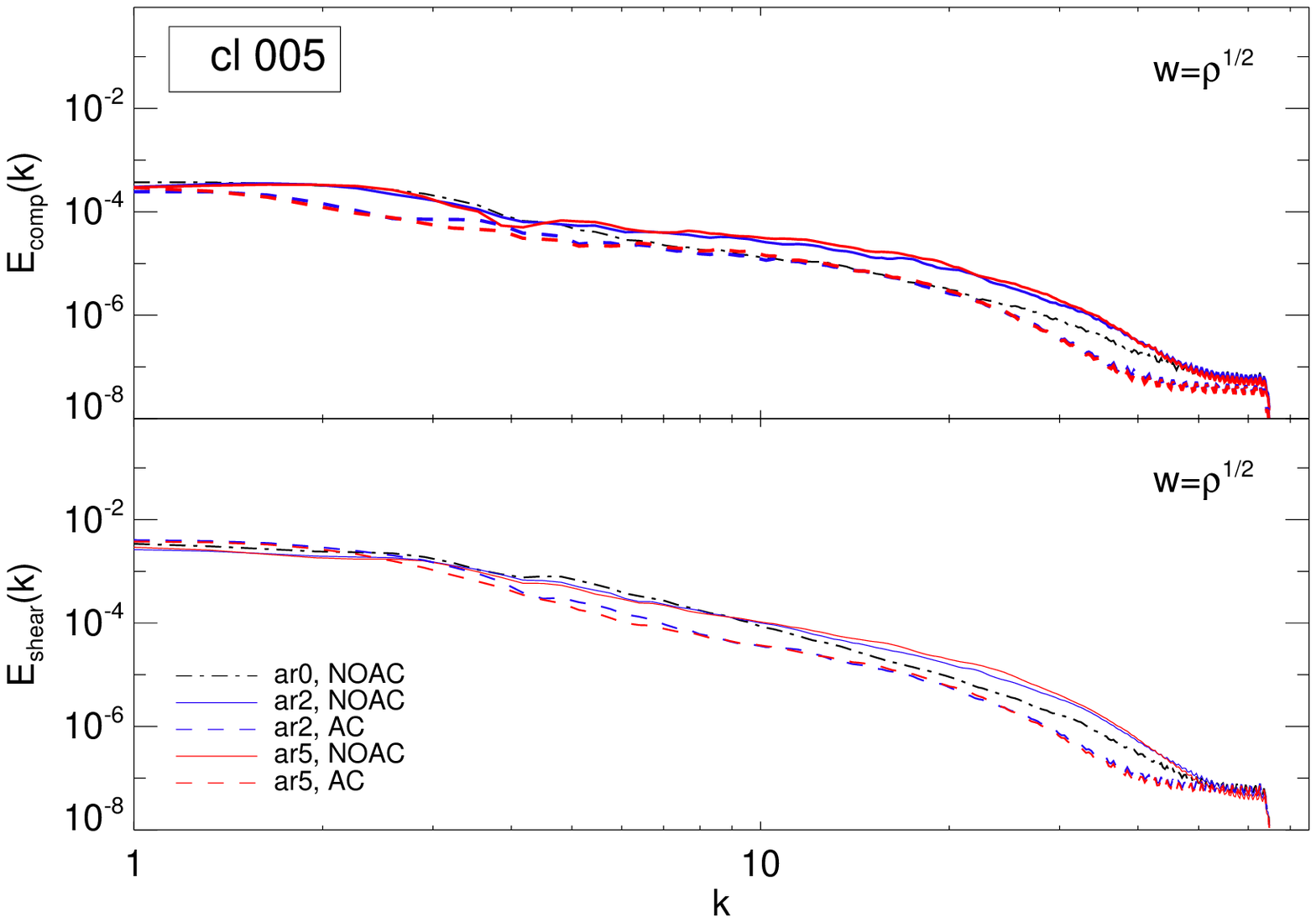}\\
\includegraphics[width=0.48\textwidth,height=0.23\textheight]{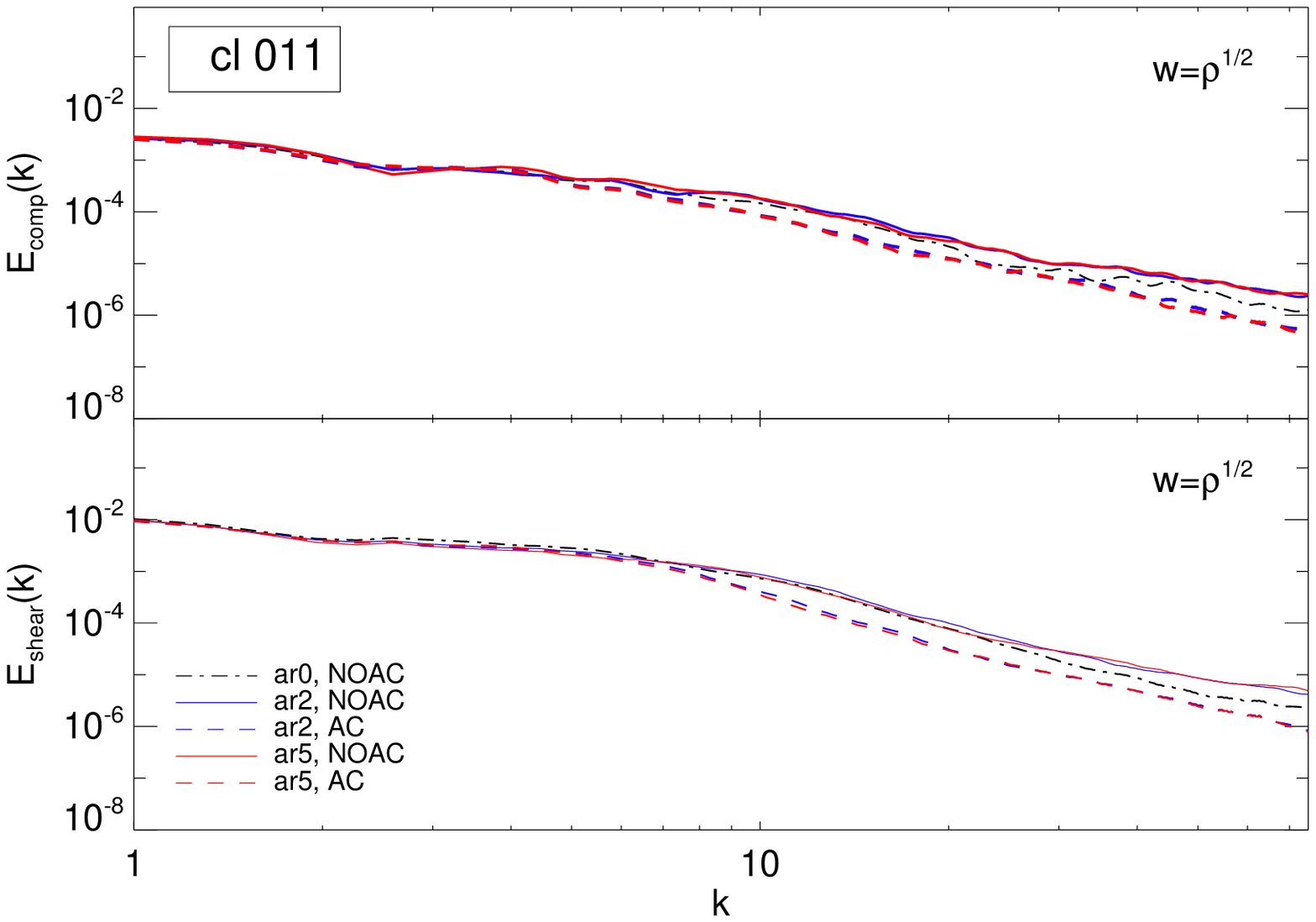}
\includegraphics[width=0.48\textwidth,height=0.23\textheight]{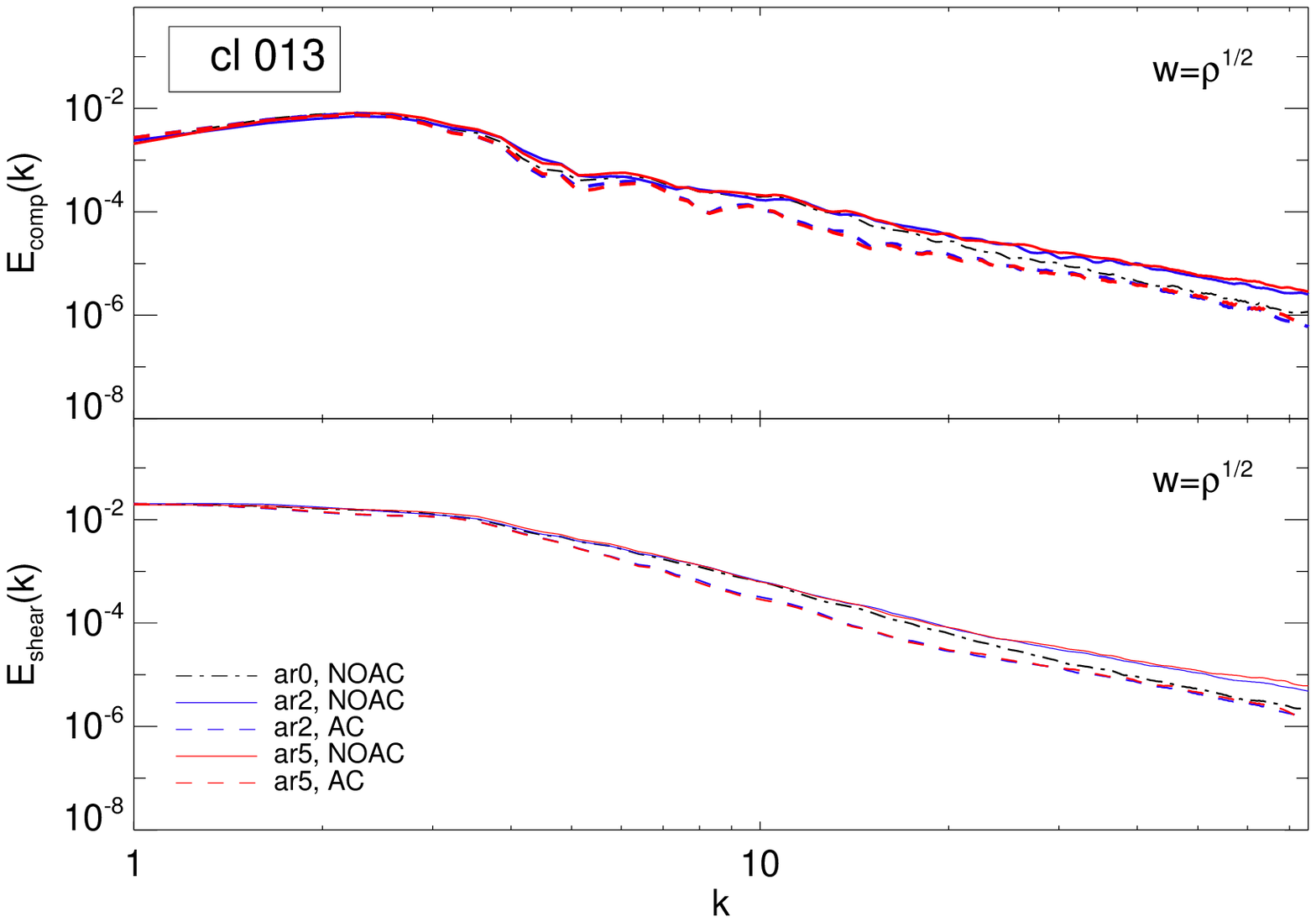}\\
\includegraphics[width=0.48\textwidth,height=0.23\textheight]{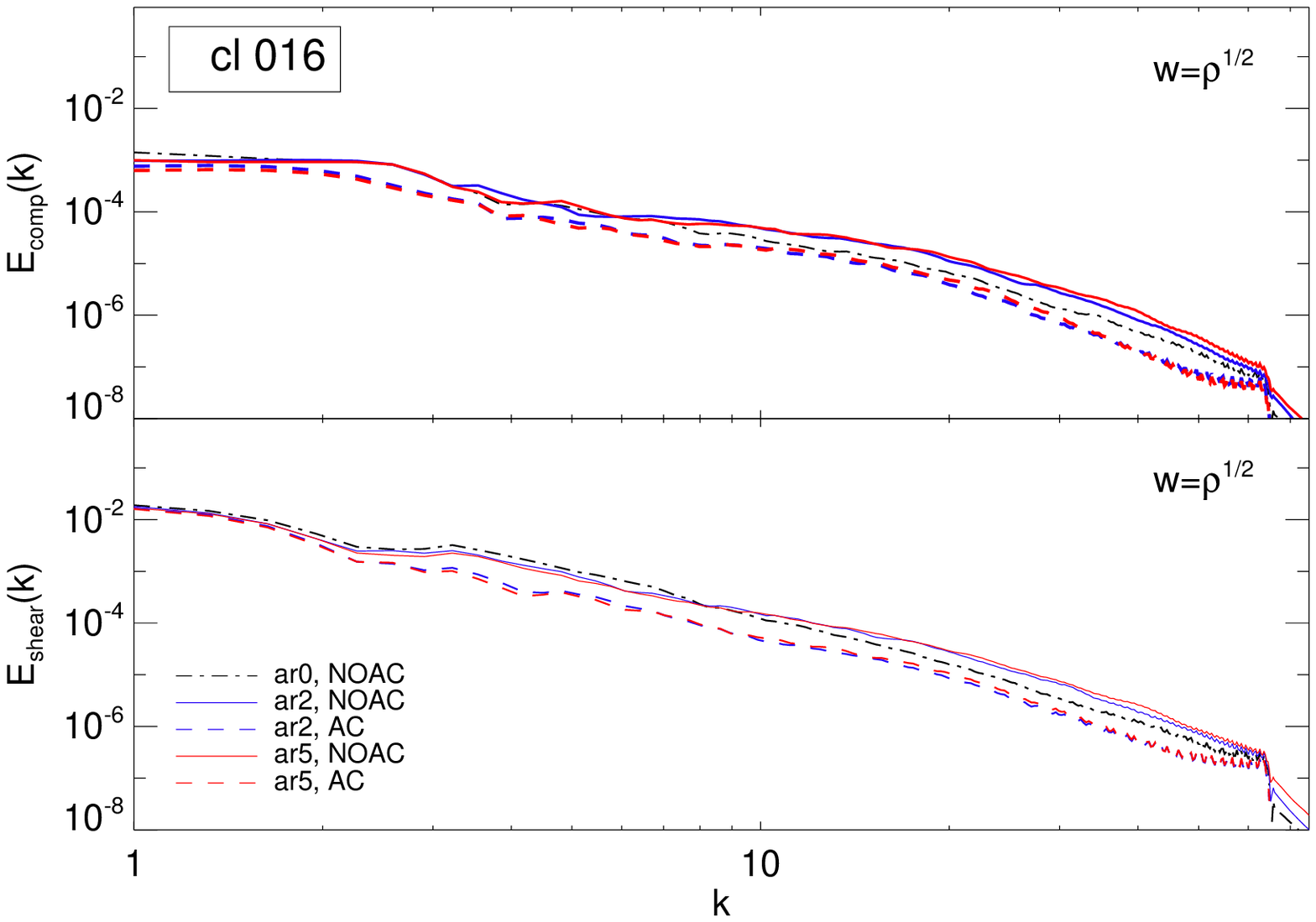}
\includegraphics[width=0.48\textwidth,height=0.23\textheight]{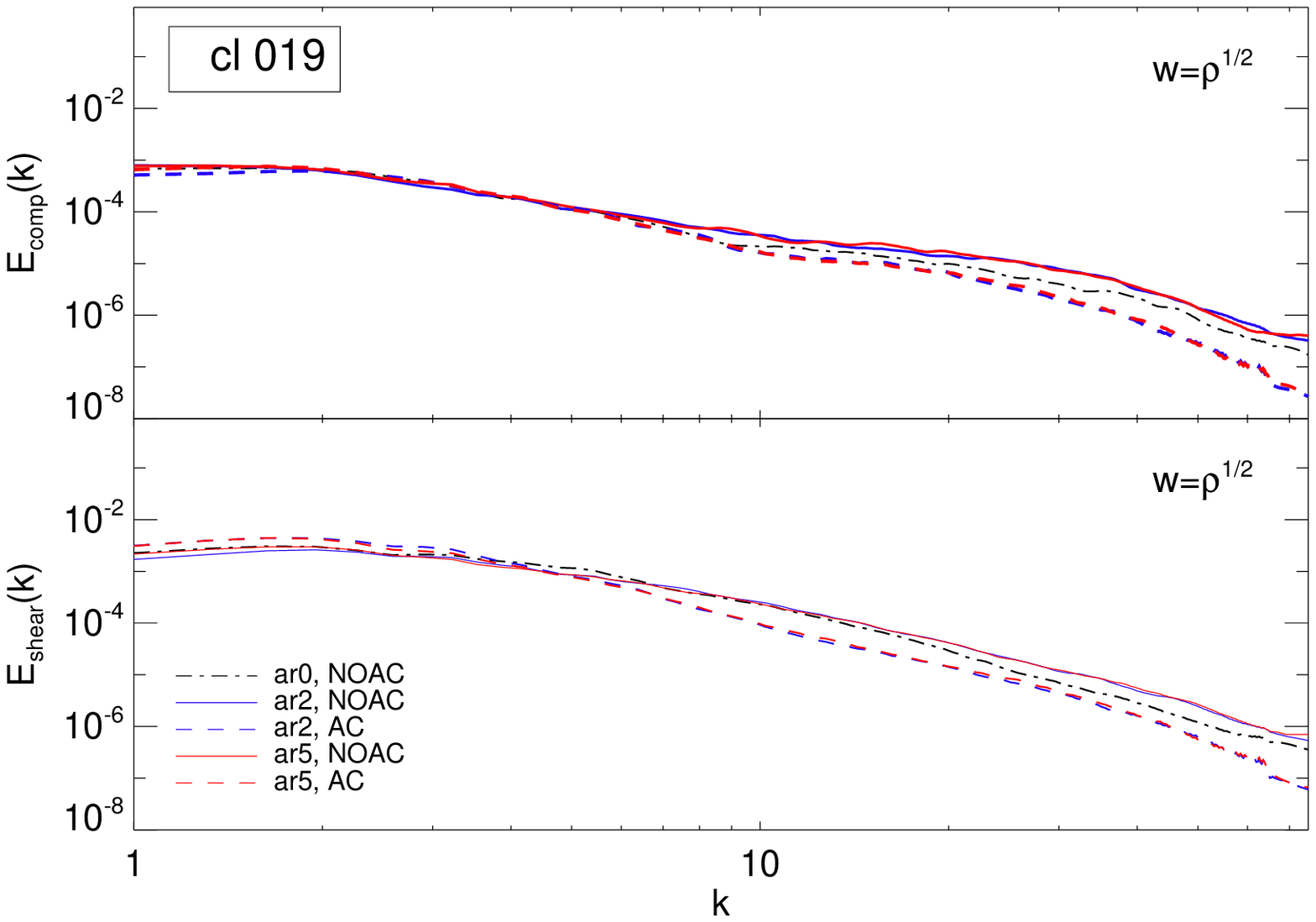}\\
\includegraphics[width=0.48\textwidth,height=0.23\textheight]{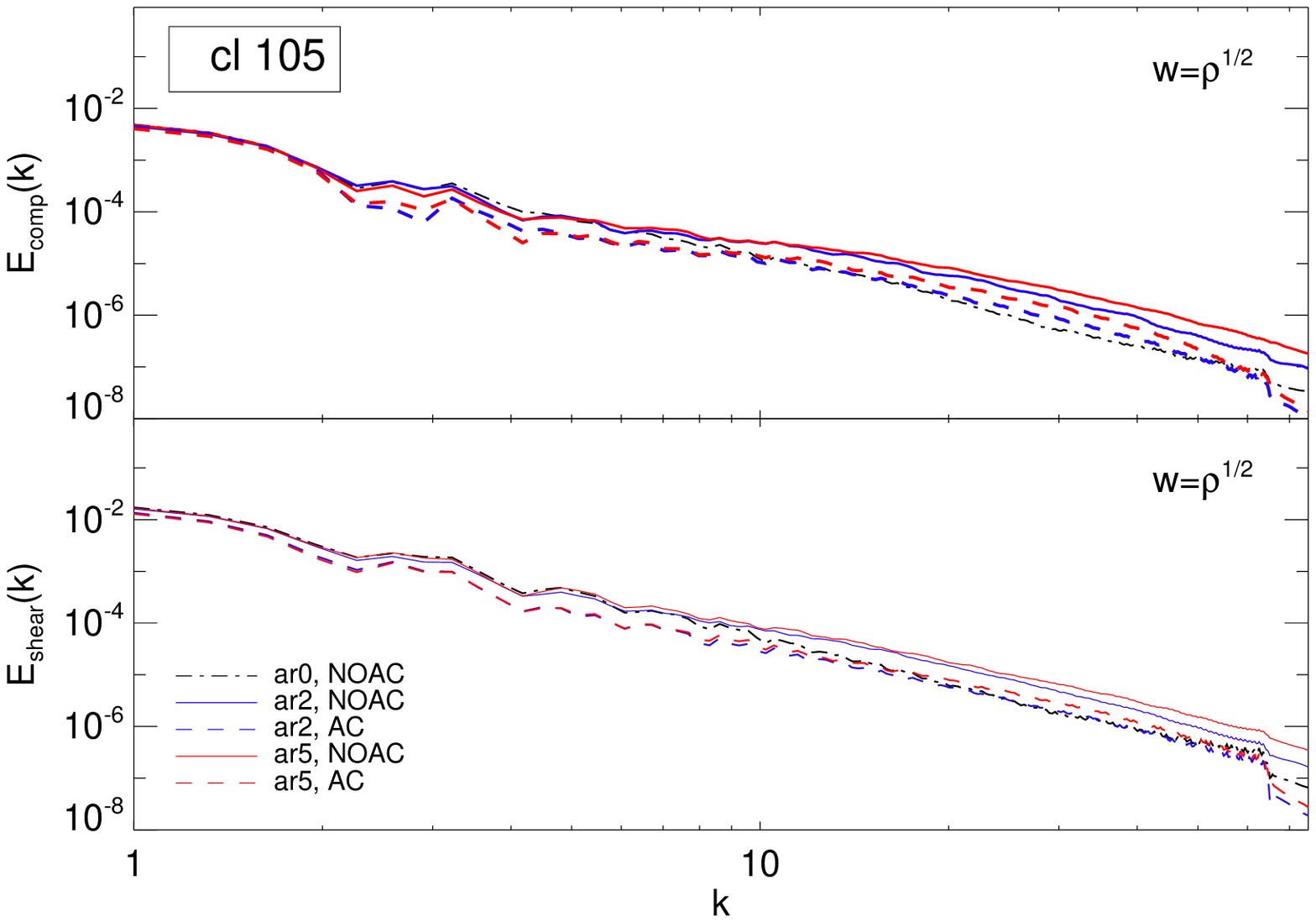}
\includegraphics[width=0.48\textwidth,height=0.23\textheight]{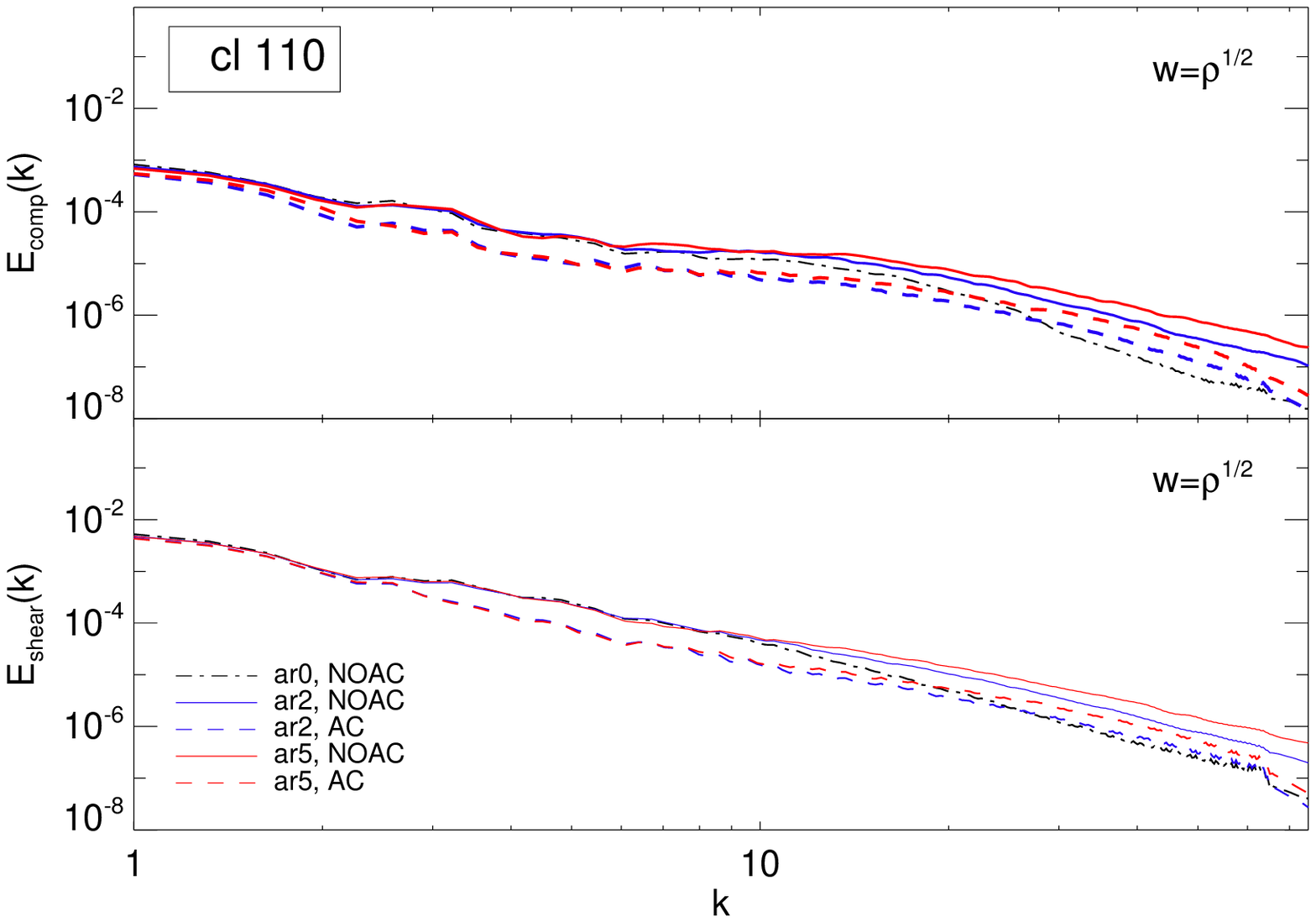}\\
%
\caption{Velocity power spectra of the clusters in the sample, for the
  adiabatic simulations at $z=0$. Line styles and colors are the same
  of Figs.~\ref{fig:entr} and~ \ref{fig:entr2}.\label{fig:velpow}}
\end{figure*}

In \fig\ref{fig:velpow} we present results on the spectral properties
of the gas velocity field for the adiabatic runs of the clusters in
the sample. The turbulent velocity field is investigated here via the
velocity power spectrum $E(k)$ (see definitions in
Section~\ref{sec:clu}).  Each panel of \fig\ref{fig:velpow} displays
the compressive and shearing components (upper and lower insets,
respectively) of the density-weighted velocity power spectrum of a
given cluster.  Using the same notation as for the radial profiles,
the solid and dashed lines refer respectively to the NOAC and AC ---
adiabatic --- simulations,
while blue and red are used to distinguish between AV$_2$ and AV$_5$,
respectively.  The reference run (ar0; standard NOAC SPH and AV$_0$
viscosity) is marked by the black, dot-dashed line.  In the figure, we
show the velocity power spectrum components as a function of the
dimensionless wavenumber $k = |\mathbf{k}|L_{sp}/2\pi$.

From the comparison, the most striking result is that the NOAC runs
show an increased spectrum amplitude for motions on the small scales
(large values of $k$) with respect to the standard ar0 case. When the
AC term is introduced, for both the viscosity schemes considered
(AV$_2$ and AV$_5$), the effect is suppressed and the power spectrum
is lower almost at all scales.  In fact, especially for the
compressive component, the AC (dashed) curves lie always below the
NOAC (solid) ones.

In general, we observe that the AV scheme and the AC term have
opposite, competing effects.  The results obtained, valid for all the
clusters analysed independently of their peculiar characteristics,
suggest that while the AV scheme contributes to enhance the coherence
of the velocity patterns, especially at small scales, therefore
augmenting the amplitude of the power spectrum (as already observed
and discussed V11), the AC term has instead the practical effect to
favour gas mixing, likely contributing to more random gas motions.
Overall, however, the effects due to the AC term seem to dominate and
have a major impact on the properties of the gas velocity field.

%
\subsubsection{Temperature distribution and global properties}\label{sec:glob}
\begin{figure*}
\centering
\includegraphics[width=0.45\textwidth]{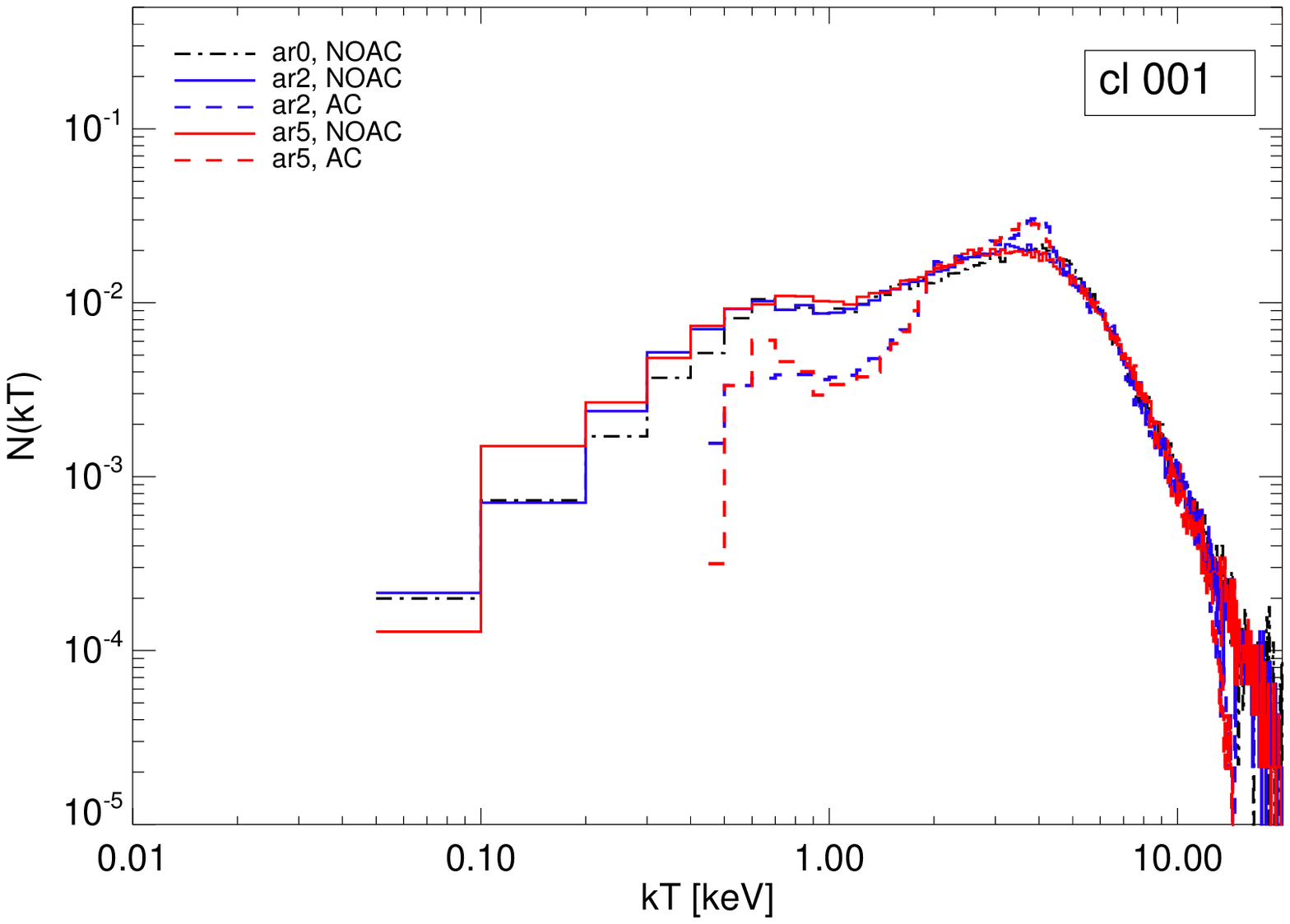}
\includegraphics[width=0.45\textwidth]{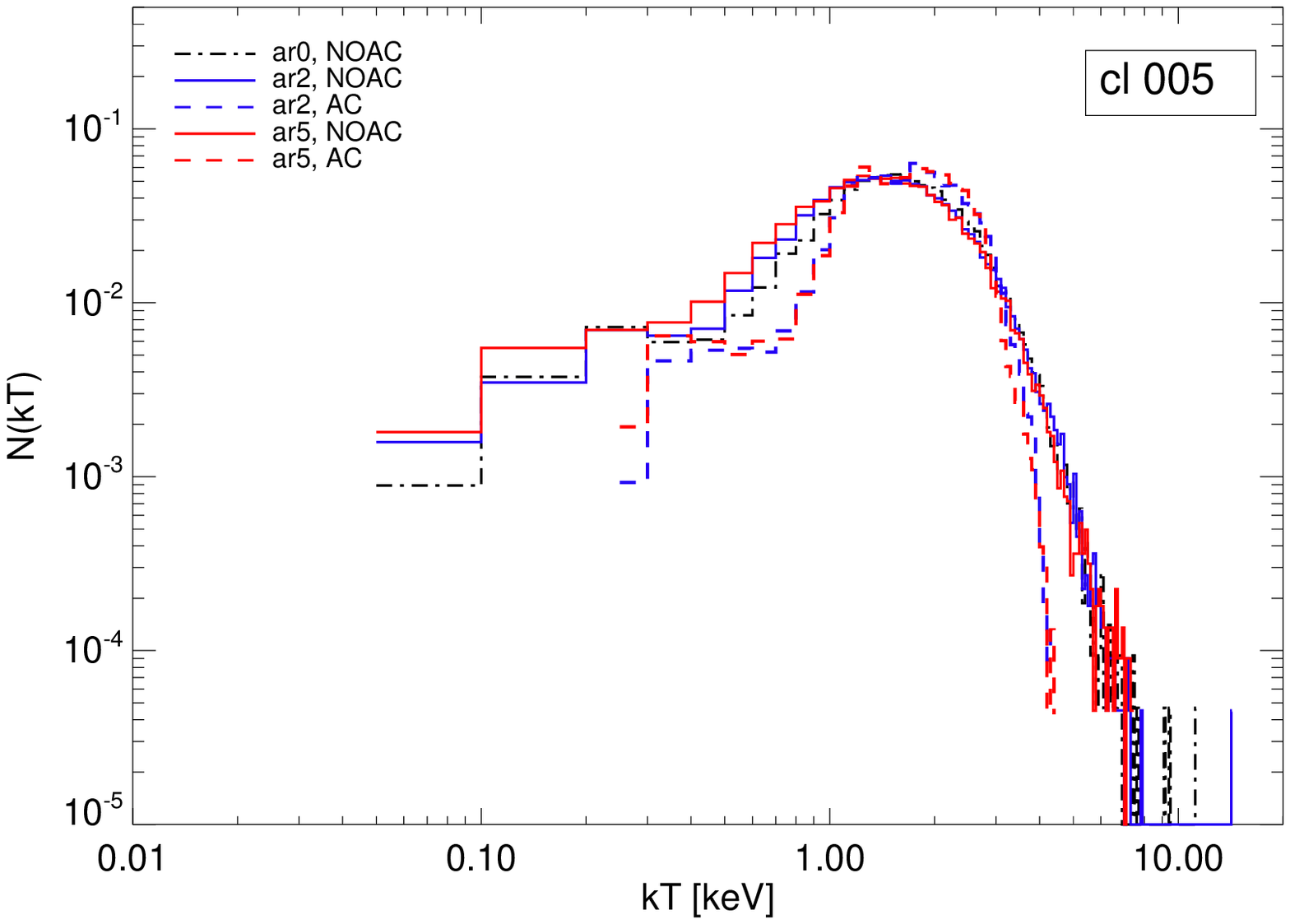}\\
\includegraphics[width=0.45\textwidth]{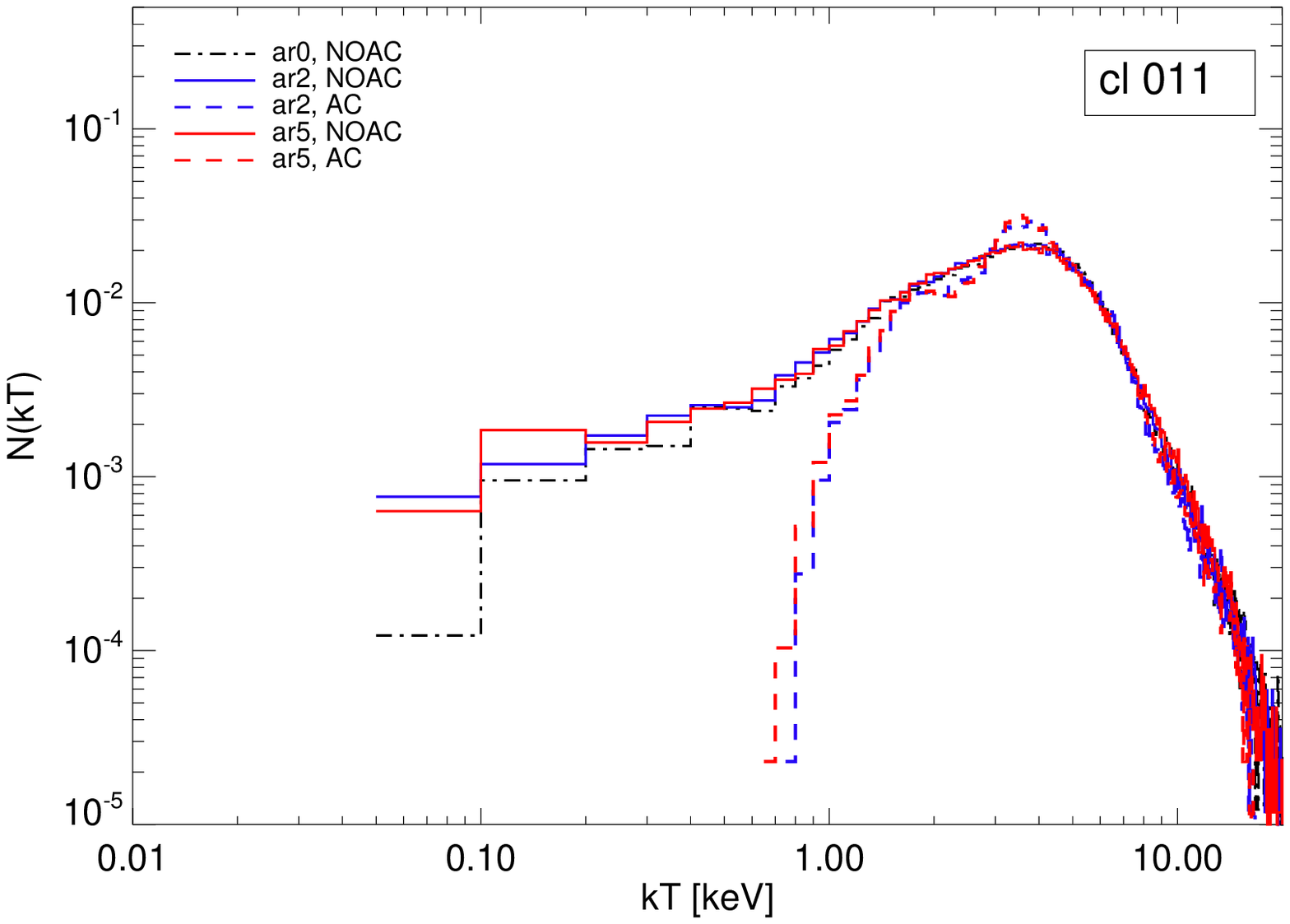}
\includegraphics[width=0.45\textwidth]{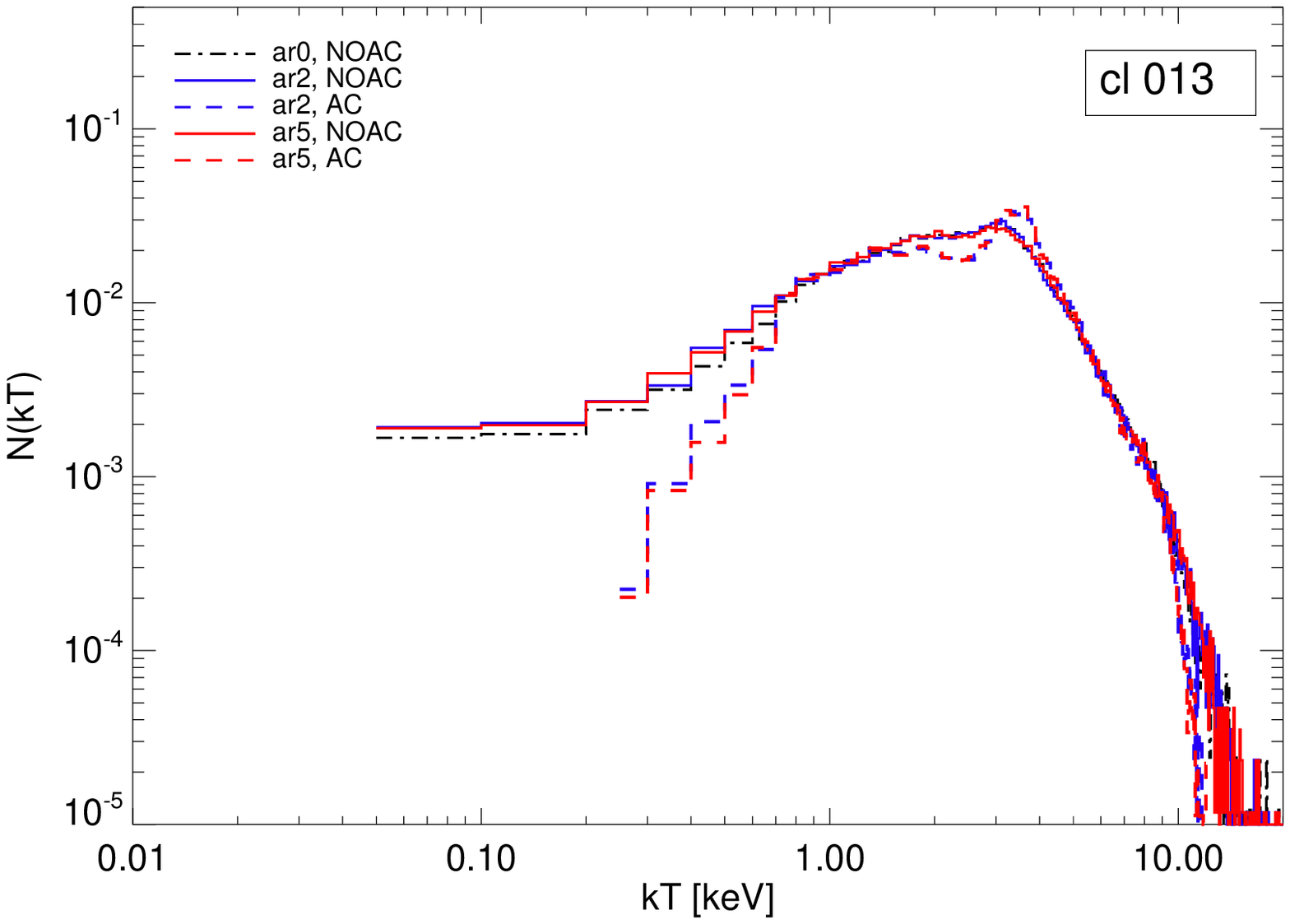}\\
\includegraphics[width=0.45\textwidth]{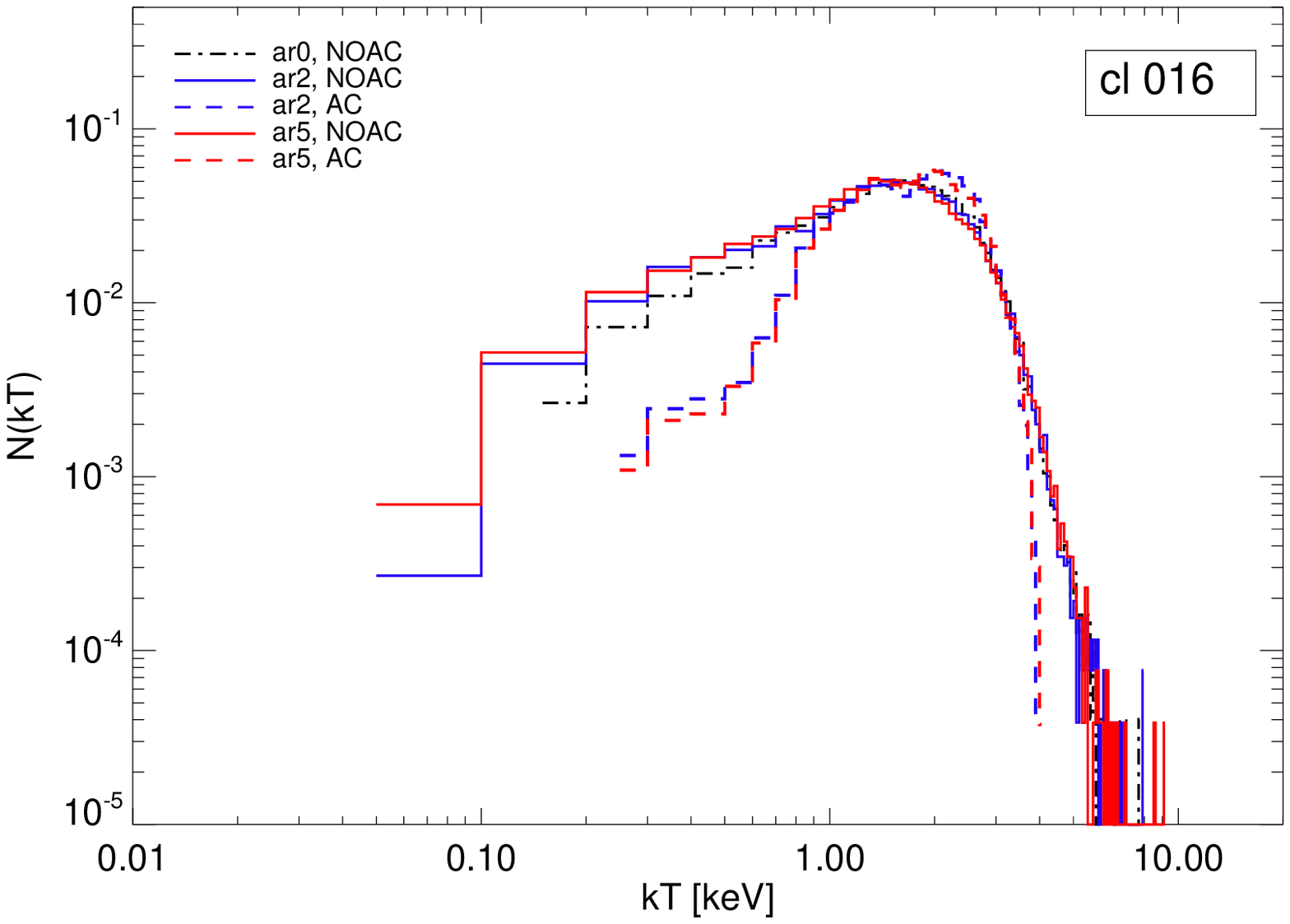}
\includegraphics[width=0.45\textwidth]{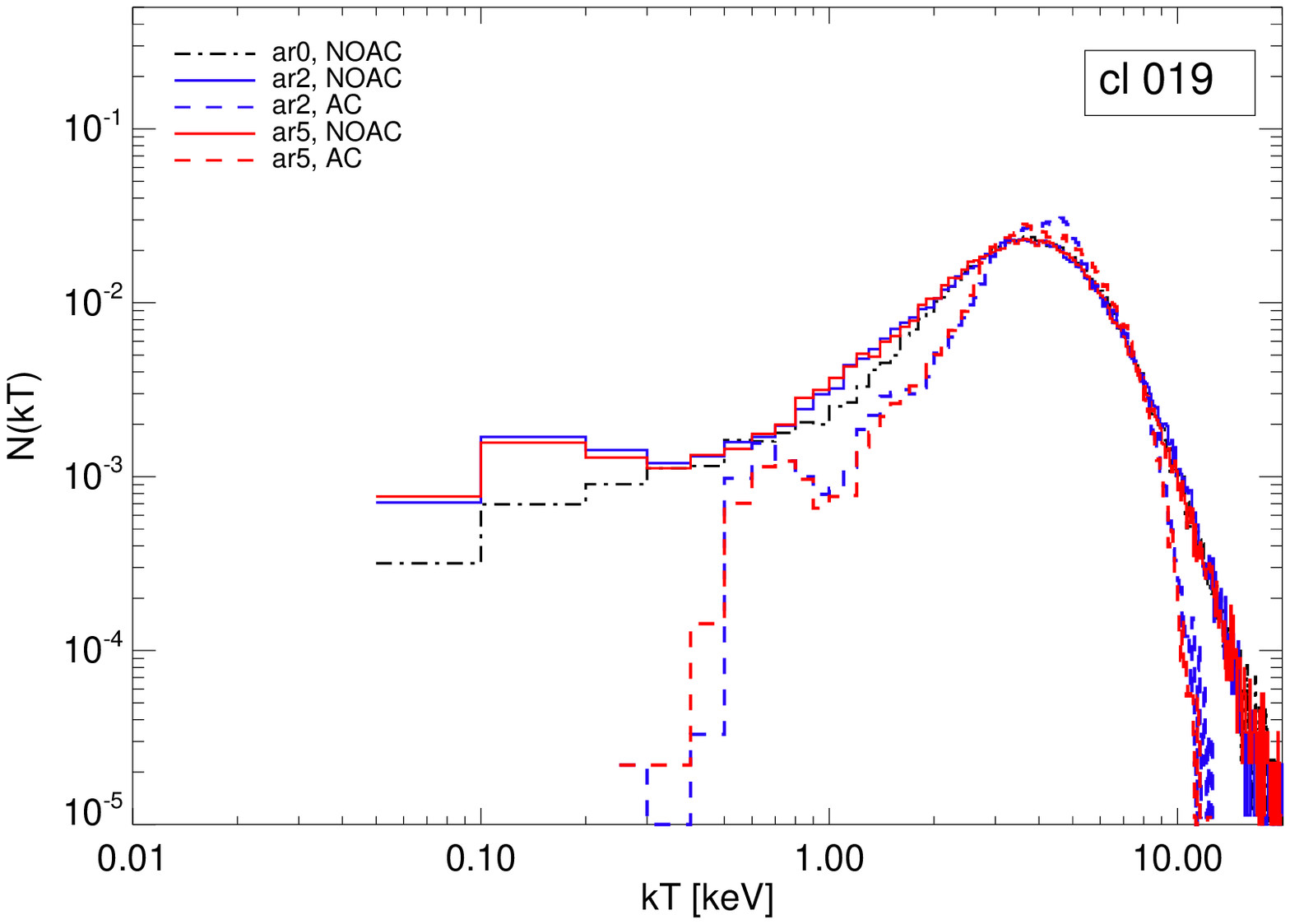}\\
\includegraphics[width=0.45\textwidth]{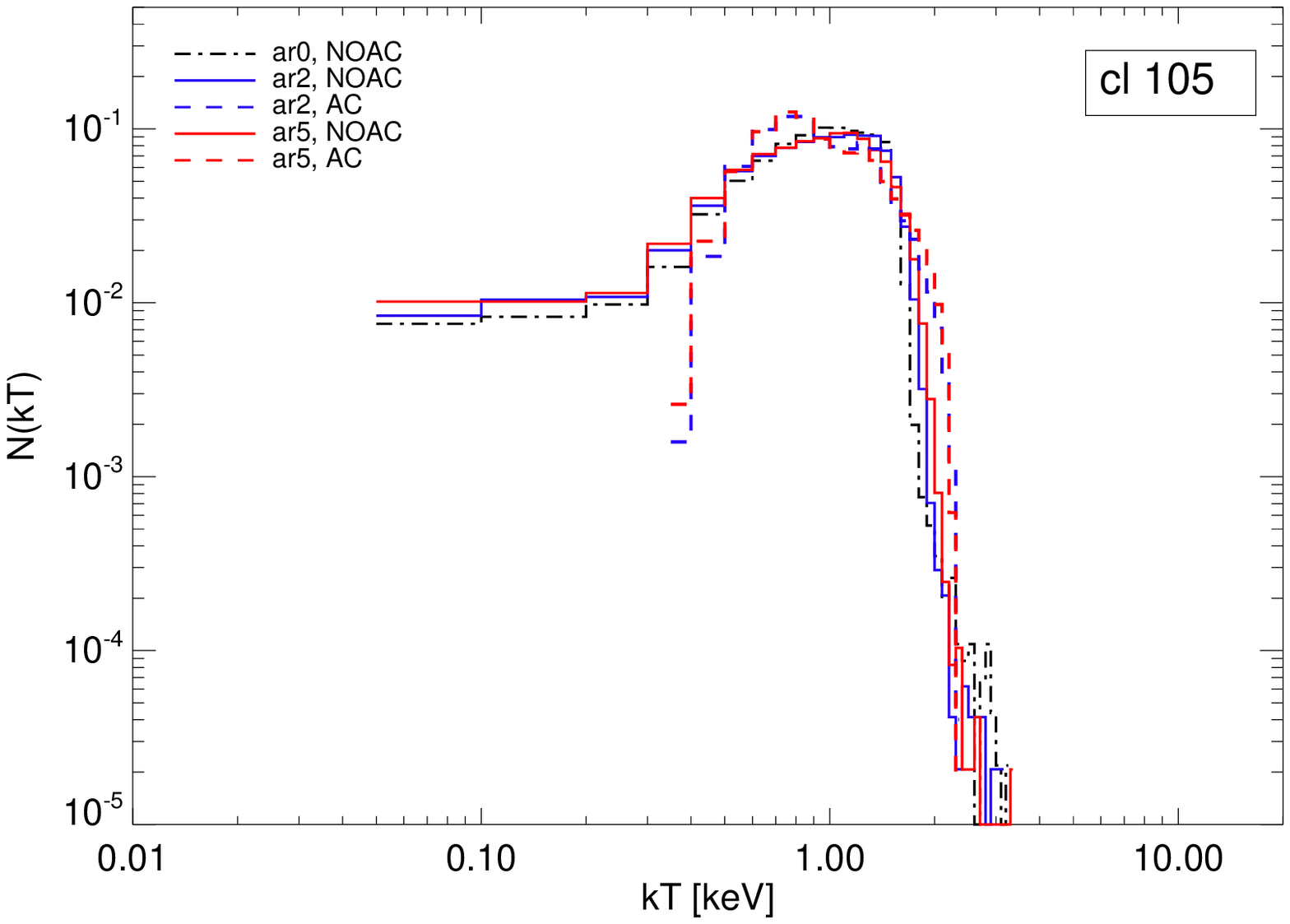}
\includegraphics[width=0.45\textwidth]{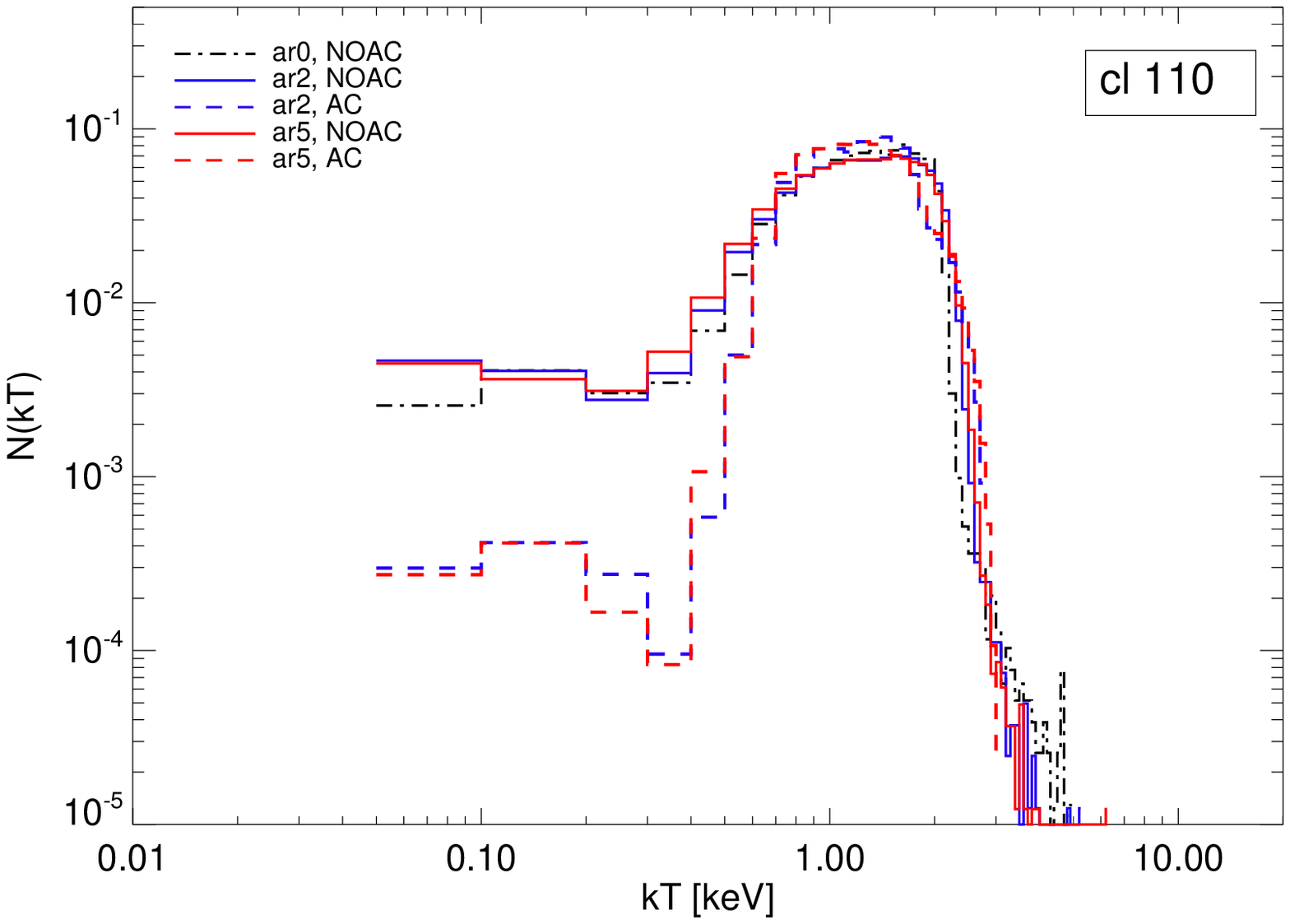}\\
\caption{Gas temperature distribution within $R_{200}$. Comparison
  among the different viscosity schemes (AV$_0$: black, dot-dashed line;
  AV$_2$: blue; AV$_2$: red) and NOAC/AC (solid/dashed lines)
  runs. Adiabatic simulations at $z=0$.}
\label{fig:kt_distrib}
\end{figure*}
\begin{figure*}
\centering
\includegraphics[width=0.95\textwidth]{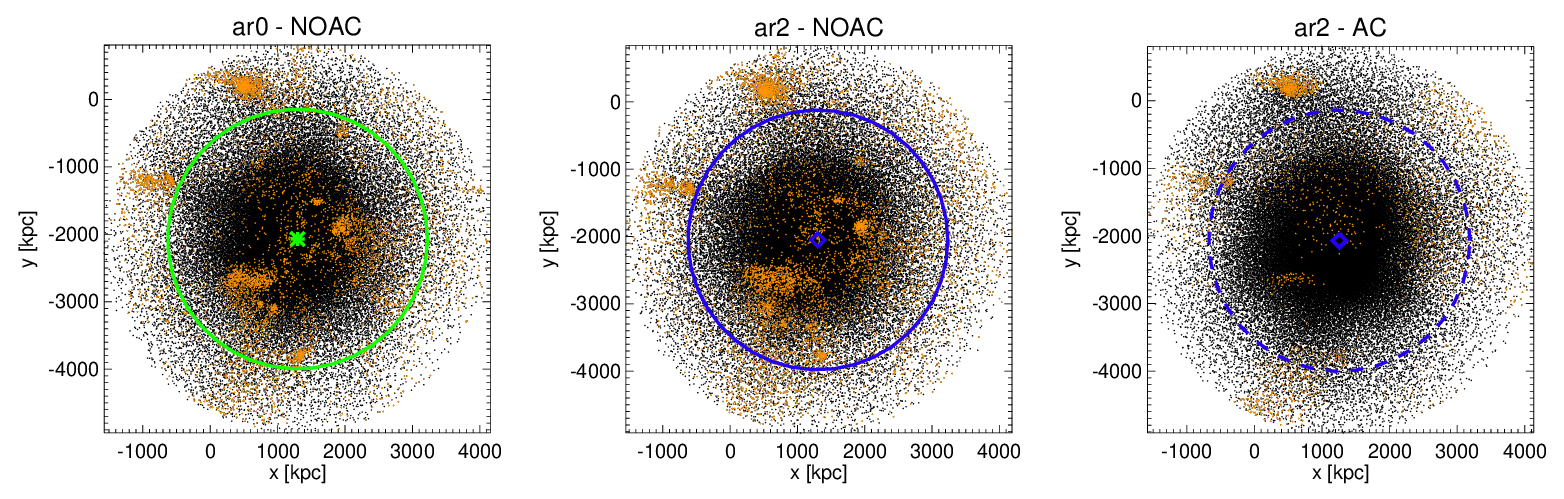}
\caption{As an example, we show the spatial distribution of the gas particles within the $R_{200}$-region of cluster cl~011, at $z=0$. The projection is made along the $z$-axis,
for the three different runs: ar0, NOAC~-~ar2 and AC~-~ar2, respectively. Overplotted are the particles with $kT<1\kev$ (orange), the center of the cluster and the $R_{200}$ radius (triangle/asterisk symbols and green/blue colors are used to remark the distinction between the AV$_0$ --- left --- and AV$_2$ --- center and right --- schemes).}
\label{fig:map_exmp}
\end{figure*}
The improvement allowed by the AC modification in the creation of
cluster entropy cores has tight connections with the ICM temperature
structure, as well.  From the temperature profiles within $\rtwo$
shown in Fig.~\ref{fig:entr}--\ref{fig:entr2} (right column) we
observe that the core of the clusters tends to be hotter in the runs
that include the AC term, consistently with an increase in the central
entropy. This has indeed the effect of improving the gas mixing,
acting as a heat diffusion term and
therefore partially heating the cold gas component.

This is confirmed by
the distribution of the gas temperature within the $\rtwo$ region,
shown in \fig\ref{fig:kt_distrib} (see also the mean temperature
  values reported in Appendix~\ref{sec:app}).  In fact, one can
note a clear suppression of the fraction of gas particles with
relatively low temperature ($kT \lesssim 1\kev$), with the strongest
differences at $kT \lesssim 0.5\kev$. For some clusters the amplitude
of the temperature distribution is damped down by up to two orders of
magnitude going from the standard NOAC runs (solid curves) to the AC
ones (dashed curves).  In \fig\ref{fig:kt_distrib} we mark the various
viscosity schemes with different colors (blue and red for AV$_2$ and AV$_5$, respectively; black, dot-dashed line for AV$_0$). From the comparison among the different
schemes, we can conclude that the impact of the AC term is dominant
with respect to the effects due to the variation in the viscosity
parameters.  Indeed, solid lines of different colors show similar
trends and the same holds, separately, for the dashed, AC curves.

From the analysis of the adiabatic runs considered here we suggest
that introducing an artificial conductivity term improves the SPH
technique, which instead might produce an over-abundance of cold gas
in its standard formulation because of the numerical approach itself.

A qualitative impression of this effect is clearly given by the maps
shown in \fig\ref{fig:map_exmp}, where the AC run (right panel) is
compared to the NOAC case, referring to both the standard and AV$_2$
viscosity schemes (left and central panel, respectively). In the
figure we show the spatial distribution of the gas in a hot cluster
(cl~011) in the sample, used as a show case.  The maps all refer to
the {\it projection} along the $z$-axis of the gas particles (black
dots) enclosed within $1.5$ times the virial radius. $\rtwo$ is
overplotted for comparison (marked as a solid circle in the NOAC cases
and as a dashed circle in the AC one) and the orange points mark the
cold-phase gas component ($kT<1\kev$).  From this, we observe that the
large-scale structure of the cluster is preserved, with very good
agreement between the runs --- the cluster resides in the same
position, similar in size and presenting the same major sub-structures
--- while the small-scale distribution of the gas is instead
perceptibly different.  In fact, we clearly see that especially in the
cluster central regions the small, cold substructures present in the
left plot, and still visible in the central one, are definitely more
diffused and less pronounced in the AC case (right panel), where some
of them are actually entirely disappeared.

Also, in order to quantify the differences related to the various
viscosity schemes and to the introduction of the artificial
conductivity term, we calculate for any given cluster the residuals of
global intrinsic properties such as total mass and radius in the
different runs, with respect to the same quantities for the cluster
reference (ar0) simulation.

\begin{figure}
\centering
\includegraphics[width=0.48\textwidth]{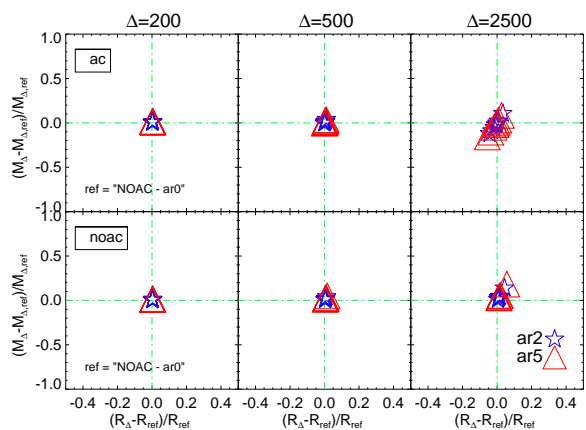}
\caption{Differences in the global properties of the clusters for the
  different runs (ar2: blue stars; ar5: red triangles; AC/NOAC
  top/bottom panels). The variations in radius and total mass are
  calculated with respect to the reference run (ar0), at
  $z=0$.\label{fig:deltar}}
\end{figure}
For the region corresponding to a given overdensity $\Delta$,
we estimate such residuals as:
\begin{equation}\label{eq:deltay}
\Delta Y = (Y - Y_{\rm ref})/Y_{\rm ref},
\end{equation}
where $Y_{\rm ref}$ corresponds to the same quantity calculated in the
standard reference run, ar0.  Results for the adiabatic simulations
are shown in \fig\ref{fig:deltar}. The panels in the upper row refer
to the AC runs and those in the lower row to the NOAC ones, while the
different columns refer to the three typical overdensities considered:
$\Delta=200,500,2500$, respectively.  The effects due to the different
SPH modifications are expected in this case to be less significant,
given the dominant role played by dark matter in shaping the global
potential well. Indeed, this is confirmed by the residuals in mass and
radius at large and intermediate radii ($\rtwo$ and $\rfive$), which
are basically consistent with zero. This indicates that, on global
scale, the main characteristics of the simulated clusters do not
depend on the treatment of viscosity or artificial conductivity as
strongly as the thermo-dynamical properties of the ICM. In the
innermost region (i.e.\ $<R_{2500}$), instead, there might be
differences introduced by the AC term or the AV scheme. In the cluster
core, in fact, the change of the ICM properties might play a more
important role. The bias in radius, and consequently in the mass
enclosed therein, is different from zero by a few percents, as seen by
the last column panels in \fig\ref{fig:deltar}. We observe that in the
AC runs the majority\footnote{One of the clusters in the sample shows
  here an opposite trend, but this is the case of a very disturbed
  system.} of the clusters present a smaller $R_{2500}$ radius (and a
smaller mass within it) with respect to the reference run (ar0). This
trend is not as evident when only the time-dependent viscosity scheme
is included in the simulations (bottom-right panel).
\begin{figure*}
\centering
\includegraphics[width=0.47\textwidth]{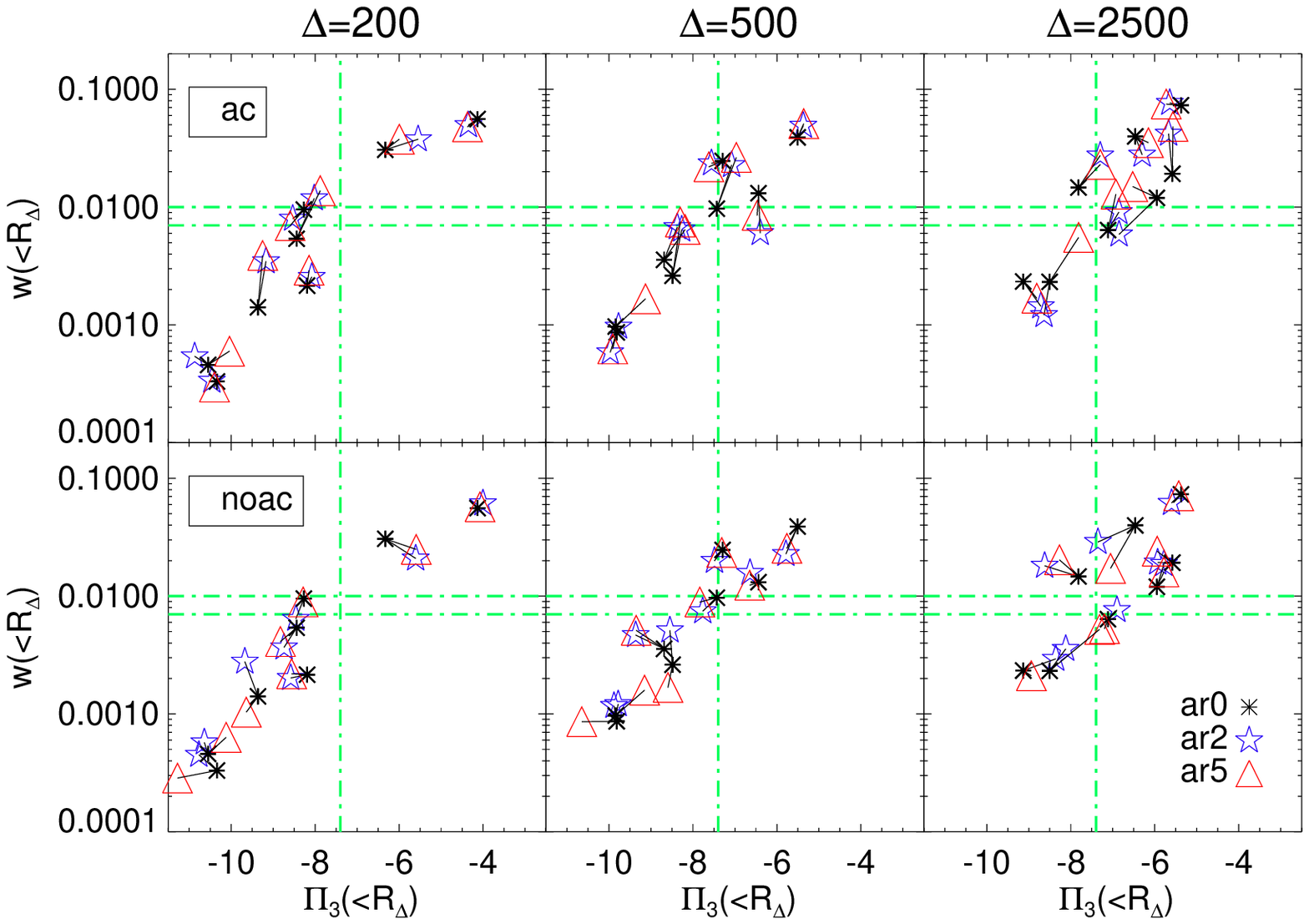}
\includegraphics[width=0.46\textwidth]{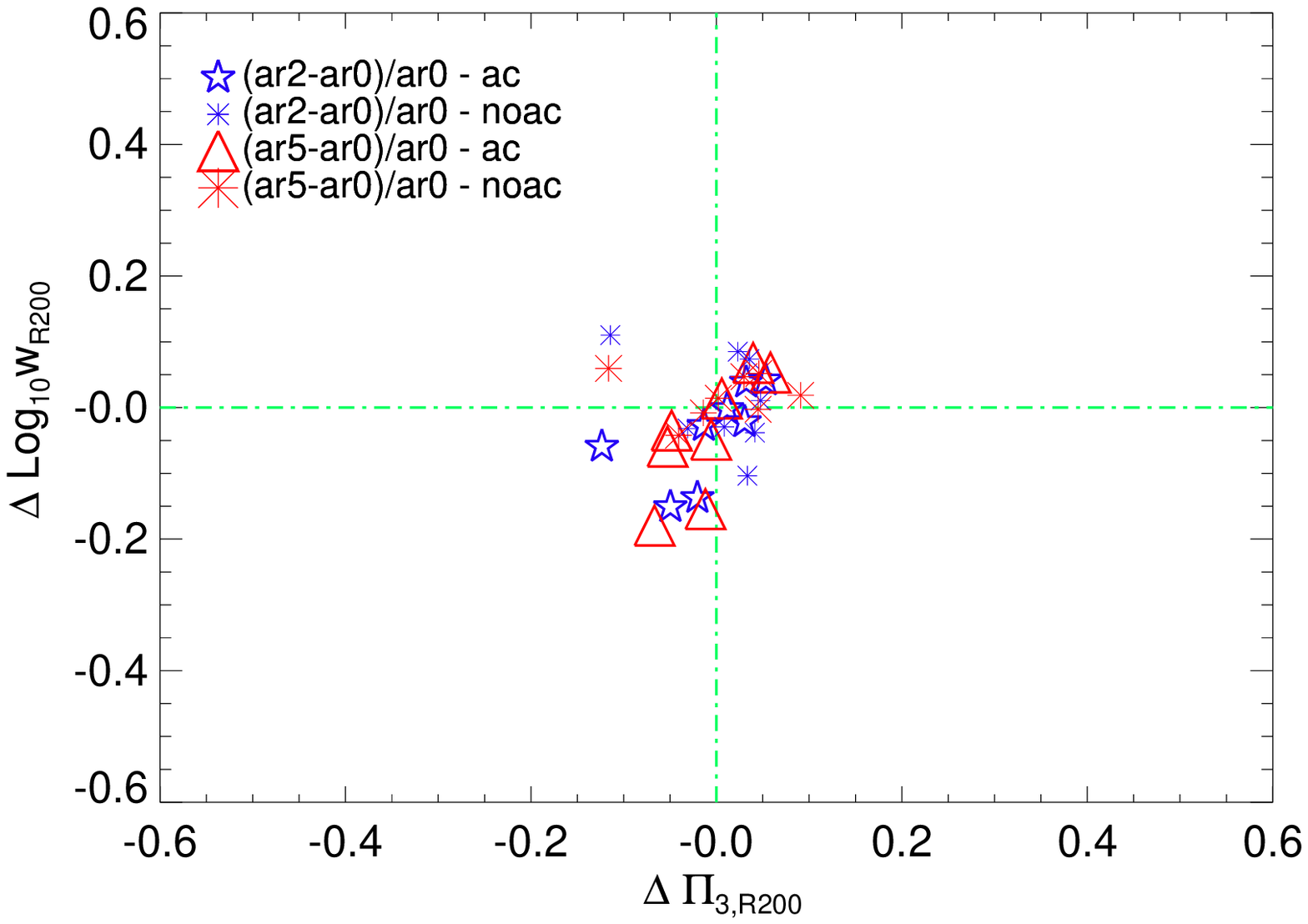}
\caption{Left-hand-side panel: power ratios and center shifts at
  $z=0$, for all the various runs and overdensities as in the
  legend. Right-hand-side panel: variation with respect to the
  reference run ``NOAC - ar0'', for $\Delta=200$.\label{fig:pw}}
\end{figure*}
%
\subsubsection{Level of substructures: centroids and power ratios}
With centroids and power ratios we quantify the level of substructures
in the clusters of the sample, at redshift $z=0$.  Here, we still
analyse the adiabatic runs (``ar''), while deferring the discussion of
radiative runs to Sec.~\ref{sec:cr}.

Interestingly, the dynamical configuration of the clusters,
marked by
the values of $w$ and $\Pi_3$ can be affected by
the change in the numerical implementation of the SPH
run, even for the same ensemble of physical processes accounted~for.

This can be seen from the left-hand-side panel of \fig\ref{fig:pw}, where we report the values of
$w$ and $\Pi_3$ for each cluster in the sample for standard (NOAC) and
AC SPH (upper and lower row, respectively), for various overdensities
($\Delta=200, 500, 2500$ in the left, central and right columns,
respectively) and for the different viscosity schemes considered
(different colors and symbols, i.e.\ black asterisks for AV$_0$, blue stars
for AV$_2$ and red triangles for AV$_5$).  In each panel of the figure we also
mark the threshold values commonly adopted to distinguish between
relaxed and disturbed clusters, corresponding to $w\sim 0.007$--$0.01$
and $P_3/P_0=4\times 10^{-8}$ (i.e.\ $\Pi_3=-7.39794$).
In the figure we notice that especially clusters with intermediate
level of substructures can undergo more relevant changes, displayed
by the lines connecting the standard case (ar0; black asterisk)
with the two AV$_2$ and AV$_5$ counterparts (blue stars and red triangles, respectively).
Despite this, we observe a fair correlation between the two
substructure estimators, basically at all overdensities, and the
systems that are either significantly relaxed or strongly disturbed
tend to remain so in all the runs, showing changes in the $w$ or
$\Pi_3$ that do not compromise their final classification, especially on
global ($<\!\rtwo$) scales.
At higher overdensities (i.e.\ $\Delta=500,2500$) changes among the
runs can also be influenced by the changes in the thermal structure of
the core, which is the region most significantly affected (as
discussed in \sec\ref{sec:rad_prof}).

In \fig\ref{fig:pw} (right) we quantify more clearly the global changes
in the cluster dynamical configuration by calculating the residuals
between the $w$ and $\Pi_3$ values in a new run with respect to the
reference (ar0) case (similarly to \eq\eqref{eq:deltay}),
\begin{eqnarray}
\Delta \Log10 w &=& \Log10 w/\Log10 w_{\rm ref} -1 \label{eq:deltaw} \\
\Delta \Pi_{3} &=& \Pi_3/\Pi_{\rm 3,ref}-1,\label{eq:deltap3}
\end{eqnarray}
for the region enclosed within $\rtwo$.

According to \eq\eqref{eq:deltaw} and \eqref{eq:deltap3}, in the case
of $\Delta \Log10 w > 0$ and $\Delta \Pi_{3} > 0$ the cluster would
show a more relaxed configuration in the considered run with respect
to the standard SPH (NOAC, AV$_0$) simulation.
From \fig\ref{fig:pw} (right) we see more quantitatively that, despite
some scatter, the clusters do not show any significant trend to become
on average neither more relaxed nor more disturbed.
Overall, we observe variations in both $\Log10 w$ and $\Pi_3$
comprised between few and $\sim 10$--$15\%$ (generally correlated), but
the distributions of the residuals are centered very close to zero, in
the logarithmic scale of the figure. Typically
the effect is even milder when only the viscosity
scheme is modified.
%
\begin{figure*}
\centering
\includegraphics[width=0.44\textwidth]{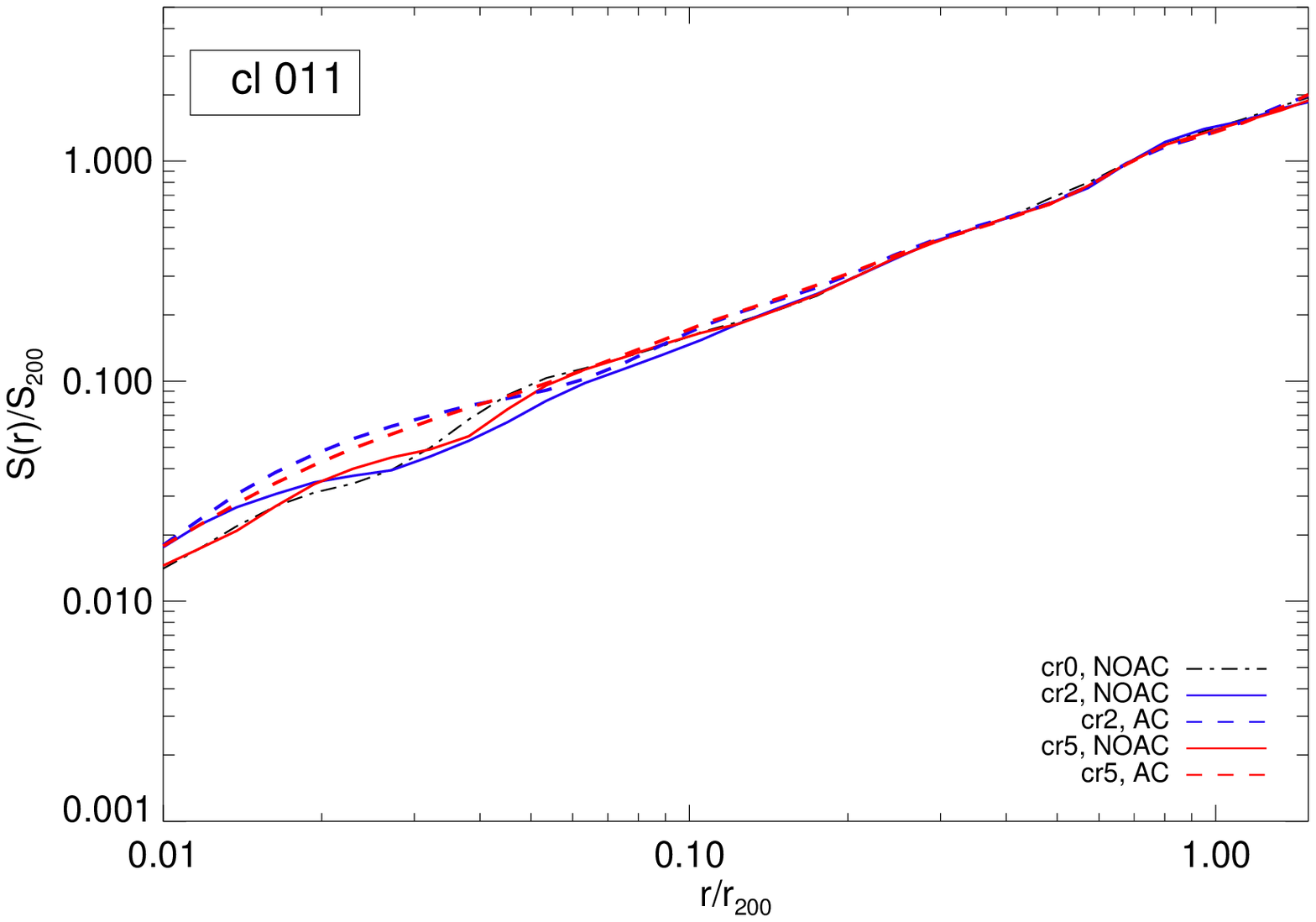}\qquad
\includegraphics[width=0.44\textwidth]{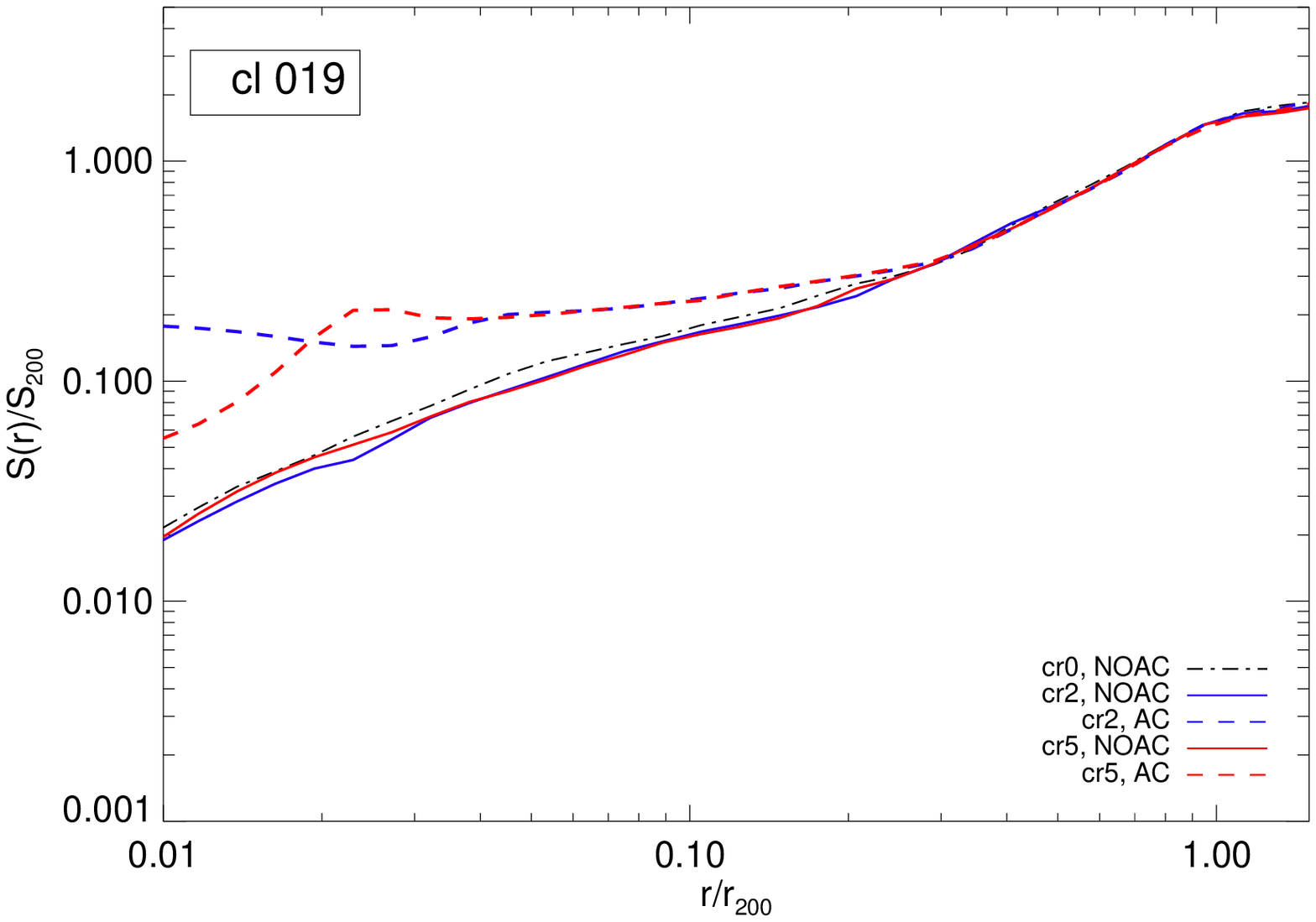}\\
\includegraphics[width=0.44\textwidth]{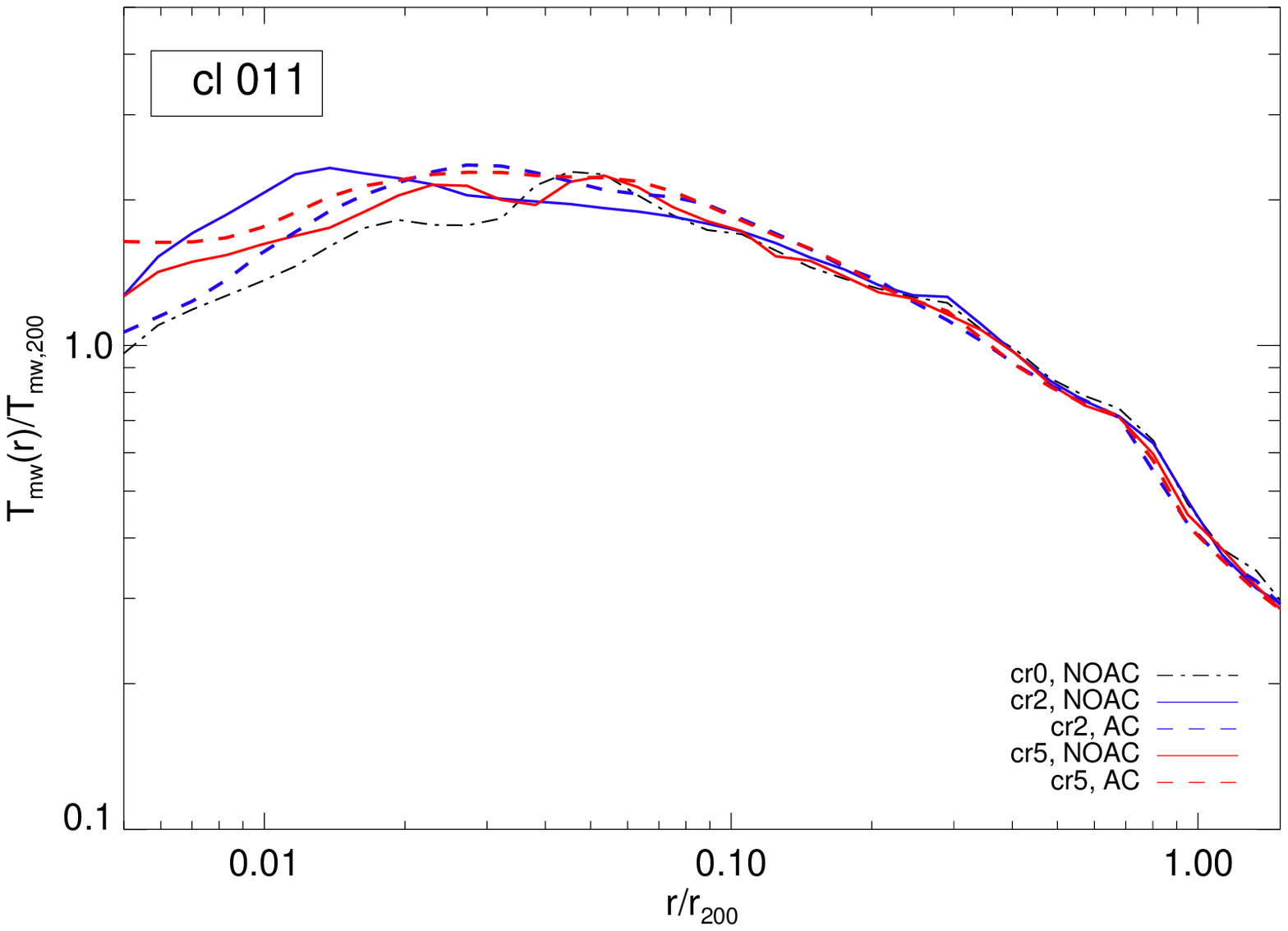}\qquad
\includegraphics[width=0.44\textwidth]{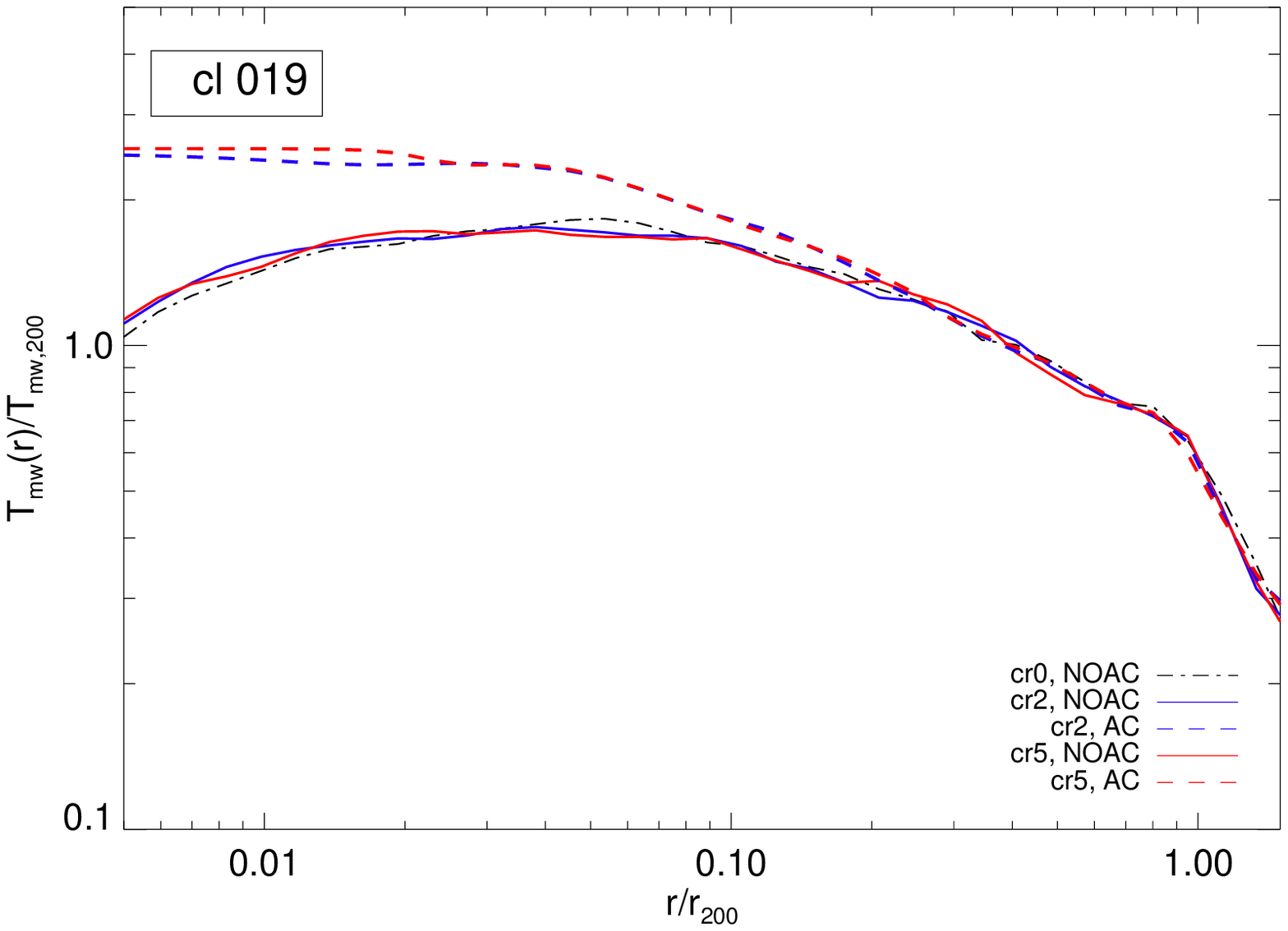}
\caption{Similar to \fig\ref{fig:entr}: entropy (top panels) and temperature (bottom panels) profiles for the two clusters chosen to present results from the radiative (``cr'') runs at $z=0$: cl~011~(left) and cl~019~(right).\label{fig:entr_cv}}
\end{figure*}
\subsection{Radiative runs}\label{sec:cr}
At variance with the adiabatic runs presented in the previous
sections, we consider here more realistic simulations of clusters, in
which not only gravity is responsible for the structure formation, but
also more complex physical processes are traced in order to describe
and model the baryonic component.
In this section we discuss results for the radiative runs, where
cooling of the gas, star formation, chemical enrichment and energy
feedback following supernova explosions are included in the
simulations.
The inclusion of these physical phenomena is expected to complicate
the picture discussed so far and to interact with the numerical
schemes, possibly affecting the impact due to the
AV and AC modifications.
In order to explore this, we
repeated here the same analysis done in the previous sections, using
the same set of simulated clusters.  In the following, we will
restrict ourselves to the comparison and discussion of the most
important results and, for simplicity, we will rather show them for a
couple of clusters only, used as representative extreme case studies
in the sample.

In this section, we will consider as reference case the radiative run:
``cr0'' (NOAC, AV$_0$).
\subsubsection{Comparison between adiabatic and radiative runs}
With respect to adiabatic simulations, radial profiles of entropy and
temperature for the radiative clusters seem to be less affected by either
the modified viscosity scheme and the AC term, although mild differences
among the various implementations can still be observed in the
innermost region ($<0.1$--$0.2\rtwo$).

In \fig\ref{fig:entr_cv} we display two clusters in the sample,
presenting very small effects in one case (cl~011, left) and showing
more significant variations of the core thermal structure in the other
one (cl~019, right). The entropy and temperature profiles of the
second cluster show in fact more similarities to the results discussed
for the adiabatic simulations and the introduction of the AC term
seems to play a more significant role in shaping the gaseous core,
despite gas cooling and star formation. Overall, however, the
majority of the clusters in the sample rather resemble the behaviour
of cl~011,
i.e.\ they typically do not present higher central entropy and temperature nor a
significant entropy-profile flattening towards the cluster center, with respect
to standard SPH simulations.

From the velocity power spectra of the gas in the cooling runs,
we can still conclude that the primary source of energy injection into the
ICM consists of gravity-driven merging and accretion processes, as in
fact the curves present a maximum at small wavenumbers $k$,
corresponding to the largest spatial scales in the systems.
Similarly to the adiabatic case, the major differences among the runs
are still visible mainly at high values of $k$, corresponding to
turbulent motions on small spatial scales, and are more prominent than
in the adiabatic counterparts for the majority of the clusters
considered.
\begin{figure*}
\centering
\includegraphics[width=0.48\textwidth,height=0.23\textheight]{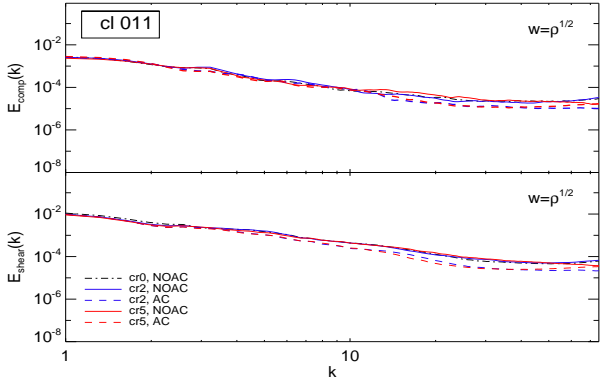}
\includegraphics[width=0.48\textwidth,height=0.23\textheight]{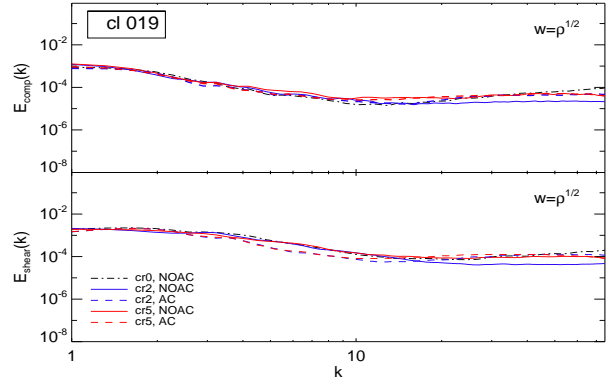}
\caption{Similar to \fig\ref{fig:velpow}: velocity power spectra for the two clusters chosen to present results from the radiative (``cr'') runs at $z=0$: cl~011~(left) and cl~019~(right).\label{fig:velpow_cv}}
\end{figure*}
The major difference here is that we observe a generally higher amplitude of the spectra
at small spatial scales, independently of the particular run
considered.  For comparison, the spectrum components for the adiabatic
clusters were typically spanning up to $\sim 6$ order of magnitudes
(see \fig\ref{fig:velpow}), while the spectra of the radiative
counterparts only span $\sim$2--3 order of magnitudes across the
$k$-range considered (see \fig\ref{fig:velpow_cv}).
In particular, after an initial similar decaying shape of the velocity
spectrum, at $k\sim 10$ the trend is basically arrested and for
$k\gtrsim 10$ both compressive and shearing components remain nearly
flat or even grow again with increasing $k$ (decreasing scale).  This
would suggest the presence of an additional source driving turbulence
at small scales, likely linked to the development of compact cold gas
clumps in the central cluster region that interact with the ambient
ICM.  With respect to the runs including only the AV modification
(solid lines in \fig\ref{fig:velpow_cv}), this finding is consistent
with the results presented in V11.

\begin{figure}
\centering
\includegraphics[width=0.48\textwidth,height=0.23\textheight]{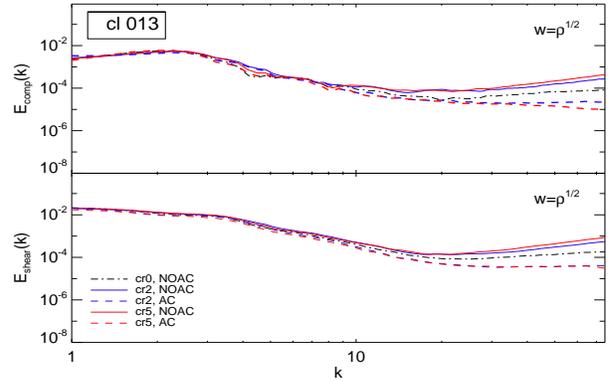}
\caption{Velocity power spectra of a cluster (cl~013) at $z=0$ where the effect of the AC term in suppressing the high-$k$ tail is significant even in the ``cr'' runs.\label{fig:velpow_cv2}}
\end{figure}

By including the additional AC term, however, we find that this effect
is moderately reduced (dashed lines in \fig\ref{fig:velpow_cv}).
Therefore, the general feature presented by the adiabatic case, where
the introduction of the AC term seems to suppress the amplitude of the
spectrum despite the opposite increasing trend due to the AV
modification, survives the inclusion of cooling and star formation.
For some clusters in the ``cr'' sample (see, as an example, the case
of cl~013 in \fig\ref{fig:velpow_cv2}) the impact of the AC term is
even stronger than in the adiabatic case, causing the amplitude of the
velocity spectrum to decrease by 1--2 orders of magnitude, at
$k\gtrsim 50$, with respect to the runs with the AV modification only.
Interestingly, this further suggests that the AC term, favouring
diffusion, tends to heat the gas, smooth out the cold substructures
and consequently soften the power of small-scale turbulent motions
originating from the
interaction between cold compact clumps and surrounding gas.

\begin{figure*}
\centering
\includegraphics[width=0.47\textwidth]{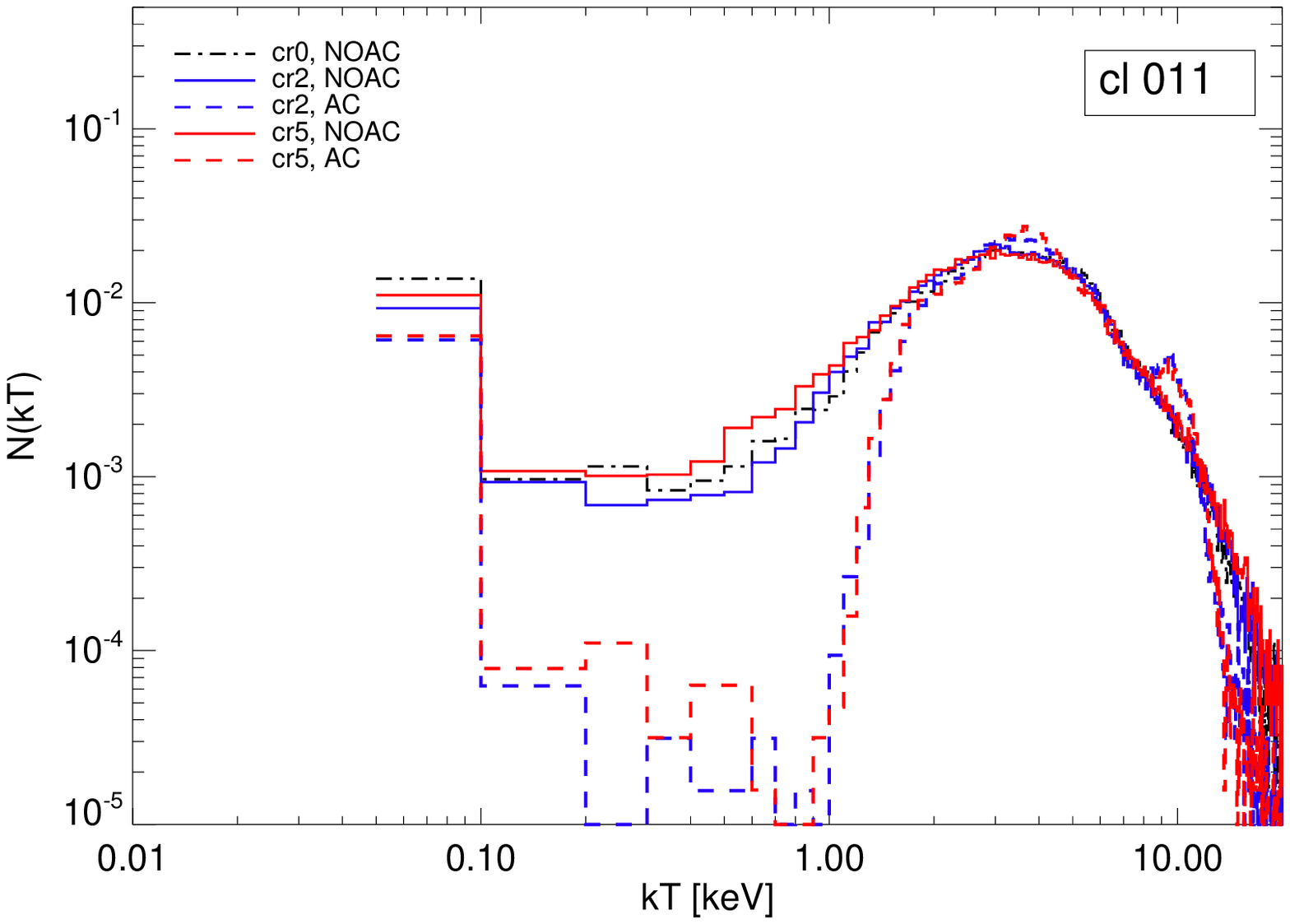}
\includegraphics[width=0.47\textwidth]{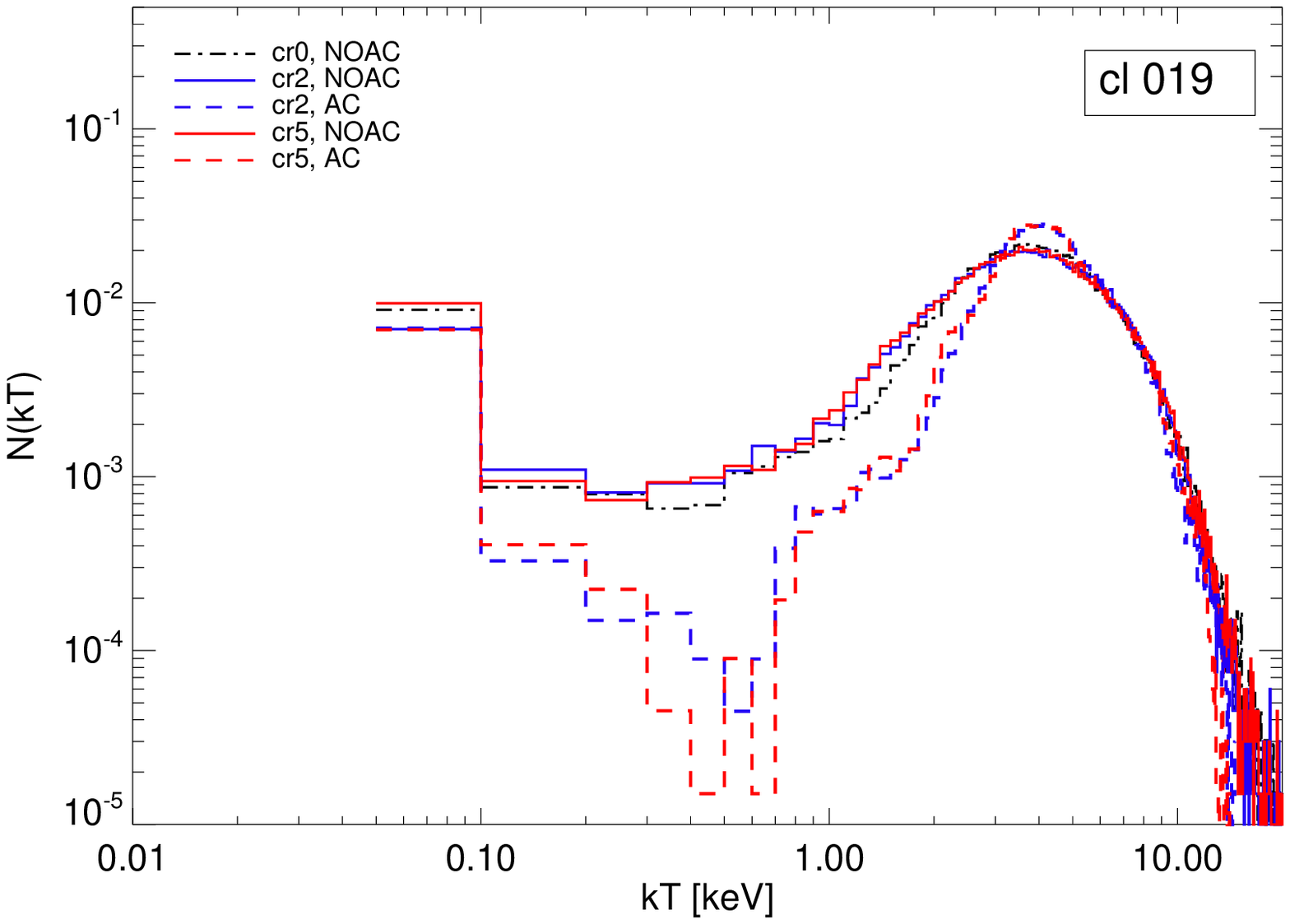}
\caption{
Similar to \fig\ref{fig:kt_distrib}: gas temperature distribution within $R_{200}$ for the two clusters chosen to present results from the radiative (``cr'') runs at $z=0$: cl~011~(left) and cl~019~(right).\label{fig:kt_distrib_cv}}
\end{figure*}

%
Although, on average, the effects of the AC term are not significantly
visible from the radial entropy and temperature profiles, the ICM
thermal structure is actually affected by it.
In fact, even in the ``cr'' simulations, this seems to have a
systematic impact on the temperature distribution of the gas within
$\rtwo$, reducing the fraction of the cold component with
$0.1<kT<1\kev$.  Similarly to the features discussed for the ``ar''
case (\sec\ref{sec:glob}, \fig\ref{fig:kt_distrib}),
the AC curves (dashed lines) in
\fig\ref{fig:kt_distrib_cv} show a
suppression up to 2 orders of magnitude at low temperatures, with
respect to the peak of the distribution.

The major difference in the ``cr'' simulations is that the
$kT$-distributions generally tend to increase again at the very low
temperature-end, i.e. for $kT \lesssim 0.2$--$0.1\kev$.  We expect
this very cold gas component to be associated to dense gas regions
that are probably about to form stars.
Nevertheless, for all the clusters of the sample the AC
simulations present a significant suppression, as displayed in
\fig\ref{fig:kt_distrib_cv}, of the cooler gas component
(see also Appendix~\ref{sec:app} for global
mass-weighted temperatures).  This confirms the robustness of the
result independently of the peculiar properties of the clusters (mass,
dynamics, thermal state).

\begin{figure*}
\centering
\includegraphics[width=0.95\textwidth]{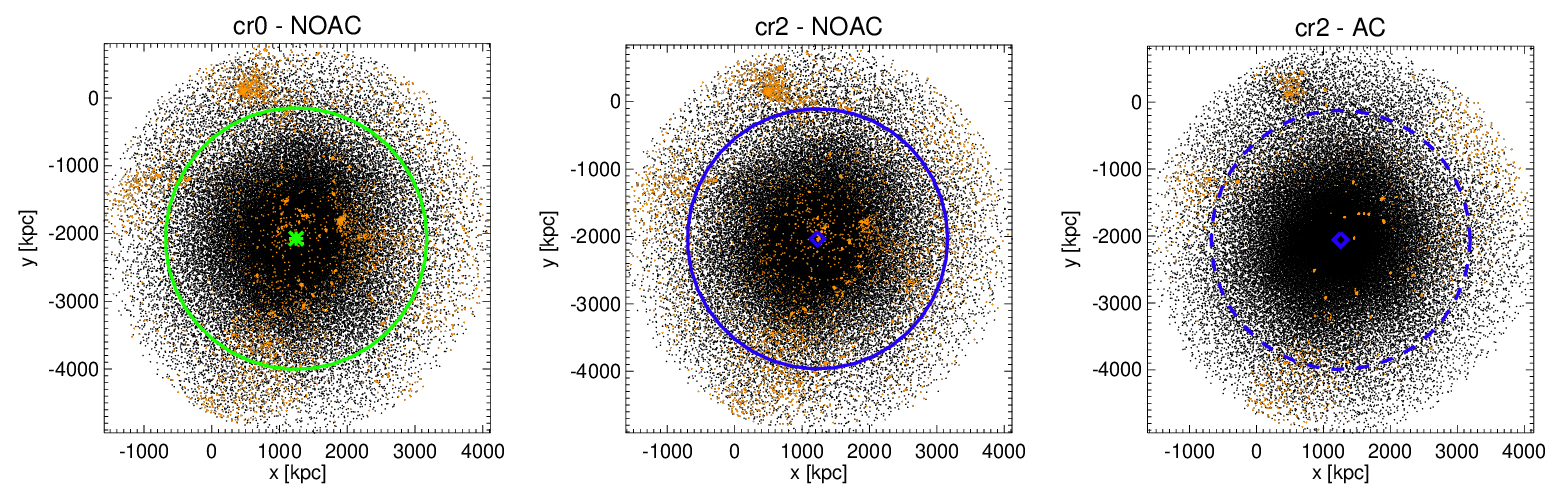}\\
\includegraphics[width=0.95\textwidth]{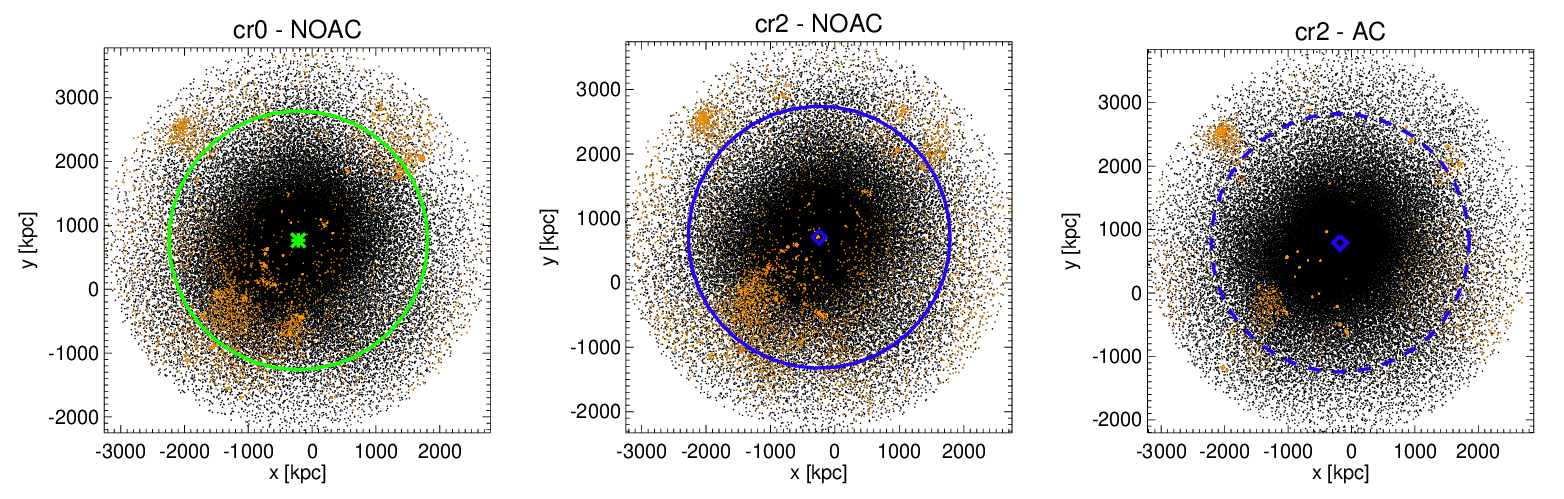}
\caption{Same as Fig.~\ref{fig:map_exmp}, but for the two example cases chosen to present radiative (``cr'') runs, at $z=0$: cl~011~(top) and cl~019~(bottom).\label{fig:map_exmp_cv}}
\end{figure*}

Similarly to \fig\ref{fig:map_exmp}, we also visualize the different
distribution of the cold ($kT<1\kev$) gas in
\fig\ref{fig:map_exmp_cv}, for the two ``cr'' show cases.  Despite the
competing effects of cooling and AC (which favours instead gas mixing and
heating), we still find visible differences between the AC map (for
AV$_2$; right panels) and the standard NOAC ones, for both viscosity
schemes (standard AV$_0$, and AV$_2$; left and central panels).  We
confirm here the tendency to preserve the large-scale structure and
the main gas substructures, while the small-scale configuration
is instead clearly modified. Unlike the adiabatic case, however, there
is here a certain persistence of the very small cold blobs even in the
AC run.

%
In general, the results from the analysis of the ``cr'' runs suggest
that on the global scale, the effects due to the highly-non-linear
baryonic processes are dominant with respect to the numerical
treatment of both viscosity and conductivity for the gas.
Nonetheless, some noticeable effects from both of these improved SPH
modifications can still be appreciated even in the presence of radiative physics.

\begin{figure*}
\centering
\includegraphics[width=0.47\textwidth]{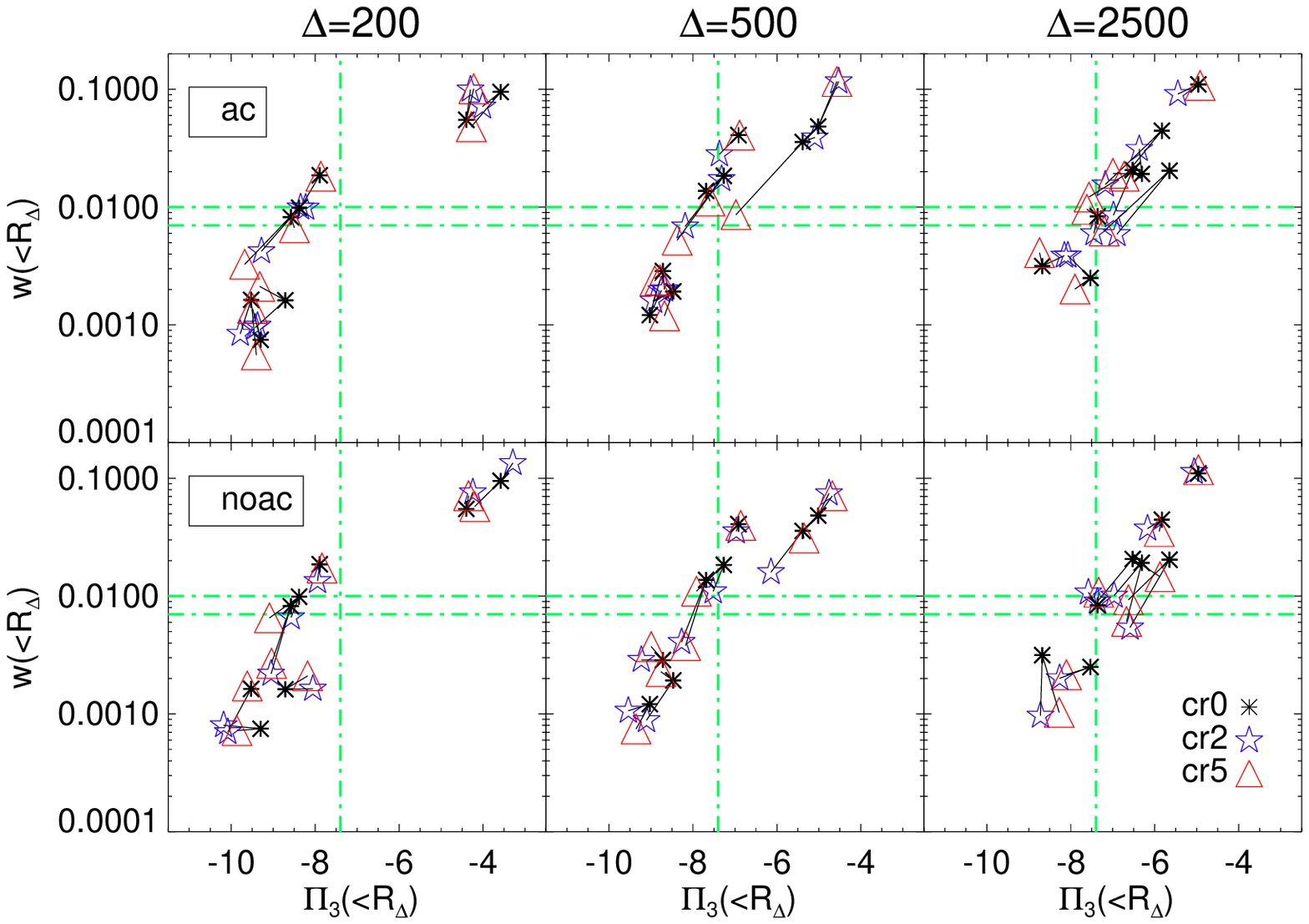}
\includegraphics[width=0.46\textwidth]{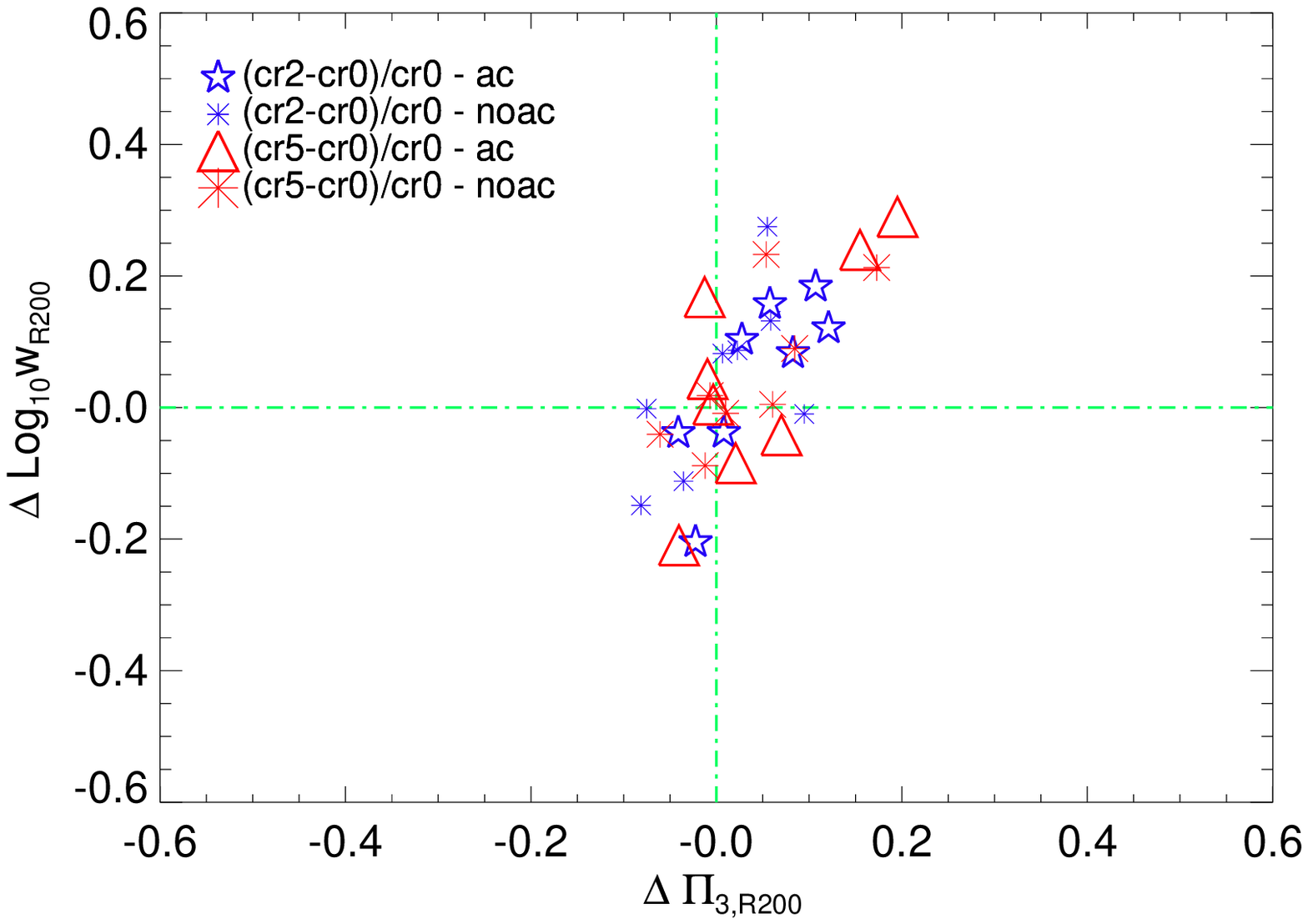}
\caption{Same as Fig.~\ref{fig:pw}, done for ``cr'' runs.\label{fig:pw_cv}}
\end{figure*}

For instance, the centroid and power-ratio diagnostics for the ``cr''
clusters show that the AV$_2$/AV$_5$ and AC modifications appear to
produce, on average, more regular clusters than the standard
SPH-AV$_0$ implementation. This is displayed in \fig\ref{fig:pw_cv} (left-hand-side panel)
for the three usual overdensities and, more quantitatively, in
the right-hand-side panel for the $\rtwo$-region.  The residuals
calculated as in \eq\eqref{eq:deltaw} and \eqref{eq:deltap3} show here
a mild shift towards $\Delta \Log10 w > 0$ and $\Delta \Pi_{3} > 0$, with
positive variations up to $\sim 30\%$ of both $\Delta \Log10 w$ and
$\Delta \Pi_{3}$.
This result can still be connected to the change in the thermal
properties of the gas, mainly to the heating contributed especially by
the improved artificial conductivity.
Interestingly, this would provide further support to the role played
by turbulence in heating the ICM and therefore compensating radiative
cooling \cite[see discussion in V11 and
,e.g.,][]{fujita2004,fujita2005}.

\begin{figure}
\centering
\includegraphics[width=0.48\textwidth]{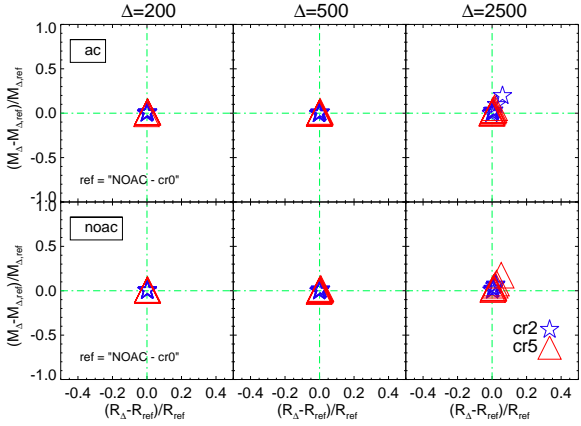}
\caption{Same as Fig.~\ref{fig:deltar}, done for ``cr'' runs.\label{fig:deltar_cv}}
\end{figure}

%
Other global properties, such as total mass and radius, instead, show
basically no evident dependence on the particular run (see
\fig\ref{fig:deltar_cv}). Therefore, we conclude that the differences
in the temperature distribution or in the dynamical indicators (both
for the $\rtwo$-region) cannot be due to any relevant difference in
the spatial selection of the gas.  Only the innermost region,
corresponding to an overdensity $\Delta=2500$, presents very mild
variations with respect to the reference run (cr0), similarly to what
we observed for the adiabatic sample.
\subsection{Observable effects on the X-ray spectra}\label{sec:phox}
The effects of the numerical treatment of SPH on the ICM properties
can also affect the expected X-ray features,
which can be highly useful for a direct comparison between simulated
and observed clusters.
Even though major discrepancies between numerical findings and X-ray observations
are expected to relate mainly to the physical processes accounted for in the
simulations, also the numerical approach itself can play a role in
this.

In order to explore the effects of the AV and AC modifications of SPH
on the velocity field and thermal structure of the ICM,
we generate and analyse X-ray synthetic observations of the cluster
core adopting the high energy-resolution spectrometer planned
to be on board the upcoming Athena mission \cite[][]{athena2013}. The
motivation for this choice resides in the possibility to reliably
constrain the gas (line-of-sight --- l.o.s.) velocity dispersion from the
detailed spectral analysis of well resolved emission lines from heavy
ions (e.g. iron), for which the line broadening can be dominated by
non-thermal motions rather than by thermal ones \cite[see,
e.g.,][]{inogamov2003,sunyaev2003,rebusco2008}.  Such studies, in
fact, require very good energy resolution of the X-ray spectra,
unreached so far with current telescopes but definitely achievable by
next-generation instruments, like the Athena spectrometer.

For the present study, Athena-like spectra were obtained by means
of the X-ray photon simulator PHOX\footnote{See
  http://www.mpa-garching.mpg.de/kdolag/Phox/.} \cite[for details on
the code, we refer the reader to][]{biffi2012phox,biffi2013}.

As a first step, we focus on the adiabatic sample, for which the
possible degeneracies due to radiative physics, that might compromise
the X-ray appearance of the clusters, are avoided.  As such, the pure
effects due to the numerical modifications of SPH can be more clearly
singled out.  In particular, we considered the adiabatic simulations
for a single viscosity scheme (namely, the ar2 runs) and generated the
synthetic spectra for both the standard NOAC clusters and for their AC
counterparts. Here, we consider the simulation outputs at
$z\sim0.05$\footnote{The redshift adopted in this analysis is not
  exactly equal, albeit very close, to $z=0$ in order to avoid the
  divergence of the luminosity distance and allow for realistic
  calculations of the X-ray emission.}.
An average metallicity of $0.3\Zsun$ was assumed for all the SPH gas
particles in the simulations, with the solar abundances assigned
following \cite{angr1989}. The X-ray synthetic photons were generated
from each gas element assuming a typical absorbed APEC model
\cite[][]{apec2001} for a collisionally-ionized plasma. The equivalent
hydrogen column density parameter for the WABS model
\cite[][]{wabs1983} was fixed to the value of $10^{20}\cm^{-2}$.
The final X-ray spectra were obtained by assuming a fiducial exposure
time of $50\ks$ and the response matrix file (RSP) of the Athena X-IFU
instrument\footnote{Available on the
  Athena mission website http://www.the-athena-x-ray-observatory.eu/.}.

In order to evaluate the observable impact on the X-ray spectrum due
to the changes in the ICM thermal structure and velocity field due for different numerical
implementations, we want to focus on the central region of the
clusters. For the given cosmological parameters and the very low
redshift considered here, however, the field of view (FoV) of the
X-IFU spectrometer is relatively small ($5'\times5'$) and only
covers a region with $\sim 300\kpc$-diameter (physical
units). Therefore, despite convolving the ideal photon list generated
with PHOX with the response of the chosen instrument, we assume to
cover instead a larger cluster-centered projected region, enclosed
within a radius of $\sim 300\kpc$ --- corresponding to
$\sim15$--$34\%$ of $\rtwo$, for the selected
clusters. Observationally, this would require a multiple-pointing
observation of the clusters with the X-IFU instrument.

\begin{figure}
\centering
\includegraphics[width=0.45\textwidth]{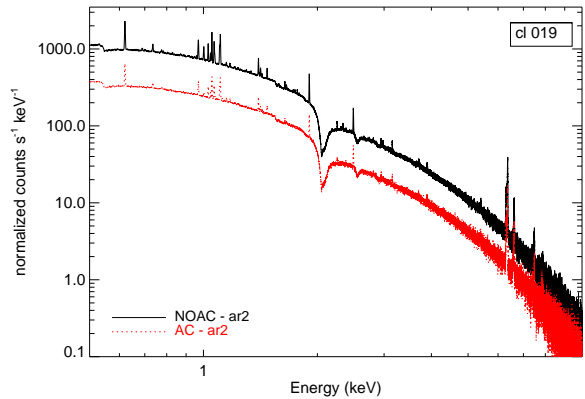}
\caption{Athena X-IFU high-resolution spectrum of a massive cluster (cl~019)
  in the sample simulated with the standard SPH code (NOAC --- black, solid line) and
  with the AC term (red, dotted line). Here, as a test case, we only compare the
  two adiabatic runs (ar2) for the same viscosity scheme (AV$_2$). The projected
  region considered is that enclosed within $300\kpc$ from the cluster
  center.}\label{fig:019xifu}
\end{figure}

In \fig\ref{fig:019xifu} we display a case example in order
to compare the mock spectra for the NOAC (black, solid) and AC (red, dotted) ar2 runs
of the same cluster. Qualitatively, one can already appreciate the
main difference between the two datasets, namely the lower
normalization of the AC spectrum.  For a better visualization, we
choose the case of a massive system (cl~019) where the difference
between the runs is particularly evident.  This is the likely
consequence of the reduced amount of cold gas in the AC runs, already
discussed in \sec\ref{sec:rad_prof} and \sec\ref{sec:glob} (see
Figs.~\ref{fig:entr}--\ref{fig:entr2} for the differences in the
cluster central region from the temperature profiles and, for global
scales, \fig\ref{fig:kt_distrib}), as the suppression of cold and
dense X-ray-emitting gas eventually causes the
number of photons to be smaller in the AC spectra.
%
More quantitatively, always referring to the ar2 runs, the original
generation of photons from the simulation output provides for the NOAC
clusters, on average, a factor of $\sim2$ more photons than for their
AC counterparts. By applying the projection unit of PHOX and
restricting the selection to the (projected) virial radius, this ratio
is still roughly preserved across the sample.  Finally, when also the
convolution with the instrumental response is taken into account and
only the central region is selected to generate the synthetic spectra,
the suppression of observed photons in the AC runs is conserved for
the smaller systems and it is even stronger for the hotter clusters,
for which we obtain spectra with a total number of counts smaller by a
factor up to 2.5--3 with respect to the NOAC case.

We expect these observable differences to depend only on the details
of the ICM thermal structure and velocity field, mainly determined by
the inclusion or omission of the AC term. In fact, for any given
cluster:
(i) we consider a single viscosity scheme; (ii) we consider the same
spatial selection for the mock observation, and for the same viscosity
parameters the sizes of the NOAC cluster and its AC counterpart are not
dramatically different; and (iii) we generate the ideal emission
assuming fixed values for gas metallicity, redshift and Galactic
absorption, reducing therefore the possible additional degeneracies.
Overall, this can have an impact on the global luminosity. Despite
these mock observations are not best-suited for such a study, we can
expect also the global X-ray luminosity of the adiabatic clusters to be partially
reduced by the changes in the ICM thermal structure due to the
AC term.
In other words, this would suggest
that the tendency of adiabatic simulations to produce
``over-luminous'' clusters \cite[see, e.g., discussions
in][]{biffi2014,planelles2014} cannot be entirely ascribed to the role
played by the physical processes not accounted for, as also the
standard SPH method itself can have an impact on the results.

\subsubsection{Velocity broadening of spectral lines}
Moreover, the spectral analysis of the restricted $6$--$7\kev$ band
(comprising the Fe lines at 6.4\kev\ and 6.7\kev) specifically allowed
us to derive estimates for the average l.o.s.\ velocity dispersion of
the gas across the central projected $600\kpc$.  To do this, we made
use of the publicly available XSPEC package \cite[v.12.7.1;
see][]{xspec1996} and fitted the spectra with an absorbed,
velocity-broadened APEC model (BAPEC model), which assumes the
distribution of the gas non--thermal velocity along the l.o.s.\ to be
Gaussian and the broadening is quantified by the standard deviation,
$\sigma,$ of the distribution.  In the fit, normalization,
temperature, $\sigma$ and redshift were free to vary.  From the
variation in redshift, known precisely from the simulation output, one
can ultimately evaluate the mean l.o.s.\ bulk motion of the gas as an
energy shift in the center of the iron emission line.

We remark here that the motions detectable in this way, however, are
those along the line of sight through the cluster, so that they also
include the contribution from global motions on the large scale (e.g.\
due to streaming patterns, rotation, etc.), in addition to the
small-scale turbulence.
Nonetheless, even though the whole cluster size along the l.o.s.\
contributes to the emission, we expect a relatively low contamination
from the foreground and background gas residing in the cluster
outskirts. In fact, by considering the Fe emission line we mainly
focus on the motions of the hot-gas component, which resides
principally in the central cluster region, especially for the low- and
intermediate-mass systems.

\begin{figure}
\centering
\includegraphics[width=0.45\textwidth]{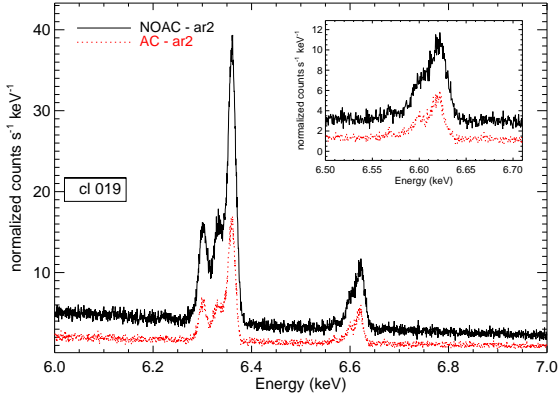}
\caption{Visualization of the Fe complex at $6.4\kev$ and zoom on the
  $6.7\kev$ line for a hot system (cl~019). The standard NOAC (black,
  solid line) and AC (red, dotted line) runs are compared, for the same
  AV$_2$ viscosity scheme.}\label{fig:019_fe}
\end{figure}

As an example, we show in \fig\ref{fig:019_fe} the case of the hot
system cl~019, where the Fe complex lines at $\sim 6.4\kev$ and $\sim
6.7\kev$ (rest-frame-energies) are very well visible in the X-ray
spectrum. Comparing the NOAC (black, solid line) and AC (red, dotted
line) spectra we observe immediately the aforementioned different
normalization.
Furthermore, we can also appreciate qualitatively the
different shape and broadening of the lines, especially for the
$6.7\kev$ one (inset panel).
In fact, the red, dotted (AC) line profile presents more sharp features, while
the broadening of the black profile is clearly more significant.
This difference in the line broadening is quantitatively estimated in
the spectral fit via the $\sigma$ value, which differ by about $\sim
50\kms$:
\begin{eqnarray*}
\sigma^{\rm NOAC}(\rm cl~019) &=& (349.710 \pm 2.24464) \kms\\
\sigma^{\rm AC}(\rm cl~019) &=& (297.460 \pm 3.05533) \kms.
\end{eqnarray*}
Remarkably, given the very good spectral resolution of the Athena
spectrometer at the energies of interest ($\sim {\rm few} eV$), the difference between the
$\sigma$ values for the two different cluster simulations considered
(NOAC and AC, both for AV$_2$) can be distinguished down to few tens
of kilometers per second.
Such result, however, could not be easily tested for the colder
systems in the sample, where the Fe lines in the mock X-ray spectra
are not very bright features and a reliable fit cannot therefore be
pursued.

In general, the result of this exercise suggests very promising
prospects about the performance of the next-generation
high-resolution X-ray instruments (like Athena, but also ASTRO-H) in
the study of cluster gas motions.  Here, however, we note that this
test was performed with idealized conditions, mainly due to the use of
adiabatic simulations, fixed average metallicity --- which would
otherwise represent an additional free parameter in the fit --- and no
inclusion of an X-ray background.  The purpose of the present
analysis, in fact, solely concerns the effects of the proposed numerical
modifications onto the X-ray observable properties of
simulated clusters, which can be very sensitive to the gas
thermo-dynamical structure and are nonetheless crucial
in order to properly compare simulations and observations.
\subsubsection{Comparison with radiative cluster simulations}
Similarly to the results presented in the previous sections, also the
X-ray synthetic properties of the radiative cluster simulations are
evaluated, and the comparison with the adiabatic case can be
discussed.  Synthetic X-ray data have been generated with PHOX, by
adopting exactly the same set of observational parameters and
instrumental responses used for the adiabatic clusters, except for the
physical properties of the systems, such as the center of the gas
distribution on which the pointing is centered.

Despite the possibility to account for the specific metallicity of the
gas elements followed within the radiative simulations, we decide to
artificially impose a fixed metallicity here, $0.3\Zsun$, like for the
adiabatic clusters.  The motivation for this choice is that we intend
to study and single out especially the effects of the numerical SPH
implementation --- of the AC term, in particular --- reducing as much
as possible any additional source of degeneracy.  Here, we limit the
discussion to the interplay between the AC term and the
baryonic physics (gas cooling and star formation).  In
fact, a metallicity value or metal abundances that vary for each gas
element would further increase the level of complexity of the
spectrum, affecting the comparison of different clusters and between
radiative and adiabatic runs.
\begin{figure*}
\centering
\includegraphics[width=0.45\textwidth]{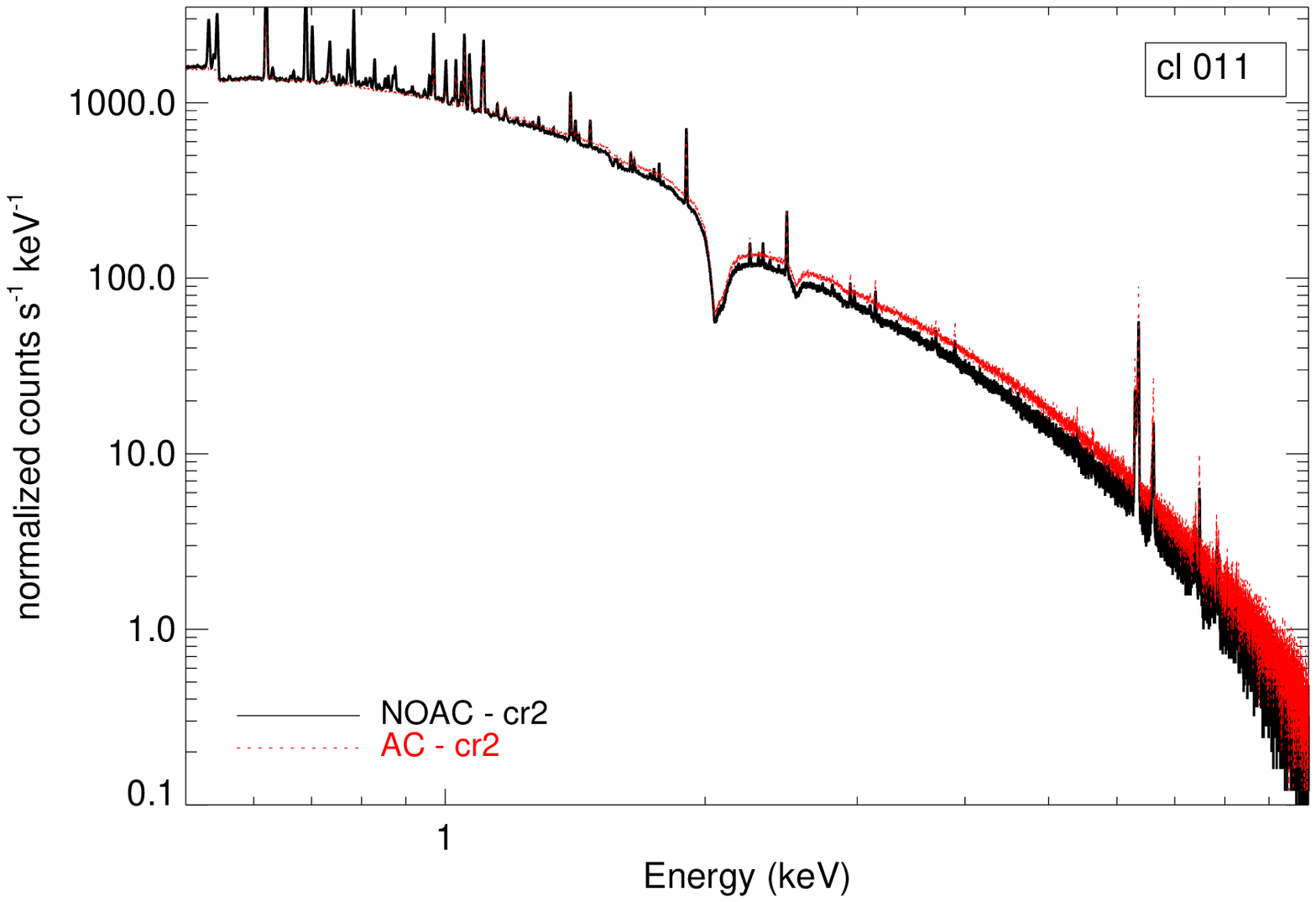}
\includegraphics[width=0.45\textwidth]{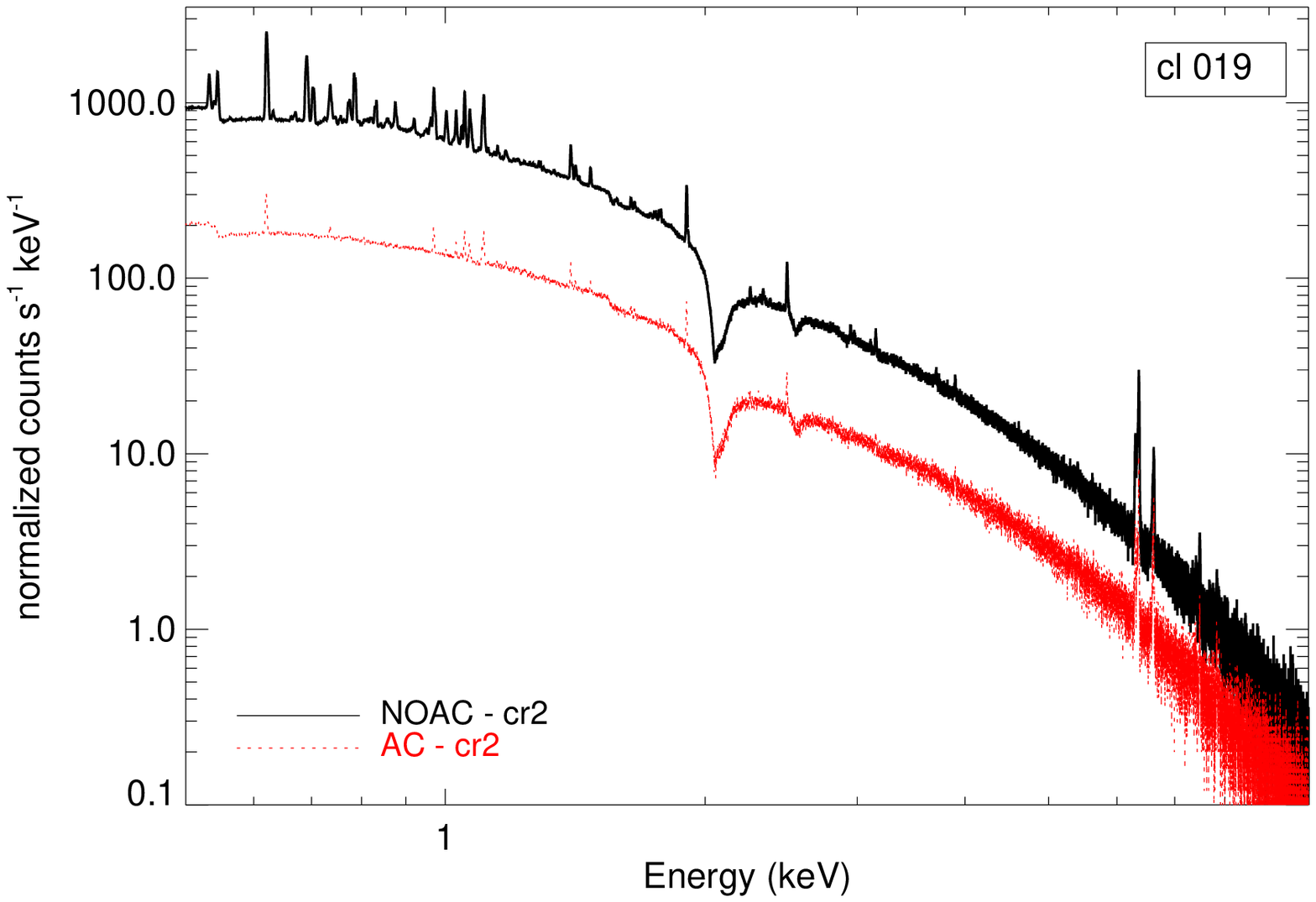}\\
\includegraphics[width=0.45\textwidth]{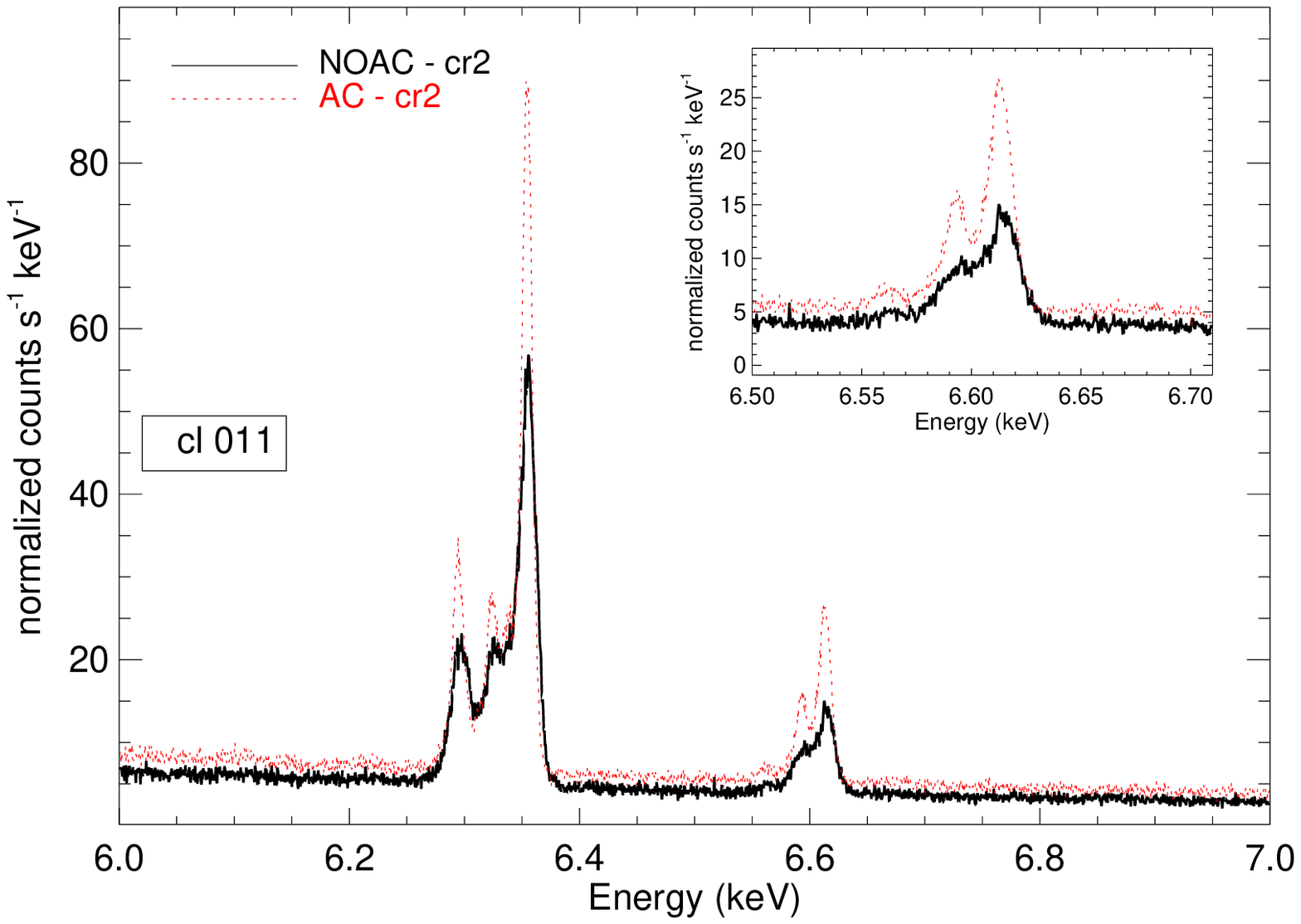}
\includegraphics[width=0.45\textwidth]{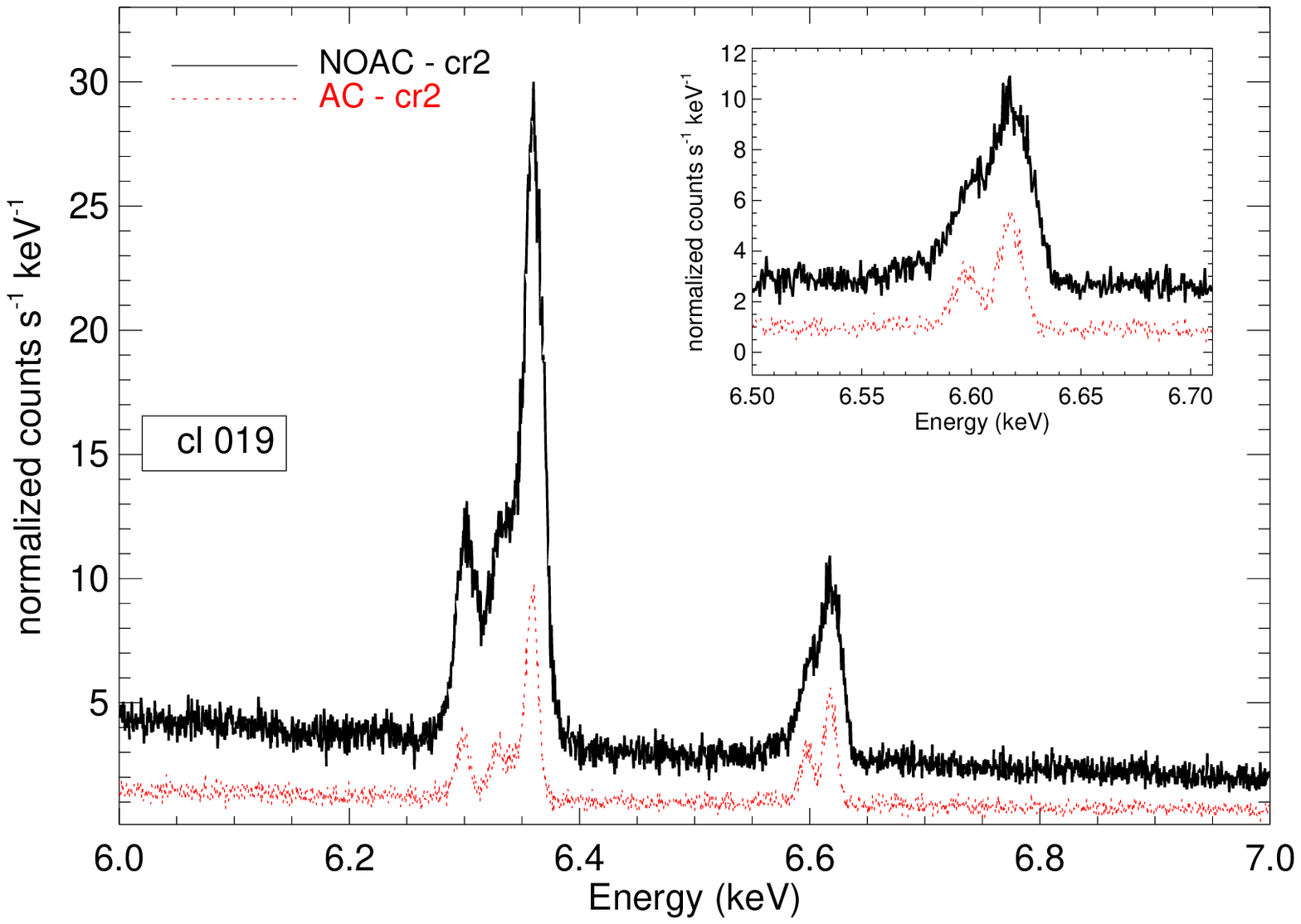}
\caption{Same as Figs.~\ref{fig:019xifu} and~\ref{fig:019_fe}, but for the two example clusters used to show results from the radiative (``cr'') runs: cl~011~(left) and cl~019~(right).}\label{fig:cv_fe}
\vspace{-0.3cm}
\end{figure*}

We find that the results from the synthetic X-ray observations of the
cooling-run clusters basically mirror the effects discussed in
Section~\ref{sec:cr}.
The major differences introduced by the AC term are substantially
smoothed out, with respect to the adiabatic case, by the account for
baryonic physical processes that dominate in shaping the cluster final
thermo-dynamical properties.
The core region observed here shows in fact that the impact of the
numerical scheme (mainly the AC term) can vary from cluster to
cluster, overall presenting no systematic formation of an entropy
core, except for few exceptions (see, e.g., \fig\ref{fig:entr_cv}).
Similarly, the suppression of the intermediate-low temperature gas
can influence differently the final X-ray spectrum depending
on the particular cluster considered.  Essentially, while such
cold-gas suppression appears to be a common characteristics also for
all the clusters in the radiative set (Section~\ref{sec:cr}), the
importance of this effect in the core can vary from case to case.  We
remind, in fact, that the innermost cluster region is more
significantly affected by cooling and star formation processes, to
which the very cold gas ($kT < 0.1\kev$) is still associated in spite
of the partial mixing and heating driven by the AC term.

More quantitatively, we find that the NOAC clusters still tend to
typically produce more (on average by $30\%$) photons than the AC
counterparts. The precise factor, however, varies
significantly across the sample, ranging from $\sim 1.1$ up to
$\sim 2.3$, with a single case where the amount of photons in the NOAC case
is even less than in the AC one.  This remarks how important are the
specific ICM properties, mainly determined by the baryonic physics.

Following Section~\ref{sec:cr}, we propose the same examples to
represent two opposite behaviours: the clusters in \fig\ref{fig:cv_fe}
show, in fact, very similar NOAC/AC mock spectra in one case (cl~011,
left) and very different ones in the other (cl~019, right).
\fig\ref{fig:cv_fe} also shows the zoom onto the $6$--$7\kev$ energy
band, comprising the iron emission lines targeted to measure the
amplitude of l.o.s.\ gas motions.
Concerning this diagnostics, we already discussed for the adiabatic
case the impossibility to draw a common conclusion for all the
clusters in the sample, given the importance of their particular
characteristics. When the cooling runs are concerned, this caveat is
still valid. Nevertheless, for the most prominent iron lines visible in
some of the cluster spectra we still generally observe a sharper line
profile in the AC case and a more broadened shape in the NOAC ones
(red-dotted and black-solid lines, respectively; lower-row panels).
Again, this can be related in some degree to the enhanced gas mixing
and heating in AC clusters, which reduces the amount of cold dense clumps
and, in turn, the turbulence driven by the hydrodynamical
instabilities originating from their interaction with the ambient ICM.

\section{Discussion}

The generation of mock X-ray observations of the simulated clusters
has further confirmed the effective impact of the particular
implementation of SPH on the physical properties of the ICM and,
consequently, on their observable features.

The main result of the different gas thermal distribution, found when
the AC term is included, is that the amount of X-ray photons produced from
the adiabatic clusters is significantly reduced with respect to the
standard-SPH runs. This is the consequence of the suppression of the
cold dense gas component, which is responsible especially for the
soft-X-ray emission.  This effect is visible already before taking
into account the projection and instrumental response, but survives
after these.

In the cooling runs, X-ray synthetic spectra confirm that the
radiative physics does play a major role, softening the differences
induced by the numerical treatment of SPH. Nevertheless, a general
decrease of the total amount of X-ray photons produced for the AC
clusters is still present, although less systematic.
Focusing the analysis on the cluster core, moreover, contributes to
milden the differences between AC and standard SPH runs, since the
central part of the cluster is particularly affected by baryonic
processes such as cooling, star formation and feedback, likely to
happen on shorter time-scales than the diffusion favoured by the
artificial conductivity.

The instrument adopted to derive mock X-ray observations, the
high-resolution spectrometer (X-IFU) designed for the up-coming Athena
mission, is chosen here for testing the possibility to constrain ICM
bulk and turbulent velocities from well-resolved bright emission lines
in X-ray spectra \cite[see also][]{biffi2013,biffi2013AN}.  From the
spectral analysis of the Iron lines in the 6--7$\kev$ band we find
velocity differences from cluster to cluster and between AC and NOAC
runs down to few tens of $\kms$.  With the high spectral resolution
achievable with next-generation of X-ray instruments such as ASTRO-H
and Athena, observational studies of this sort will become possible
and motions in the ICM will be measured with unprecedented level of
detail \cite[][]{biffi2013,bianconi2013,nagai2013,gaspari2014}.
Combining this with high spatial resolution (e.g. with Athena), the
ultimate goal will be to spatially constrain the plasma velocity field
\cite[][]{athenaEttori}.

Even though turbulent and bulk motions can
not be easily disentangled, progresses in this direction are needed in
order to observationally constrain ICM turbulence models, bracket the
possible sources of deviation from hydrostatic equilibrium and, for
instance, better understand the bias in the X-ray mass
reconstructions, likely affected by non-thermal pressure contributions and temperature inhomogeneities
\cite[e.g.][]{fang2009,lau2009,biffi2011,Rasia14}.

These observational uncertainties on the amount of turbulence present in the ICM
pose the issue to ascertain whether the numerical resolution employed in our
simulations is adequate to capture the main features of ICM turbulent motion.
In the absence of direct measurements of turbulent velocities indirect
constraints on the magnitude of ICM viscosity can be extracted by measuring
some ICM properties.

One possible method is to investigate the structure of Cold Fronts
\cite[CFs:][]{Ma07} in the sloshing scenario
\cite[][]{ZuH10,Roe11,Roe2013a,Roe2013b}, since the growth of KHI is suppressed in
presence of sufficient ICM viscosity.
Along this line of reasoning \cite{Roe2013a} investigated the structure of sloshing CFs
in the Virgo cluster, using surface brightness profiles from XMM-Newton data,
showing that for a Spitzer-like viscosity a suppression factor $f_V\sim 0.1$
is needed to avoid suppression of KHI at CFs.
The corresponding Reynolds number $Re=UL/\nu$ can be written as \cite[][]{Roe2013a}:
 \begin{equation}
Re\sim \frac{140}{f_V} \left( \frac {U}{300\kms} \right)
\left( \frac{L}{100\kpc} \right) \left( \frac{\rho_g}{10^3 \rho_c} \right)
 \left( \frac {T}{3\kev} \right)^{-5/2},
  \label{reicm.eq}
 \end{equation}
 where $U$ is the characteristic injection scale,  $V$ is the characteristic  velocity,
$\rho_g$ the gas density and $\nu\propto T^{5/2} n^{-1}$  is
  the kinematic viscosity of the medium.

Another indirect approach to probe the presence of turbulence in the ICM is to
analyze the power spectrum of gas density fluctuations
\cite[][]{Schu04,gas13,gaspari2014}.
 In a recent paper \cite{gas13}  studied the effects of thermal conduction and
turbulence in the ICM using high-resolution  hydrodynamical simulations of the
Coma cluster. By comparing the gas density  fluctuation spectrum of the
simulations against  deep Chandra observations,  they were able to put
strong upper limits on the thermal conduction suppression  factor $f_C$.
 The best accord is obtained for simulations with $f_C \simlt 10^{-3}$  and
mild subsonic turbulence with Mach number $M\sim 0.45$. For these simulations the
Reynolds number is $Re \simlt 500$.

The Reynolds number $Re_{\rm sim}$ for our SPH simulations  can be derived by equating
the AV terms to a corresponding physical Navier-Stokes viscosity
\cite[][]{PR12b,Price2012}, so to obtain
 \begin{equation}
\nu_{\rm SPH}\sim \frac{1}{10} \alpha v^{AV} h~.
  \label{nusph.eq}
 \end{equation}
 To estimate  $Re_{\rm sim}$ we  set $\alpha$
to the floor value   $\alpha_{\rm min}\sim0.01$  of the AV$_5$ run, $v^{\rm AV}$
to the sound velocity  from Eq.~(\ref{vsig.eq}) and $h$ is computed using
Eq. (\ref{hzeta.eq}). We then obtain
 \begin{equation}
Re_{\rm sim}\sim 3500 \left( \frac{M}{0.3} \right) \!\!
\left( \frac{0.01}{\alpha} \right) \!\!
\left( \frac{L}{100\kpc} \right) \!\!
\bigg( \frac{\rho_g}{10^3 \rho_c}
 \frac{10^8\msun}{m_g} \bigg)^{1/3}.
  \label{resph.eq}
 \end{equation}
A comparison between Eqs.~(\ref{reicm.eq}) and~(\ref{resph.eq}) shows that
the two Reynolds numbers are of comparable size.
Note hovewever that
setting  $\alpha_{\rm min}$  to $0.1$, as in the AV$_2$ run, would have
reduced $Re_{\rm sim}$ by an order of magnitude, whilst the simulation results
 indicate little variations between the AV$_2$ and AV$_5$ runs.

 We therefore conclude that our findings are not affected in a
significant way by the numerical resolution of the simulations.

 In particular, despite the success of our AC implementation in alleviating the
problem of standard SPH in treating the gas mixing,
our results from radiative runs
suggest that the effects seen in adiabatic clusters are less pronounced when gas cooling and star formation are accounted for. Radiative physics basically dominates the resulting ICM properties, mainly acting on time-scales smaller than those set by diffusion.
Moreover, the possibility to rely on physical conduction to overcome the cooling-flow problem seems very unlikely \cite[see, e.g.,][]{dolag2004,gas13}
and we do not expect a very significant impact due to physical thermal conductivity in real clusters.

On the numerical simulation side, the effort to investigate and
predict the observable signatures of the ICM thermal and dynamical
state carried by its X-ray emission is crucial for a faithful
comparison to observational findings.

Nonetheless, from the present analysis, it turns out that the numerical
implementation itself of the SPH method can impact the ICM state in a
way that is not negligible, as in fact the effects can be still
appreciated from the final X-ray mock observations. Namely, they must
be profound enough to survive the complication due to 2-D projection
and convolution with instrumental effects. Standard SPH, in
particular, seems to overpredict {\it per se} the X-ray emission from
the ICM and this can be visibly alleviated by introducing a
heat-diffusion term (AC), which contributes to smooth out small-scale
substructures of cold dense gas.
By increasing the entropy mixing and heating the colder gas substructures,
such diffusion term basically suppresses the sources of hydrodynamical
instabilities, mainly generated by cold clumps moving in the ambient
ICM.
The resulting damping of small-scale motion amplitude, found to be
a systematic result for all our clusters in AC-SPH (both adiabatic and
radiative) runs, is also mirrored by the different broadening of
the Fe emission lines, which typically present sharper features than
for standard SPH simulations.

Overall, the combined implementation of the time-dependent artificial
viscosity and artificial conductivity schemes points in the direction
of bridging the gap between SPH and Eulerian codes, by partially
alleviating the commonly known problems of the Lagrangian approach in
resolving the turbulent field and, simultaneously, reducing the amount
of over-abundant, cold gas clumps.

%
\section{Conclusions}\label{sec:discuss}
In this paper we have presented results extracted from a suite of
hydrodynamical simulations of galaxy clusters.
The ensemble has been constructed from various runs with different
hydrodynamical parameters, starting from the same set of initial conditions.
The code we use is an
improved SPH scheme which employs a time-dependent artificial
viscosity (AV) parameter and an artificial thermal diffusion term
(AC). The latter has been introduced with the purpose of resolving the
lack of entropy mixing which in standard SPH inhibits the growth of
instabilities. These improvements in the SPH method are motivated by
the difficulty in achieving consistent results between hydrodynamical
simulations produced by AMR and SPH codes, an issue on which recently
there has been a significant debate by many authors, as already
outlined in Section~\ref{sec:intro}.

Our study is aimed at assessing the impact on ICM properties of the
simulated clusters due to the new AC-SPH scheme.
In the following, we summarize our main results.
\begin{itemize}
\item Introducing the AC term has a significant impact of the
  thermodynamical properties of the ICM, in particular the cores of
  the simulated clusters exhibit higher entropies and temperatures
  than in the standard scheme. This is a consequence of the gas mixing
  driven by the new diffusion term.
\item In particular, for the adiabatic runs, the level of core
  entropies is now consistent with those predicted by AMR codes,
  pointing towards numerical convergence between the two approaches.
\item The temperature structure of the ICM is significantly altered in
  simulations that incorporate the AC term, with central temperatures
  higher than in the standard runs.
  We find that the cooler part ($T\lesssim 1$--$0.5\kev$) of the
  temperature distribution is strongly suppressed.
\item The time-dependent AV scheme is effective in reducing viscous
  damping of velocities associated with AV and spectral analysis shows
  at high wavenumbers velocity power spectra amplitudes which are
  higher than in the no-AV case.  However, this effect is
  significantly reduced when the AC term is introduced.  This follows
  because the gas mixing induced by the diffusion term favours
  the evaporation of cold dense substructures, which are responsible
  for the instabilities originating at the surface between the
  moving clumps and the ambient ICM and in turn for the generation of
  turbulence.
\item The morphological estimators we use to quantify the level of
  substructures do not show significant changes among different
  runs. This is consistent with the finding (see
  Figures~\ref{fig:map_exmp} and \ref{fig:map_exmp_cv}) that the
  presence of the AC term modifies the small-scale thermal structure
  of the gas, while leaving unaffected the scales probed by our estimators
  ($\gtrsim R_{2500}$).
\item For the cooling runs these findings are still present, although
  not with the strength exhibited by adiabatic simulations. In
  particular, the suppression of the cooler part of the temperature
  distribution is less pronounced and very cold blobs survive in the
  inner parts of the clusters.  This is clearly indicative that the
  time-scales set by the diffusion term are much larger than the ones
  set by the cooling rate.
\end{itemize}

About the first item, it is worth noting that the convergence between
the final levels of core entropies produced in grid-based cluster
simulations and the ones obtained with the AC-SPH code used here is
strongly indicative of the effectiveness of the adopted numerical
scheme in solving some of the inconsistencies present in standard
SPH. However, this still leaves open the issue of determining the {\it
  correct} level of central entropy which is produced in non-radiative
simulations of galaxy clusters.
In Eulerian mesh codes, there is a fluid mixing which occurs at the
cell level and is inherent to the method itself. Therefore,
a certain level of over-mixing is present in AMR simulations
\cite[][]{Springel2010b}. This suggests that the correct results lies
between the range bracketed by standard SPH and AMR.  To overcome the
difficulties present in both SPH and AMR methods,
\cite[][]{Springel2010b} proposed a new scheme in which the
hydrodynamical equations are solved on a moving unstructured mesh using
a Godunov method with an exact Riemann solver (AREPO). The new code
has been tested in a variety of test cases \cite[][]{Sijacki2012}.
To further investigate this issue, it would be interesting to directly
compare ICM entropy profiles from non-radiative cluster simulations
obtained with such different numerical codes.

The above findings are also consistent with the results presented in a
recent paper by \cite{Rasia14}, in which the authors investigate the
differences in thermal structure of the ICM in cluster sets simulated
using either the standard SPH or the AMR method. In particular,
non-radiative simulations performed with the AMR code show in the
cluster inner regions variations in temperature much smaller than in
the SPH runs.
This is due to the absence in standard SPH of gas stripping of cold
blobs and substructure evaporation owing to its limitations in
handling fluid instabilities, whilst the AC-SPH scheme has been proved
to be successful in passing the blob test \cite[see sect. 3.4
of][]{valdarnini2012}.  However, these differences are strongly
reduced when radiative losses are taken into account, since cold dense
gas is more quickly removed.

In summary, the most important aspects of our results is that in
hydrodynamical simulations of galaxy clusters, incorporating
the AC term within SPH has a direct influence on the global thermal structure of
the ICM, and not only in cluster cores.
The changes are
significant in adiabatic simulations and still present, though
with a lesser extent, in cooling runs.
We expect these improvements in the physical modeling of the ICM to be
rich of implications in many aspects which exploit results of
hydrodynamical simulations of galaxy clusters as tools to test
cosmological models.

Firstly, introducing the AC-SPH scheme to perform the simulations is
likely to have a significant impact on the so-called overcooling
problem \cite[][]{BK11}, in which the fraction of cooled gas which is
turned into stars is higher than observed. The new term is expected to
strongly reduce the survival of cold blobs in cluster inner regions
because gas stripping is much more efficient, thereby reducing the
amount of cold dense gas available to form stars.

Secondly, the amplitude of the thermal Sunyaev-Zel'dovich (tSZ) power
spectrum depends on the electron pressure profile. This will be
affected by the modifications in entropy and temperature of the ICM
expected in cluster simulations performed using the AC-SPH code. It is
not straightforward to predict the impact of these changes on the
final tSZ spectrum since the spectrum itself is redshift- as well as
scale-dependent. Nonetheless, the changes should go in the direction
of a lower spectral amplitude, since the raise in temperature is
accompanied by a decrease in central density and in turn by a lower
hot-gas pressure profile at the cluster center.
These modifications in the predictions of the tSZ power spectrum
amplitude might eventually
reduce the tension between the best-fit
values of cosmological parameters, such as $\sigma_\mathrm{8}$ and
$\Omega_\mathrm{m}$, extracted from current measurements
\cite[][]{MC14}.

Lastly, the modifications in the thermal structure of the ICM will
also affect the bias of the
X-ray mass estimator based on hydrostatic equilibrium.
\cite{Rasia14} find that at $\sim R_{500}$ the mass bias
in standard SPH simulations can be a factor $\sim 2$ larger than in
the corresponding AMR runs. We expect these differences to be
reduced in the new AC-SPH scheme.  We plan to
investigate these issues in a more systematic way in a forthcoming
paper in which we will use the new numerical scheme to construct a
large set of hydrodynamical simulations of galaxy clusters,
designed to cover more than a decade in cluster masses.

Efforts to study the role of numerics in shaping the final appearance
of simulated galaxy clusters are definitely necessary in order to
fully single out the effects purely due to the baryonic physics and
finally constrain its modelling, which is essential to match and
interpret observations of real~clusters.

\section*{Acknowledgments}
The authors acknowledge the anonymous referee for helpful comments
that contributed to improve the presentation of our results and the
quality of this paper.
%
\bibliographystyle{mn2e}
\bibliography{bibl.bib}
%

\appendix
\section{Cluster global temperatures}\label{sec:app}
In Tables~\ref{av_clu.tab} and~\ref{cv_clu.tab} we report the global
mass-weighted ($T_{\rm mw}$) temperatures of the clusters in the
sample (referring respectively to adiabatic and cooling runs), for the
regions enclosed within $R_{200}$ and $R_{2500}$.

In particular, we choose to focus on one single viscosity scheme, the
AV$_2$ run, and rather compare the standard NOAC implementation of SPH
against the AC modification.  The effects on the thermal properties of
the ICM discussed in Section~\ref{sec:results} are also reflected in
the differences for the global temperatures.
\begin{table}
\caption{\label{av_clu.tab}%
Cluster temperatures
for both standard NOAC and AC runs (with AV$_2$ viscosity scheme) --- {\it adiabatic runs} (ar2).}
\centering
\begin{tabular}{ccc}
\hline
index & $T_{200,\rm mw}$\,[keV] & $T_{2500,\rm mw}$\,[keV] \\
      &  NOAC / AC &  NOAC / AC \\
\hline
       1 & 4.01 / 4.15 & 5.02 / 6.26 \\
       5 & 1.82 / 1.90 & 2.25 / 2.68 \\
      11 & 4.20 / 4.32 & 5.35 / 5.85 \\
      13 & 3.13 / 3.24 & 3.84 / 4.11 \\
      16 & 1.75 / 1.92 & 2.14 / 2.55 \\
      19 & 4.49 / 4.58 & 5.51 / 6.28 \\
     105 & 0.98 / 1.02 & 1.24 / 1.32 \\
     110 & 1.35 / 1.34 & 1.71 / 1.71 \\
\hline
\end{tabular}
\caption{\label{cv_clu.tab}
Same as Table~\ref{av_clu.tab} but for {\it radiative runs} (cr2).}
\begin{tabular}{ccc}
\hline
index & $T_{200,\rm mw}$\,[keV] & $T_{2500,\rm mw}$\,[keV] \\
      &  NOAC / AC &  NOAC / AC \\
\hline
       1 & 4.20 / 4.47 & 5.74 / 6.42 \\
       5 & 2.08 / 2.19 & 2.92 / 3.09 \\
      11 & 4.58 / 4.79 & 6.55 / 7.11 \\
      13 & 3.37 / 3.52 & 4.66 / 5.03 \\
      16 & 1.94 / 2.11 & 2.70 / 3.20 \\
      19 & 4.81 / 4.89 & 6.47 / 7.01 \\
     105 & 1.07 / 1.14 & 1.49 / 1.60 \\
     110 & 1.49 / 1.51 & 2.13 / 2.26 \\
\hline
\end{tabular}
\end{table}



\bsp

\label{lastpage}
\end{document}